\pdfoutput=1
\def\notikzex{1}
\def\twocolumnmode{1}
\def\usecentralfigs{1}

\documentclass[%
  aip,%
  jcp,%
  amsmath,%
  amsfonts,%
  amssymb,%
  citeautoscript,%
  \ifdefined\twocolumnmode
    reprint,%
  \else
    preprint,%
  \fi
  floatfix,%
]{revtex4-2}

\usepackage[version=4]{mhchem}

\usepackage{siunitx}
\DeclareSIUnit{\atomicunit}{a.u.}

\usepackage{%
  amsthm,%
  amssymb,%
  mathrsfs,%
  mathtools,%
  gensymb,%
  braket,%
  mleftright,%
  nicefrac,%
  interval%
}
\intervalconfig{%
  soft open fences%
}

\mleftright

\usepackage{bm}

\usepackage[usenames, svgnames, x11names, rgb]{xcolor}

\usepackage{graphicx}
\usepackage{grffile}

\usepackage{pgfplots}

\usetikzlibrary{external}
\newif\iftikzex
\ifdefined\notikzex
  \tikzexfalse
\else
  \tikzextrue
\fi

\iftikzex
  \pgfplotsset{compat=1.17}

  \usetikzlibrary{
    calc,
    luamath,
    positioning,
    decorations.pathreplacing,
    backgrounds,
    arrows.meta,
    pgfplots.groupplots,
    pgfplots.patchplots,
    3d,
  }

  \usepgfplotslibrary[colormaps]

  \pgfplotscreateplotcyclelist{coloronly}{
    {red},
    {blue},
    {black!60!green},
    {black!20!orange},
    {black!30!blue},
    {green!30!brown},
    {blue!40!red},
    {IndianRed},
    {black!40!yellow},
    {Magenta4},
    {DodgerBlue4},
    {Maroon4},
    {Firebrick4},
    {Cyan4},
  }
\fi
\tikzexfalse

\makeatletter
\newcommand*{\useexternalfile}[5]{
  \iftikzex
    \tikzsetnextfilename{tikzoutput/#4-output}
    \scalebox{#1}{\input{\tikzexternal@filenameprefix#4.tikz.tex}}
  \else
    \ifdefined\usecentralfigs
      \includegraphics[scale=#1, trim=#2 0 #3 0]{./central-figures/#5.pdf}
    \else
      \includegraphics[scale=#1, trim=#2 0 #3 0]{\tikzexternal@filenameprefix tikzoutput/#4-output.pdf}
    \fi
  \fi
}
\makeatother

\usepackage{xifthen}

\usepackage[caption=false]{subfig}

\usepackage{%
  booktabs,%
  array,%
  adjustbox,%
  makecell,%
  multirow,%
  bigdelim%
}

\usepackage[inline]{enumitem}

\theoremstyle{plain}
\newtheorem{proposition}{Proposition}
\theoremstyle{remark}
\newtheorem*{remark}{Remark}

\usepackage[%
  symbols,%
  abbreviations,%
  nomain,%
  nonumberlist,%
  nogroupskip,%
  nopostdot%
]{glossaries-extra}

\setabbreviationstyle{long-short}

\glsaddkey
  {hyphenated}        
  {\relax}            
  {\glsentryhyphx}    
  {\Glsentryhyphx}    
  {\glshyphx}         
  {\Glshyphx}         
  {\GLShyphx}         
\newcommand{\GENglspostlinkhook}{%
  \ifglsused{\glslabel}{}{ (\glsentryshort{\glslabel})}\glsunset \glslabel}
\makeatletter
\newcommand\metadef[1]{%
  \expandafter\newcommand\csname gls#1\endcsname{%
    \@ifstar{\csname sgls#1\endcsname}{\csname ngls#1\endcsname}%
  }
  \@namedef{sgls#1}##1{{\let\glspostlinkhook \GENglspostlinkhook\expandafter\csname gls#1x\endcsname*{##1}}}%
  \@namedef{ngls#1}##1{{\let\glspostlinkhook \GENglspostlinkhook\expandafter\csname gls#1x\endcsname{##1}}}%
  \expandafter\newcommand\csname Gls#1\endcsname{%
    \@ifstar{\csname sGls#1\endcsname}{\csname nGls#1\endcsname}%
  }
  \@namedef{sGls#1}##1{{\let\glspostlinkhook \GENglspostlinkhook\expandafter\csname Gls#1x\endcsname*{##1}}}%
  \@namedef{nGls#1}##1{{\let\glspostlinkhook \GENglspostlinkhook\expandafter\csname Gls#1x\endcsname{##1}}}%
  \expandafter\newcommand\csname GLS#1\endcsname{%
    \@ifstar{\csname sGLS#1\endcsname}{\csname nGLS#1\endcsname}%
  }
  \@namedef{sGLS#1}##1{{\let\glspostlinkhook \GENglspostlinkhook\expandafter\csname GLS#1x\endcsname*{##1}}}%
  \@namedef{nGLS#1}##1{{\let\glspostlinkhook \GENglspostlinkhook\expandafter\csname GLS#1x\endcsname{##1}}}%
}
\makeatother

\metadef{hyph}

\DeclareMathOperator{\tr}{tr}
\DeclareMathOperator{\id}{id}
\DeclareMathOperator{\Ln}{Ln}
\DeclareMathOperator{\Arg}{Arg}
\newcommand*{\T}{\mathsf{T}}
\newcommand*{\D}{\mathrm{d}}

\usepackage[%
  hyperfootnotes=false,%
  hypertexnames=false,%
  bookmarksnumbered=true,%
  colorlinks=true,%
  citecolor=teal,%
  linkcolor=Blue4,%
]{hyperref}

\newglossaryentry{symbcat:gen}{%
  name = {General},%
  description = {\nopostdesc},%
  sort = {A.gen},%
}
\newglossaryentry{gen:Ne}{%
  name = {$N_{\mathrm{e}}$},%
  text = {N_{\mathrm{e}}},%
  description = {total number of electrons},%
  sort = {1Ne},%
  parent = symbcat:gen,%
  type = symbols,%
}
\newglossaryentry{gen:Nn}{%
  name = {$N_{\mathrm{n}}$},%
  text = {N_{\mathrm{n}}},%
  description = {total number of nuclei},%
  sort = {2Nnuc},%
  parent = symbcat:gen,%
  type = symbols,%
}
\newglossaryentry{gen:Ndet}{%
  name = {$N_{\mathrm{det}}$},%
  text = {N_{\mathrm{det}}},%
  description = {number of determinants in a set},%
  sort = {3Ndet},%
  parent = symbcat:gen,%
  type = symbols,%
}
\newglossaryentry{gen:Nindept}{%
  name = {$N_{\mathrm{indept}}$},%
  text = {N_{\mathrm{indept}}},%
  description = {number of linearly independent wavefunctions in a set},%
  sort = {4Nindept},%
  parent = symbcat:gen,%
  type = symbols,%
}
\newglossaryentry{symbcat:struct}{%
  name = {Structures},%
  description = {\nopostdesc},%
  sort = {B.struct},%
}
\newglossaryentry{struct:hilbert}{%
  name = {$\mathcal{H}$},%
  text = {\mathcal{H}},%
  description = {generic Hilbert space},%
  sort = {1H},%
  parent = symbcat:struct,%
  type = symbols,%
}
\newglossaryentry{struct:hilbert1e}{%
  name = {$\gls*{struct:hilbert}[_1]$},%
  text = {\gls*{struct:hilbert}[_1]},%
  description = {one-electron complex Hilbert space},%
  sort = {2H1e},%
  parent = symbcat:struct,%
  type = symbols,%
}
\newglossaryentry{struct:hilbertNe}{%
  name = {$\gls*{struct:hilbert}[_{\gls*{gen:Ne}}]$},%
  text = {\gls*{struct:hilbert}[_{\gls*{gen:Ne}}]},%
  description = {$\gls*{gen:Ne}$-electron complex Hilbert space},%
  sort = {3HNe},%
  parent = symbcat:struct,%
  type = symbols,%
}
\newglossaryentry{struct:hilbertNep}{%
  name = {$\gls*{struct:hilbert}'_{\gls*{gen:Ne}}$},%
  text = {\gls*{struct:hilbert}'_{\gls*{gen:Ne}}},%
  description = {$\gls*{gen:Ne}$-electron complex Hilbert space},%
  sort = {3HNe},%
  parent = symbcat:struct,%
  type = symbols,%
}
\newglossaryentry{struct:rep}{%
  name = {$\Gamma$},%
  text = {\Gamma},%
  description = {subrepresentation of $\gls*{struct:hilbertNe}$},%
  sort = {4group0.gamma},%
  parent = symbcat:struct,%
  type = symbols,%
}
\newglossaryentry{struct:gengroup}{%
  name = {$\mathcal{G}$},%
  text = {\mathcal{G}},%
  description = {generic group},%
  sort = {4group1.gen},%
  parent = symbcat:struct,%
  type = symbols,%
}
\newglossaryentry{struct:pointgroup}{%
  name = {$\mathcal{B}$},%
  text = {\mathcal{B}},%
  description = {generic spatial point group},%
  sort = {4group2.spatialpoint},%
  parent = symbcat:struct,%
  type = symbols,%
}
\newglossaryentry{struct:spinrotgroup}{%
  name = {$\mathcal{S}$},%
  text = {\mathcal{S}},%
  description = {spin rotation group, identical to $\symsfup{SU}(2)$},%
  sort = {4group3.spinrot},%
  parent = symbcat:struct,%
  type = symbols,%
}
\newglossaryentry{struct:timerevgroup}{%
  name = {$\mathcal{T}$},%
  text = {\mathcal{T}},%
  description = {time-reversal group},%
  sort = {4group4.timerev},%
  parent = symbcat:struct,%
  type = symbols,%
}
\newglossaryentry{struct:timerevdoublegroup}{%
  name = {$\mathcal{T}^*$},%
  text = {\mathcal{T}^*},%
  description = {time-reversal double group},%
  sort = {4group5.timerevdouble},%
  parent = symbcat:struct,%
  type = symbols,%
}
\newglossaryentry{symbcat:op}{%
  name = {Operators},%
  description = {\nopostdesc},%
  sort = {C.op},%
}
\newglossaryentry{op:hamil}{%
  name = {$\hat{\mathscr{H}}$},%
  text = {\hat{\mathscr{H}}},%
  description = {electronic Hamiltonian},%
  sort = {11H},%
  parent = symbcat:op,%
  type = symbols,%
}
\newglossaryentry{op:holoconj}{%
  name = {$\hat{\kappa}$},%
  text = {\hat{\kappa}},%
  description = {holomorphic conjugation},%
  sort = {31conjugation},%
  parent = symbcat:op,%
  type = symbols,%
}
\newglossaryentry{op:complexconj}{%
  name = {$\hat{K}$},%
  text = {\hat{K}},%
  description = {complex conjugation},%
  sort = {33conjugation},%
  parent = symbcat:op,%
  type = symbols,%
}
\newglossaryentry{op:wildcard}{%
  name = {$\lozenge$},%
  text = {\lozenge},%
  description = {wildcard operator for conventional/holomorphic theories},%
  sort = {34wildcard},%
  parent = symbcat:op,%
  type = symbols,%
}
\newglossaryentry{op:perm}{%
  name = {$\hat{P}$},%
  text = {\hat{P}},%
  description = {electron label permutation},%
  sort = {41P},%
  parent = symbcat:op,%
  type = symbols,%
}
\newglossaryentry{op:timerev}{%
  name = {$\hat{\Theta}$},%
  text = {\hat{\Theta}},%
  description = {time reversal},%
  sort = {52T},%
  parent = symbcat:op,%
  type = symbols,%
}
\newglossaryentry{op:piyspinrot}{%
  name = {$\hat{Q}$},%
  text = {\hat{Q}},%
  description = {spin rotation through $\pi$ about the $y$-axis},%
  sort = {62q},%
  parent = symbcat:op,%
  type = symbols,%
}
\newglossaryentry{symbcat:wf}{%
  name = {Wavefunctions},%
  description = {\nopostdesc},%
  sort = {D.wf},%
}
\newglossaryentry{wf:gen}{%
  name = {$\Psi$},%
  text = {\Psi},%
  description = {generic electronic wavefunction},%
  sort = {1Psi},%
  parent = symbcat:wf,%
  type = symbols,%
}
\newglossaryentry{wf:gencc}{%
  name = {$\bar{\Psi}$},%
  text = {\bar{\Psi}},%
  description = {complex-conjugated generic electronic wavefunction},%
  sort = {1Psi},%
  parent = symbcat:wf,%
  type = symbols,%
}
\newglossaryentry{wf:gentilde}{%
  name = {$\tilde{\Psi}$},%
  text = {\tilde{\Psi}},%
  description = {\nopostdesc},%
  sort = {1Psitilde},%
  parent = symbcat:wf,%
  type = symbols,%
}
\newglossaryentry{wf:exact}{%
  name = {$\Psi_{\mathrm{exact}}$},%
  text = {\Psi_{\mathrm{exact}}},%
  description = {exact electronic wavefunction},%
  sort = {2Psiexact},%
  parent = symbcat:wf,%
  type = symbols,%
}
\newglossaryentry{wf:det}{%
  name = {$\Psi_{\mathrm{det}}$},%
  text = {\Psi_{\mathrm{det}}},%
  description = {single-determinantal electronic wavefunction},%
  sort = {4Psidet},%
  parent = symbcat:wf,%
  type = symbols,%
}
\newglossaryentry{wf:hf}{%
  name = {$\Psi^{\mathrm{HF}}$},%
  text = {\Psi^{\mathrm{HF}}},%
  description = {Hartree--Fock electronic wavefunction},%
  sort = {5Psihf},%
  parent = symbcat:wf,%
  type = symbols,%
}
\newglossaryentry{wf:uhf}{%
  name = {$\Psi^{\mathrm{UHF}}$},%
  text = {\Psi^{\mathrm{UHF}}},%
  description = {unrestricted Hartree--Fock electronic wavefunction},%
  sort = {6Psiuhf},%
  parent = symbcat:wf,%
  type = symbols,%
}
\newglossaryentry{symbcat:bas}{%
  name = {Basis Expansion},%
  description = {\nopostdesc},%
  sort = {E.bas},%
}
\newglossaryentry{bas:ssbasis}{%
  name = {$\Sigma$},%
  text = {\Sigma},%
  description = {spin-orbital direct-product basis set spanning $\gls*{struct:hilbert1e}$},%
  sort = {11ssbasis},%
  parent = symbcat:bas,%
  type = symbols,%
}
\newglossaryentry{bas:spinspatialcoord}{%
  name = {$\boldsymbol{x}$},%
  text = {\boldsymbol{x}},%
  description = {spin-spatial coordinates},%
  sort = {14spinspatialcoord},%
  parent = symbcat:bas,%
  type = symbols,%
}
\newglossaryentry{bas:spincoord}{%
  name = {$s$},%
  text = {s},%
  description = {spin coordinates},%
  sort = {15spincoord},%
  parent = symbcat:bas,%
  type = symbols,%
}
\newglossaryentry{bas:spatialcoord}{%
  name = {$\boldsymbol{r}$},%
  text = {\boldsymbol{r}},
  description = {spatial coordinates},%
  sort = {16spatialcoord},%
  parent = symbcat:bas,%
  type = symbols,%
}
\newglossaryentry{bas:spinbasis}{%
  name = {$\omega\left(\gls*{bas:spincoord}\right)$},%
  text = {\omega},%
  description = {spin basis function},%
  sort = {12spinbasis},%
  parent = symbcat:bas,%
  type = symbols,%
}
\newglossaryentry{bas:spatialbasis}{%
  name = {$\varphi\left(\gls*{bas:spatialcoord}\right)$},%
  text = {\varphi},%
  description = {spatial basis function},%
  sort = {13spatialbasis},%
  parent = symbcat:bas,%
  type = symbols,%
}
\newglossaryentry{bas:spinorb}{%
  name = {$\chi\left(\gls*{bas:spinspatialcoord}\right)$},%
  text = {\chi},%
  description = {one-electron spin-orbital},%
  sort = {17spinorb},%
  parent = symbcat:bas,%
  type = symbols,%
}
\newglossaryentry{bas:spinorbcc}{%
  name = {$\bar{\chi}\left(\gls*{bas:spinspatialcoord}\right)$},%
  text = {\bar{\chi}},%
  description = {complex-conjugated one-electron spin-orbital},%
  sort = {17spinorb},%
  parent = symbcat:bas,%
  type = symbols,%
}
\newglossaryentry{bas:spinorbbar}{%
  name = {$\bar{\chi}\left(\gls*{bas:spinspatialcoord}\right)$},%
  text = {\bar{\chi}},%
  description = {spin-flipped one-electron spin-orbital},%
  sort = {17spinorb},%
  parent = symbcat:bas,%
  type = symbols,%
}
\newglossaryentry{bas:spinpart}{%
  name = {$\tau\left(\gls*{bas:spincoord}\right)$},%
  text = {\tau},%
  description = {spin part of a one-electron spin-orbital},%
  sort = {18spinpart},%
  parent = symbcat:bas,%
  type = symbols,%
}
\newglossaryentry{bas:spatialpart}{%
  name = {$\psi\left(\gls*{bas:spatialcoord}\right)$},%
  text = {\psi},%
  description = {spatial part of a one-electron spin-orbital},%
  sort = {19spatialpart},%
  parent = symbcat:bas,%
  type = symbols,%
}
\newglossaryentry{bas:atomicorb}{%
  name = {$\tilde{\varphi}\left(\gls*{bas:spatialcoord}\right)$},%
  text = {\tilde{\varphi}},%
  description = {atomic-orbital basis function},%
  sort = {21atomicorb},%
  parent = symbcat:bas,%
  type = symbols,%
}
\newabbreviation[sort={1.ao}, hyphenated={atomic-orbital}]{acr:ao}{AO}{atomic orbital}
\newabbreviation[sort={1.mo}, hyphenated={molecular-orbital}]{acr:mo}{MO}{molecular orbital}
\newabbreviation[sort={2.0.scf}]{acr:scf}{SCF}{self-consistent-field}
\newabbreviation[sort={2.0.csf}]{acr:csf}{CSF}{configuration state function}
\newabbreviation[sort={2.1.ci}]{acr:ci}{CI}{configuration interaction}
\newabbreviation[sort={2.1.cis}]{acr:cis}{CIS}{single-excitation configuration interaction}
\newabbreviation[sort={2.2.noci}]{acr:noci}{NOCI}{non-orthogonal configuration interaction}
\newabbreviation[sort={2.3.casscf}]{acr:casscf}{CASSCF}{complete active space self-consistent-field}
\newabbreviation[sort={2.3.casscf-sa}]{acr:sacasscf}{SA-CASSCF}{state-averaged complete active space self-consistent-field}
\newabbreviation[sort={2.3.rasscf}]{acr:rasscf}{RASSCF}{restricted active space self-consistent-field}
\newabbreviation[sort={2.4.caspt2}]{acr:caspt2}{CASPT2}{complete active space perturbation theory}
\newabbreviation[sort={2.4.raspt2}]{acr:raspt2}{RASPT2}{restricted active space perturbation theory}
\newabbreviation[sort={2.4.nevpt2}]{acr:nevpt2}{NEVPT2}{$n$-electron valence-state perturbation theory}
\newabbreviation[sort={2.5.mp2}]{acr:mp2}{MP2}{M\o{}ller--Plesset second-order perturbation theory}
\newabbreviation[sort={2.5.nocimp2}]{acr:nocimp2}{NOCI-MP2}{M\o{}ller--Plesset second-order non-orthogonal configuration interaction}
\newabbreviation[sort={2.5.nocipt2}]{acr:nocipt2}{NOCI-PT2}{non-orthogonal configuration interaction perturbation theory}
\newabbreviation[sort={3.0.hf}]{acr:hf}{HF}{Hartree--Fock}
\newabbreviation[sort={3.1.ghf}]{acr:ghf}{GHF}{generalised Hartree--Fock}
\newabbreviation[sort={3.2.cghf}]{acr:cghf}{cGHF}{complex generalised Hartree--Fock}
\newabbreviation[sort={3.3.uhf}]{acr:uhf}{UHF}{unrestricted Hartree--Fock}
\newabbreviation[sort={3.4.rhf}]{acr:rhf}{RHF}{restricted Hartree--Fock}
\newabbreviation[sort={3.5.tdhf}]{acr:tdhf}{TDHF}{time-dependent Hartree--Fock}
\newabbreviation[sort={4.0.dft}]{acr:dft}{DFT}{Density Functional Theory}
\newabbreviation[sort={4.1.ks}]{acr:ks}{KS}{Kohn--Sham}
\newabbreviation[sort={4.1.tddft}]{acr:tddft}{TDDFT}{time-dependent Density Functional Theory}
\newabbreviation[sort={5.svd}]{acr:svd}{SVD}{singular value decomposition}
\newabbreviation[sort={5.diis}]{acr:diis}{DIIS}{direct inversion in the iterative subspace}
\newabbreviation[sort={6.bkk}]{acr:bkk}{BKK}{Bernstein--Khovanskii--Kushnirenko}
\newabbreviation[sort={7.adgp}]{acr:adgp}{aDGP}{assigned Distance Geometry Problem}

\begin{document}


\title{On Symmetry and the Reality of Holomorphic Hartree--Fock Wavefunctions}

\author{Bang C. Huynh}
  \email{bang.huynh@nottingham.ac.uk}
  \affiliation{School of Chemistry, University of Nottingham, University Park, Nottingham NG7 2RD, United Kingdom}
  \affiliation{Yusuf Hamied Department of Chemistry, Lensfield Road, Cambridge CB2 1EW, United Kingdom}
\author{Alex J. W. Thom}
  \affiliation{Yusuf Hamied Department of Chemistry, Lensfield Road, Cambridge CB2 1EW, United Kingdom}

\date{\today}


\begin{abstract}
  The coalescence and disappearance of Hartree--Fock (HF) solutions as the molecular structure varies have been a common source of criticism for the breakdown of the HF approximation to the potential energy surfaces.
  However, recent developments in holomorphic HF theory show that this disappearing behavior is only a manifestation of the way conventional HF equations prevent solutions from being analytically continued, but it is unclear what factors govern the existence and the locations of these disappearances.
  In this work, we explore some of these factors from the perspective of spatial symmetry by introducing a classification for symmetry constraints on electronic-structure calculations.
  This forms a framework for us to systematically investigate several analytic holomorphic HF solutions of a model \ce{[H4]^{2+}} system in STO-3G and demonstrate that, under appropriate conditions, spatial symmetry imposes strict requirements on the reality of certain solutions.
  The implications for self-consistent-field HF search algorithms are then discussed.
  Throughout this article, the term \emph{reality} means the quality of a holomorphic HF solution having real molecular orbitals.
\end{abstract}

\maketitle 

\tikzsetexternalprefix{./introduction/tikz/}

\section{Introduction}
  \label{sec:intro}

  It has long been known that the non-linearity of the \gls*{acr:hf} equations necessitates the existence of multiple \gls*{acr:scf} solutions\cite{article:Mjolsness1968, article:Stanton1968, article:King1969, article:Redondo1989, article:Pulay1990}.
  Over the last decade, these multiple \gls*{acr:scf} solutions have re-captured a fair amount of our and some others' attention, for they are believed to present a feasible alternative pathway to recover electron correlation in ground and excited states\cite{article:Thom2009b, article:Sundstrom2014b, article:Mayhall2014, article:Jensen2018, article:Huynh2020}.
  However, since chemistry is hardly ever static, it is imperative that the behaviors of these \gls*{acr:scf} solutions must be followed as the underlying system configuration changes, so that the nature of the potential energy surfaces generated by them can be understood.

  In a number of our previous studies, we investigated the behaviors of these solutions upon the variation of molecular geometry or along certain electron-transfer reaction trajectories as part of an attempt to characterize and understand their properties\cite{article:Thom2008, article:Jensen2018, article:Huynh2020}.
  In more than one instance, we observed that the multiple \gls*{acr:hf} solutions that are close to each other in energy can become energetically degenerate at certain points in the configuration space.
  These points of degeneracy are of particular interest to us.
  If, at one of these points, the solutions remain physically distinct (\textit{i.e.}, they differ from one another by more than a phase factor), then they are connected by some symmetry operations of the underlying symmetry group of the system.
  On the other hand, if they become identical at this point, we say that they have \emph{coalesced}.

  The simplest and most well-known example for the coalescence of solutions is encountered in the stretching of \ce{H2} where the lowest doubly-degenerate and spin-symmetry-broken \gls*{acr:uhf} solutions coalesce with the lowest non-degenerate \gls*{acr:rhf} solution at \ce{H}---\ce{H} bond length of around \SI{1.20}{\angstrom}\cite{article:Coulson1949, article:Burton2016}.
  From the point of coalescence, if one follows the coalescing solutions in different directions, one can expect to get different behaviors depending on the nature of the solutions.
  In particular, the lowest doubly-degenerate \gls*{acr:uhf} solutions in \ce{H2} exist at bond lengths larger than \SI{1.20}{\angstrom} but fail to be located by conventional methods below \SI{1.20}{\angstrom}.
  On the other hand, the lowest non-degenerate \gls*{acr:rhf} solution persists at all bond lengths.
  We thus say that the \gls*{acr:uhf} solutions have \emph{disappeared} from the conventional \gls*{acr:scf} landscape past the coalescence point at \SI{1.20}{\angstrom} as the \ce{H}---\ce{H} bond length decreases.

  When the conventional theory of \gls*{acr:hf} is reformulated in such a way that the \gls*{acr:hf} equations become holomorphic\cite{article:Hiscock2014}, it turns out that the disappearing solutions can be analytically continued past the coalescence points, although the \glshyph*{acr:mo} coefficients will have to become non-real\cite{article:Burton2016}.
  In the framework of holomorphic \gls*{acr:hf} theory, the points of coalescence therefore mark the configurations at which certain \gls*{acr:hf} solutions switch from being real solutions of both conventional and holomorphic \gls*{acr:hf} equations to being non-real solutions of the latter only.
  However, the fundamental factors that govern the existence of such coalescence points are not well understood.
  Without a knowledge of where these points are located in the \gls*{acr:scf} landscape, one cannot predict when certain \gls*{acr:scf} \gls*{acr:hf} solutions, perhaps of significant chemical importance, have been missed out by \gls*{acr:scf} search algorithms that are based only on conventional \gls*{acr:hf} theory.
  And more importantly, the lack of a satisfactory account for the occurrence of such coalescences renders them to be unfairly considered as shortfalls of the \gls*{acr:hf} theory\cite{article:Mori-Sanchez2014a} rather than features that are the consequences of the theory itself.
  In fact, Fukutome notes in a detailed investigation of the \gls*{acr:uhf} theory for chemical reactions\cite{article:Fukutome1973} that the disappearance of \gls*{acr:uhf} solutions upon the system becoming asymmetrical constitutes part of a promising theoretical basis to understand the mechanism underlying radical \textit{vs.} ionic character of chemical reactions.
  Unfortunately, to the best of our knowledge, there has not been much further effort along this line of inquiry since Fukutome's work.

  The \ce{H2} stretching example above is useful because its simplicity enables a straightforward analytic description of the coalescence point\cite{article:Coulson1949}.
  However, it is \emph{too} simple for any useful patterns to be elucidated, since the linear geometry of the diatomic molecule means that the configuration space is only one-dimensional and that the molecular symmetry remains $\mathcal{D}_{\infty h}$ throughout---the position of the coalescence point therefore appears rather random in connection to the structure of \ce{H2}.
  Nevertheless, Figure~5 in Ref.~\citenum{article:Burton2018} reveals that, if one of the two hydrogen nuclei is replaced by a fictitious nucleus \ce{Z} with a variable nuclear charge $Q_{\ce{Z}}$ such that the true \ce{H2} molecule corresponds to $Q_{\ce{Z}} = \SI{1.00}{\atomicunit}$, then, as $Q_{\ce{Z}}$ deviates from this value, the symmetry of the system descends from $\mathcal{D}_{\infty h}$ to $\mathcal{C}_{\infty v}$ and some of the \gls*{acr:rhf} solutions that exist at $Q_{\ce{Z}} = \SI{1.00}{\atomicunit}$ begin to coalesce with each other and disappear.

\begin{figure*}[h]
    \centering
    \ifdefined\twocolumnmode
      \useexternalfile{1}{44.6802}{11.22185}{H42p.STO-3G.UHF.a1.058350}{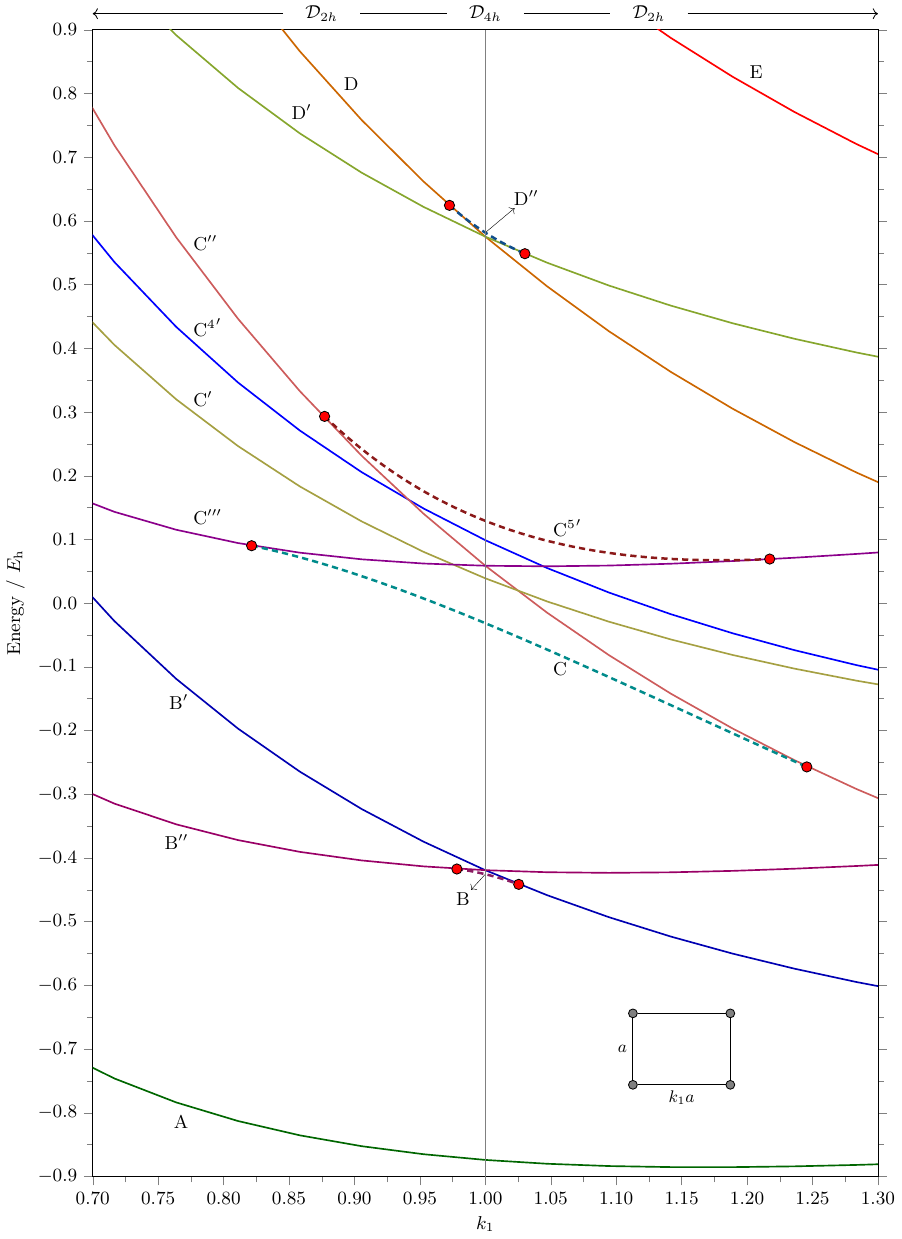}
    \else
      \useexternalfile{0.85}{44.6802}{11.22185}{H42p.STO-3G.UHF.a1.058350}{figure1}
    \fi
    \caption{%
      Low-lying $M_S = 0$ conventional \gls*{acr:uhf} solutions of \ce{[H4]^{2+}} (STO-3G) in the vicinity of square geometry for $a = \SI{1.058350}{\angstrom}$.
      Solutions that persist as the rectangular factor $k_1$ varies are represented by solid curves.
      Solutions that coalesce with others and disappear as $k_1$ deviates from unity are shown with dashed curves.
      Coalescence points are highlighted with red dots.
      Solutions are labeled alphabetically in ascending order of their energy at $\mathcal{D}_{4h}$.
      Degenerate or nearly degenerate solutions at $\mathcal{D}_{4h}$ share the same letter but are distinguished by dashes.
    }
    \label{fig:h42p-prelim}
  \end{figure*}

  This simple observation inspires the conjecture that molecular symmetry plays a role in dictating the existence of \gls*{acr:hf} solutions that would otherwise be non-locatable in the conventional \gls*{acr:scf} \gls*{acr:hf} formalism.
  In fact, this conjecture is further strengthened by the observations of similar coalescing and disappearing behaviors of multiple \gls*{acr:rhf} and \gls*{acr:uhf} solutions in the vicinity of high-symmetry molecular structures in other systems.
  The simplest non-trivial example that we consider is the side-way compression and elongation of the hypothetical square two-electron dication \ce{[H4]^{2+}} such that the molecular symmetry interchanges between $\mathcal{D}_{4h}$ and $\mathcal{D}_{2h}$.
  We show in Figure~\ref{fig:h42p-prelim} all low-lying $M_S = 0$ conventional \gls*{acr:uhf} solutions of \ce{[H4]^{2+}} located numerically in STO-3G using \gls*{acr:scf} metadynamics\cite{article:Thom2008} in \textsc{Q-Chem} 5.3\cite{article:Epifanovsky2021}, focusing particularly on the coalescence and disappearance of certain solutions in the vicinity of the square geometry.

  There exist more complicated examples still.
  One of them concerns the \gls*{acr:rhf} solutions in the four-electron \ce{H4} as the square four-membered ring opens up into an isosceles trapezium and the point group of the system descends from $\mathcal{D}_{4h}$ to $\mathcal{C}_{2v}$ (Figure~2 in Ref.~\citenum{article:Thom2008}).
  Another more complicated example involves the \gls*{acr:rhf} solutions in ethene as the planar molecule undergoes a torsional twist about the \ce{C}=\ce{C} bond and the molecular symmetry lowers from $\mathcal{D}_{2h}$ to $\mathcal{D}_2$ (Figure~7a in Ref.~\citenum{article:Burton2018}).
  And yet another example that is even more complex involves the lowest-lying \gls*{acr:uhf} solutions upon the Jahn--Teller distortion of the octahedral \ce{[TiF6]^{3-}} anions into $\mathcal{D}_{4h}$ or $\mathcal{D}_{2h}$ geometries along the $e_g$ normal vibrational coordinates (Figure~10 in Ref.~\citenum{article:Huynh2020}).
  In all of the above examples, we notice that there are two kinds of solutions: those of the first kind persist at all geometries along the tracking path (\textit{e.g.}, the solutions corresponding to the solid curves in Figure~\ref{fig:h42p-prelim}), and those of the second kind coalesce with other solutions before ceasing to be locatable by conventional \gls*{acr:scf} searches as the system descends in molecular symmetry (\textit{e.g.}, the solutions represented by the dashed curves in Figure~\ref{fig:h42p-prelim}).

  The above empirically observed patterns of solution coalescence and disappearance near high-symmetry configurations in a variety of systems with very different structures inspire the investigation into the roles played by molecular symmetry in controlling the reality of solutions to the holomorphic \gls*{acr:hf} equations.
  A completely general approach requires results from polynomial theories to determine bounds on real solutions of systems of multivariate polynomials\cite{article:Bihan2011, book:Sottile2011}.
  Unfortunately, this is a challenging task to accomplish for arbitrary systems where it can be daunting to work out how molecular symmetry affects the structural complexity of the \gls*{acr:hf} equations.
  This leads us to believe that, as an initial investigation, it is more revealing to seek and study extensively a model system that exhibits such behaviors through an analytic approach.
  We thus require that this model system is simple enough such that the relationship between molecular symmetry and the structural complexity of the \gls*{acr:hf} equations can be elucidated, and that analytic solutions can be obtained easily and examined thoroughly.
  However, the model system must be sufficiently complex in order for molecular symmetry to be non-trivial in the sense that the nuclear framework can assume a number of point-group symmetries that are inter-convertible via well-defined pathways.

  To this end, we choose the hypothetical dication \ce{[H4]^{2+}} in a minimal basis set STO-3G: there are only two electrons in four \glshyph*{acr:ao} basis functions, offering a maximum of six degrees of freedom (after accounting for normalization), and the possible high symmetries that can be adopted by the nuclear framework and that are of interest to us are $\mathcal{T}_d$, $\mathcal{D}_{4h}$, $\mathcal{D}_{2h}$, and $\mathcal{D}_2$, all of which can be easily inter-converted.
  Furthermore, as pointed out earlier, the numerically located \gls*{acr:uhf} $M_S = 0$ solutions for \ce{[H4]^{2+}} in STO-3G plotted in Figure~\ref{fig:h42p-prelim} show that this system does indeed exhibit the conjectured behaviors around the $\mathcal{D}_{4h}$ geometry along the sideways compression/elongation pathway.
  The analytic model we employ for this system can therefore be expected to provide enough richness to cast some light on the roles of symmetry in dictating the observed behaviors while remaining tractable so that the interpretation and analysis of the analytic \gls*{acr:hf} solutions do not become impossibly complicated.

  Before delving into the analytic details, we must first explain what it is that we seek to achieve in this article, and why.
  We mentioned earlier that the empirical observations thus far let us classify \gls*{acr:hf} solutions into two kinds.
  From our prior experience with wavefunction symmetry analysis using representation theory\cite{article:Huynh2020}, we know that the solutions of the second kind are symmetry-broken (we will discuss this in greater depth later) and can thus be used to form multi-determinantal wavefunctions that recover static correlation.
  It is therefore of great chemical interest to understand the conditions for the existence of these solutions in the real Hilbert space where they can be located quite easily with most contemporary \gls*{acr:scf} methods.
  Hence, throughout this article, we shall be guided by two main questions:
  \begin{enumerate*}[label=(\roman*)]
    \item What are the qualitative and quantitative differences between the two kinds of solutions observed? and
    \item How does symmetry control the reality of these solutions?
  \end{enumerate*}

  \tikzexternalenable
  \tikzexternaldisable

  This article is structured as follows.
  In Section~\ref{sec:constraints}, we discuss the roles of spin and spatial symmetry constraints on the structure of the Fock matrix and make a fundamental distinction between two types of constraints which we call \emph{intrinsic} and \emph{extrinsic}.
  We then detail the symmetry constraints applicable to the \ce{[H4]^{2+}} system that are of main interest to us in Section~\ref{sec:h4analmodel-impositionofconstraints} before formulating the corresponding algebraic holomorphic \gls*{acr:hf} equations in Section~\ref{sec:h4analmodel-algebraicequations}.
  The solutions to these equations are classified and their reality behaviors examined in Section~\ref{sec:h4analmodel-solutions} so as to map out the different real/non-real regimes exhibited by certain solutions and the transition boundaries between them.
  Section~\ref{sec:h4analmodel-symmetry} then provides an examination of the group-theoretic symmetry of the solutions and relates that to their reality behaviors.
  The connections between the different imposed constraints are subsequently presented in Section~\ref{sec:h4analmodel-constraintconnections} in an attempt to explore different local and global \gls*{acr:scf} landscapes of the \ce{[H4]^{2+}} system.
  Finally, we conclude in Section~\ref{sec:discussion} with a few discussional remarks on the implications of our findings and chart out possible directions to generalize the analysis in this work.

\tikzsetexternalprefix{./constraints/tikz/}

\section{Symmetry Constraints}
\label{sec:constraints}

  Every \gls*{acr:scf} \gls*{acr:hf} procedure that is performed in a basis set of \glspl*{acr:ao} inevitably involves the diagonalization of the Fock matrix expressed in this basis whose form strongly dictates the nature of the \gls*{acr:scf} solutions obtained.
  We therefore begin by examining the form of the Fock matrix to gain an understanding of the various kinds of constraints that symmetry can impose on the system.

  Let us consider an $\gls*{gen:Ne}$-electron single determinant,
  \begin{equation}
    \label{eq:singledet}
    \gls*{wf:det}
      = \lvert \gls*{bas:spinorb}_{1} \ldots \gls*{bas:spinorb}_{i} \ldots \gls*{bas:spinorb}_{\gls*{gen:Ne}} \rvert %
      = \hat{\mathscr{A}}
        \left[
          \prod_{i}^{\gls*{gen:Ne}}
          \gls*{bas:spinorb}_{i}(\gls*{bas:spinspatialcoord}_i)
        \right],
  \end{equation}
  where $\gls*{bas:spinorb}_{i}$ denotes the $i$\textsuperscript{th} spin-orbital, $\gls*{bas:spinspatialcoord}_i$ the spin-spatial coordinates of the $i$\textsuperscript{th} electron, and $\hat{\mathscr{A}}$ the antisymmetrizer that acts on the electron labels.
  Let us also define an antilinear conjugation operator $\gls*{op:holoconj}$ on a determinant as
  \begin{equation}
    \gls*{op:holoconj}\gls*{wf:det}
    = \hat{\mathscr{A}}
      \left[
        \prod_{i}^{\gls*{gen:Ne}}
        \gls*{bas:spinorb}_{i}^*(\gls*{bas:spinspatialcoord}_i)
      \right],
    \label{eq:kappaPsi}
  \end{equation}
  where $\gls*{bas:spinorb}_{i}^*$ will be defined in Equation~\ref{eq:chiconj}.
  We assume that the spin-orbitals $\gls*{bas:spinorb}_{i}$ optimize either the conventional energy functional,
  \begin{equation}
    E[\gls*{wf:gen}] = \frac{
      \braket{
        \gls*{wf:det} |
        \gls*{op:hamil} |
        \gls*{wf:det}
      }
    }{
      \braket{
        \gls*{wf:det} |
        \gls*{wf:det}
      }
    },
  \end{equation}
  or the holomorphic energy functional\cite{article:Hiscock2014, article:Burton2016},
  \begin{equation}
      \tilde{E}[\gls*{wf:det}] = \frac{
        \braket{
          \gls*{op:holoconj}\gls*{wf:det} |
          \gls*{op:hamil} |
          \gls*{wf:det}
        }
      }{
      \braket{
        \gls*{op:holoconj}\gls*{wf:det} |
        \gls*{wf:det}
        }
      }.
    \end{equation}
  In either case, the spin-orbitals are eigenfunctions of the Fock operator:
  \ifdefined\twocolumnmode
    \begin{equation}
      \label{eq:fockop}
      \begin{multlined}[b]
        \hat{f}(\gls*{bas:spinspatialcoord}_m) =\\
        \hat{h}(\boldsymbol{r}_m)
        + \sum_{j=1}^{\gls*{gen:Ne}}
        \int \D\gls*{bas:spinspatialcoord}_n
        \ \gls*{bas:spinorbcc}_j^{\gls*{op:wildcard}}(\gls*{bas:spinspatialcoord}_n)
        \frac{1-\gls*{op:perm}_{(mn)}}{\lvert \boldsymbol{r}_m - \boldsymbol{r}_n \rvert}
        \gls*{bas:spinorb}_j(\gls*{bas:spinspatialcoord}_n),
      \end{multlined}
    \end{equation}
  \else
    \begin{equation}
      \label{eq:fockop}
        \hat{f}(\gls*{bas:spinspatialcoord}_m) =
        \hat{h}(\boldsymbol{r}_m)
        + \sum_{j=1}^{\gls*{gen:Ne}}
            \int \D\gls*{bas:spinspatialcoord}_n
        \ \gls*{bas:spinorbcc}_j^{\gls*{op:wildcard}}(\gls*{bas:spinspatialcoord}_n)
        \frac{1-\gls*{op:perm}_{(mn)}}{\lvert \boldsymbol{r}_m - \boldsymbol{r}_n \rvert}
        \gls*{bas:spinorb}_j(\gls*{bas:spinspatialcoord}_n),
    \end{equation}
  \fi
  where $\hat{h}$ is the one-electron core Hamiltonian operator, $\gls*{bas:spinorbcc}$ the typically complex-conjugated spin-orbital, and $\gls*{op:perm}_{(mn)}$ the permutation operator corresponding to the transposition $(mn)$ of electron labels.
  We also use the wildcard operator $\gls*{op:wildcard}$ as a generic placeholder which can be either the identity for conventional \gls*{acr:hf} or the conjugation $*$ for holomorphic \gls*{acr:hf} (to be defined in Equation~\ref{eq:chiconj}).
  If we now introduce a covariant\cite{article:Head-Gordon1998} spin--spatial direct-product basis,
  \begin{equation}
    \label{eq:directproductbasis}
    \gls*{bas:ssbasis} = \left\lbrace
      \gls*{bas:spinbasis}_{\delta},\gls*{bas:spinbasis}_{\varepsilon},\ldots \vphantom{\gls*{bas:spatialbasis}_{\mu}}
    \right\rbrace \otimes
    \left\lbrace
      \gls*{bas:spatialbasis}_{\mu},\gls*{bas:spatialbasis}_{\nu},\ldots
    \right\rbrace,
  \end{equation}
  with $\gls*{bas:spinbasis}$ representing covariant spin basis functions and $\gls*{bas:spatialbasis}$ covariant spatial basis functions, such that each spin-orbital can be expanded using the contravariant\cite{article:Head-Gordon1998} \glshyph*{acr:mo} coefficients $G_i^{\delta\mu,\cdot}$ as
  \begin{subequations}
    \begin{equation}
      \label{eq:spinorbexpansion}
      \gls*{bas:spinorb}_i(\gls*{bas:spinspatialcoord}_m) = \gls*{bas:spinbasis}_{\cdot \delta}(\gls*{bas:spincoord}_m) \gls*{bas:spatialbasis}_{\cdot \mu}(\gls*{bas:spatialcoord}_m) G_i^{\delta\mu,\cdot},
    \end{equation}
  where $\delta\mu$ are double indices such that any twice-occurring Greek indices are implicitly contracted over and $\gls*{bas:spincoord}_m$ and $\gls*{bas:spatialcoord}_m$ are the spin and spatial coordinates of the $m$\textsuperscript{th} electron.
  We can write this more succinctly as
    \begin{equation}
      \label{eq:spinorbexpansion_mat}
      \gls*{bas:spinorb}_i =
        (\boldsymbol{{\gls*{bas:spinbasis}}}
        \otimes
        \boldsymbol{{\gls*{bas:spatialbasis}}})^{\T}
        \boldsymbol{G}_i,
    \end{equation}
  \end{subequations}
  where $\boldsymbol{{\gls*{bas:spinbasis}}}$ and $\boldsymbol{{\gls*{bas:spatialbasis}}}$ are column vectors containing the covariant spin and spatial basis functions respectively, and $\boldsymbol{G}_i$ the $i$\textsuperscript{th} column of the contravariant \glshyph*{acr:mo} coefficient matrix $\boldsymbol{G}$ of dimensions $\lvert\gls*{bas:ssbasis}\rvert \times \gls*{gen:Ne}$.
  Assuming that the spatial basis functions are all real, we then \emph{define}
  \begin{equation}
    \label{eq:chiconj}
    \gls*{bas:spinorb}_i^{*}
    \equiv \gls*{bas:spinbasis}_{\cdot \delta}
      (\gls*{op:complexconj}\gls*{bas:spatialbasis}_{\cdot \mu})
      (G_i^{\delta\mu,\cdot})^{*}
    = (\boldsymbol{{\gls*{bas:spinbasis}}}
      \otimes
      \gls*{op:complexconj}\boldsymbol{{\gls*{bas:spatialbasis}}})^{\T}
      \boldsymbol{G}_i^{*}
  \end{equation}
  where $\gls*{op:complexconj}$ is the typical complex-conjugation operator acting only on the spatial basis functions and the coefficients in $\boldsymbol{G}_i$ are complex-conjugated.
  The action of the $*$ conjugation on spin-orbitals is therefore not that of an actual complex conjugation.
  Since the spin functions can be chosen to be orthonormal, it can be shown that this definition of $\gls*{bas:spinorb}_i^{*}$ ensures that $\gls*{op:holoconj}$ in Equation~\ref{eq:kappaPsi} is indeed a conjugation operator (see Refs.~\citenum{article:Riss1998} and \citenum{incoll:Garcia2007} for definition).
  We require in addition that, if the spatial basis functions $\gls*{bas:spatialbasis}_{\mu}$ are real-valued, then $\gls*{op:complexconj}\gls*{bas:spatialbasis}_{\mu} = \gls*{bas:spatialbasis}_{\mu}$.
  Thus, for real spatial basis functions, which we will consider exclusively from here on,
  \begin{equation}
    \gls*{bas:spinorb}_i^{*}
    = (\boldsymbol{{\gls*{bas:spinbasis}}}
      \otimes
      \boldsymbol{{\gls*{bas:spatialbasis}}})^{\T}
      \boldsymbol{G}_i^{*}.
  \end{equation}

  We now write the contravariant one-particle density matrix for $\gls*{wf:det}$ as
  \begin{align*}
    P^{\delta\mu, \delta'\mu'} &=
      \sum_{i=1}^{\gls*{gen:Ne}}
      G_i^{\delta\mu,\cdot}
      (G^{\dagger\gls*{op:wildcard}}_i)^{\cdot,\delta'\mu'} \\
    \Leftrightarrow
    \boldsymbol{P} &=
      \boldsymbol{G}\boldsymbol{G}^{\dagger\gls*{op:wildcard}},
  \end{align*}
  and hence the elements of the Fock matrix $\boldsymbol{F}$ as
  \begin{align}
    F_{\delta'\mu', \delta\mu}
    &= \braket{
         \gls*{bas:spinbasis}_{\delta'}(s) \gls*{bas:spatialbasis}_{\mu'}(\boldsymbol{r}) |
         \hat{f}(\gls*{bas:spinspatialcoord}) |
         \gls*{bas:spinbasis}_{\delta}(s) \gls*{bas:spatialbasis}_{\mu}(\boldsymbol{r})
       } \nonumber\\
    &= H^{\textrm{core}}_{\delta'\mu',\delta\mu}
      + \Pi_{\delta'\mu'\varepsilon'\nu', \delta\mu\varepsilon\nu} P^{\varepsilon\nu, \varepsilon'\nu'} \label{eq:fockmatelem} \\
    \Leftrightarrow
    \boldsymbol{F}
    &= \boldsymbol{H}^{\textrm{core}} + \boldsymbol{\Pi}\cdot\boldsymbol{P},
    \label{eq:fockmat}
  \end{align}
  where the one-electron contribution is
  \begin{equation}
    \label{eq:onee}
    H^{\textrm{core}}_{\delta'\mu',\delta\mu}
      = \Omega_{\delta'\delta} \braket{\gls*{bas:spatialbasis}_{\mu'} | \hat{h} | \gls*{bas:spatialbasis}_{\mu}},
  \end{equation}
  and the density-independent part of the two-electron contribution is
  \ifdefined\twocolumnmode
    \begin{multline}
      \label{eq:twoe}
        \Pi_{\delta'\mu'\varepsilon'\nu', \delta\mu\varepsilon\nu}
          =
          \Omega_{\delta'\delta}\Omega_{\varepsilon'\varepsilon} \braket{\gls*{bas:spatialbasis}_{\mu'} \gls*{bas:spatialbasis}_{\nu'} |\gls*{bas:spatialbasis}_{\mu}\gls*{bas:spatialbasis}_{\nu}}\\
          - \Omega_{\delta'\varepsilon}\Omega_{\varepsilon'\delta} \braket{\gls*{bas:spatialbasis}_{\mu'} \gls*{bas:spatialbasis}_{\nu'} |\gls*{bas:spatialbasis}_{\nu}\gls*{bas:spatialbasis}_{\mu}},
    \end{multline}
  \else
    \begin{equation}
      \label{eq:twoe}
      \Pi_{\delta'\mu'\varepsilon'\nu', \delta\mu\varepsilon\nu}
      =
      \Omega_{\delta'\delta}\Omega_{\varepsilon'\varepsilon} \braket{\gls*{bas:spatialbasis}_{\mu'}^{\gls*{op:wildcard}} \gls*{bas:spatialbasis}_{\nu'}^{\gls*{op:wildcard}} |\gls*{bas:spatialbasis}_{\mu}\gls*{bas:spatialbasis}_{\nu}}
      - \Omega_{\delta'\varepsilon}\Omega_{\varepsilon'\delta} \braket{\gls*{bas:spatialbasis}_{\mu'}^{\gls*{op:wildcard}} \gls*{bas:spatialbasis}_{\nu'}^{\gls*{op:wildcard}} |\gls*{bas:spatialbasis}_{\nu}\gls*{bas:spatialbasis}_{\mu}},
    \end{equation}
  \fi
  with $\braket{\cdot\cdot|\cdot\cdot}$ denoting a two-electron repulsion integral in physicists' notation.
  The binary dot operator in Equation~\ref{eq:fockmat} indicates a tensor contraction of the double indices $\varepsilon\nu$ and $\varepsilon'\nu'$ as in Equation~\ref{eq:fockmatelem}.
  In both Equations~\ref{eq:onee} and~\ref{eq:twoe}, $\boldsymbol{\Omega}$ is the spin-only overlap matrix,
  \begin{equation*}
    \Omega_{\delta\varepsilon} = \braket{\gls*{bas:spinbasis}_{\delta} | \gls*{bas:spinbasis}_{\varepsilon}}.
  \end{equation*}
  If we also define $\boldsymbol{S}^{\mathrm{AO}}$ as the spatial-\gls*{acr:ao}-only overlap matrix,
  \begin{equation*}
    S^{\mathrm{AO}}_{\mu\nu} = \braket{\gls*{bas:spatialbasis}_{\mu} | \gls*{bas:spatialbasis}_{\nu}},
  \end{equation*}
  then the \gls*{acr:hf} equations in the basis $\gls*{bas:ssbasis}$ are given by
  \begin{equation}
    \label{eq:hfeqnbasis}
    F_{\delta'\mu',\delta\mu} G_i^{\delta\mu,\cdot} =
      \Omega_{\delta'\delta} S^{\mathrm{AO}}_{\mu'\mu}
      G_i^{\delta\mu,\cdot}
      \varepsilon_i,
  \end{equation}
  where $\varepsilon_i$ are the eigenvalues of $\boldsymbol{F}$.

  The form of the Fock matrix in Equation~\ref{eq:fockmat} shows contributions from three terms, two of which ($\boldsymbol{H}^{\textrm{core}}$ and $\boldsymbol{\Pi}$) are independent of the \gls*{acr:mo} coefficients and hence of the actual solutions of the \gls*{acr:hf} equations, whereas the remaining one ($\boldsymbol{P}$) shows a direct dependence on the \gls*{acr:hf} solutions.
  This allows us to distinguish between two kinds of constraints imposable on the Fock matrix: \emph{intrinsic constraints} are those that arise solely from the properties of the basis functions and affect $\boldsymbol{H}^{\textrm{core}}$ and $\boldsymbol{\Pi}$ directly, and \emph{extrinsic constraints} are additional constraints imposed on the \gls*{acr:mo} coefficients that affect $\boldsymbol{P}$ but not $\boldsymbol{H}^{\textrm{core}}$ and $\boldsymbol{\Pi}$.
  We will discuss both types of constraints in turn.

  \subsection{Intrinsic Constraints}
  \label{subsec:intrinsicconstraints}
    Most basis sets used in electronic-structure calculations take the form of a direct-product basis (Equation~\ref{eq:directproductbasis}) between an implicit spin basis and a suitable real spatial basis.
    This decomposability into separate spin and spatial bases enables us to examine intrinsic constraints due to spin and spatial symmetries separately.

    \subsubsection{Spin Symmetry}
    \label{subsubsec:spinsym}
      In the familiar two-component orthonormal spinor basis $\lbrace\alpha, \beta\rbrace \equiv \lbrace\ket{\nicefrac{1}{2}, \nicefrac{1}{2}}, \ket{\nicefrac{1}{2}, -\nicefrac{1}{2}}\rbrace$, $\boldsymbol{\Omega}$ must equal the $2 \times 2$ identity matrix $\boldsymbol{I}_2$.
      This forces $\boldsymbol{H}^{\textrm{core}}$ to have only two non-zero spin blocks, $\alpha\alpha$ and $\beta\beta$, by Equation~\ref{eq:onee}.
      Similarly, by Equation~\ref{eq:twoe}, the non-zero spin blocks in the first contribution to $\boldsymbol{\Pi}$ (the so-called Coulomb term) are $\alpha\alpha\alpha\alpha$, $\alpha\beta\alpha\beta$, $\beta\alpha\beta\alpha$, and $\beta\beta\beta\beta$, and those in the second contribution to $\boldsymbol{\Pi}$ (the so-called exchange term) are $\alpha\alpha\alpha\alpha$, $\alpha\beta\beta\alpha$, $\beta\alpha\alpha\beta$, and $\beta\beta\beta\beta$, so that $\boldsymbol{\Pi}$ can only have up to six non-zero spin blocks.
      Overall, these constraints cause $\boldsymbol{F}$ to be rather sparse, and as soon as a basis set is fixed for a system (in this case, $\lbrace\alpha, \beta\rbrace$ for the spin basis), the forms of \emph{all} \gls*{acr:scf} solutions of the \gls*{acr:hf} equations in this basis are indiscriminately governed by the intrinsic constraints.

    \subsubsection{Spatial Symmetry}
      In a completely analogous manner to spin intrinsic constraints, Equations~\ref{eq:fockmatelem}, \ref{eq:onee}, and~\ref{eq:twoe} allow us to recognize that the structures of the one- and two-electron \gls*{acr:ao} integral tensors $\braket{\gls*{bas:spatialbasis}_{\mu'} | \hat{h} | \gls*{bas:spatialbasis}_{\mu}}$ and $\braket{\gls*{bas:spatialbasis}_{\mu'} \gls*{bas:spatialbasis}_{\nu'} |\gls*{bas:spatialbasis}_{\mu}\gls*{bas:spatialbasis}_{\nu}}$ introduce spatial intrinsic constraints to the Fock matrix.
      However, the sheer number of \gls*{acr:ao} basis functions in each basis set even for very small molecules and the dependence of the electron integrals on the nuclear arrangement of the system make the spatial intrinsic constraints much more complicated than their spin counterparts.
      This means that the one- and two-electron integral tensors do not possess any general sparse structures for a gross simplification of the Fock matrix.

      In spite of that, representation theory provides a way to quantify the degree of intrinsic constraints arising from the spatial basis functions.
      To this end, we first define the problem more concretely.
      Given an arrangement of nuclei that is invariant under the operations of a certain point group $\gls*{struct:gengroup}$ and a set of \gls*{acr:ao} spatial basis functions $\{\gls*{bas:spatialbasis}_{\mu}\}$ localized on these nuclei and transforming according to a (generally reducible) representation $\gls*{struct:rep}_{\gls*{struct:gengroup}}$ of $\gls*{struct:gengroup}$, we determine $n_1(\gls*{struct:rep}_{\gls*{struct:gengroup}}, \gls*{struct:gengroup})$, the number of non-vanishing independent components of the one-electron integral tensor $\braket{\gls*{bas:spatialbasis}_{\mu} | \hat{o} | \gls*{bas:spatialbasis}_{\nu}}$ where $\hat{o}$ is either the identity operator or the one-electron core Hamiltonian $\hat{h}$, and $n_2(\gls*{struct:rep}_{\gls*{struct:gengroup}}, \gls*{struct:gengroup})$, the number of non-vanishing independent components of the two-electron integral tensor $\braket{\gls*{bas:spatialbasis}_{\mu'} \gls*{bas:spatialbasis}_{\nu'} |\gls*{bas:spatialbasis}_{\mu}\gls*{bas:spatialbasis}_{\nu}}$.
      We then define the degree of spatial intrinsic constraints on the one- and two-electron integrals as
      \begin{align}
        \eta_i(\gls*{struct:rep}_{\gls*{struct:gengroup}}, \gls*{struct:gengroup}) = 1 - \frac{n_i(\gls*{struct:rep}_{\gls*{struct:gengroup}}, \gls*{struct:gengroup})}{n_i(\gls*{struct:rep}_{\mathcal{C}_1}, \mathcal{C}_1)}, \qquad i=1, 2.
        \label{eq:degreeintrinsicconstraints}
      \end{align}
      We note that $n_i(\gls*{struct:rep}_{\mathcal{C}_1}, \mathcal{C}_1) \ge n_i(\gls*{struct:rep}_{\gls*{struct:gengroup}}, \gls*{struct:gengroup}) \ \forall \gls*{struct:gengroup} \le \mathsf{O}(3)$, and hence $0 \le \eta_i \le 1$.
      Thus, the closer $\eta_i$ is to unity, the larger the extent to which the spatial symmetry of the basis functions constrains the values of the electron integrals and hence the structure of the Fock matrix and the \gls*{acr:scf} landscape.
      Detailed expressions for $n_i(\gls*{struct:rep}_{\gls*{struct:gengroup}}, \gls*{struct:gengroup})$ are given in Appendix~\ref{app:nonvanishingeints}.

  \subsection{Extrinsic Constraints}
  \label{subsec:extrinsicconstraints}

    Intrinsic constraints are, however, often still too general to restrict the \gls*{acr:scf} landscape to the appropriate regions of interest.
    This is not a problem if one seeks to explore as much of the Hilbert space as possibly allowed by the underlying \gls*{acr:scf} formalism.
    However, very often one would choose to focus on certain local regions in the \gls*{acr:scf} landscape, perhaps after considerations that are motivated by empirical observations, physical reasoning, or computational cost, and extrinsic constraints provide a way to achieve this.
    For instance, as explained in the next Section, to understand the \gls*{acr:uhf} solutions of \ce{[H4]^{2+}} in Figure~\ref{fig:h42p-prelim} analytically, we need to consider the numerical forms of their \glspl*{acr:mo} and then impose appropriate constraints on the \gls*{acr:mo} coefficients to limit ourselves to the right parts of the \gls*{acr:scf} landscape in which these solutions reside.
    Incidentally, this also simplifies the governing equations sufficiently such that analytic forms for these solutions can be obtained.
    Or more generally, by restricting the structure of the spin blocks of the \gls*{acr:mo} coefficient matrix $\boldsymbol{G}$ (Equation~\ref{eq:spinorbexpansion_mat}) in an arbitrary system, one obtains a hierarchy of \gls*{acr:hf} variants that are simultaneous eigenfunctions of some combination of the spin projection operator $\hat{S}_z$, the squared total spin operator $\hat{S}^2$, the time reversal operator $\gls*{op:timerev}$, and the complex-conjugation operator $\gls*{op:complexconj}$.
    These have been originally characterized by Fukutome\cite{article:Fukutome1981} and then later clarified by Stuber and Paldus\cite{book:Stuber2003}.

    By imposing suitable extrinsic constraints, \gls*{acr:scf} solutions that exhibit certain desirable symmetries can be targeted, but it is worth noting that these constraints can always be relaxed to allow for more symmetry-broken solutions to be found.
    For example, let $\gls*{wf:hf}[_{\gls*{struct:hilbert}[_0]}]$ be an \gls*{acr:scf} solution located within a space $\gls*{struct:hilbert}[_0]$ (such as \gls*{acr:uhf}) that exhibits less restrictive extrinsic constraints than another space $\gls*{struct:hilbert}[_1]$ (such as \gls*{acr:rhf}), then the stability of $\gls*{wf:hf}[_{\gls*{struct:hilbert}[_1]}]$ with respect to spin-orbital transformation in $\gls*{struct:hilbert}[_0]$ has already been explored in great depth for the various Fukutome classes of conventional \gls*{acr:hf}\cite{book:Thouless2014, article:Cizek1967, article:Seeger1977, article:Paldus1990, article:Goings2015}.
    From an epistemological point of view, therefore, there is much to gain from studying the high-symmetry \gls*{acr:scf} solutions that result from certain extrinsic constraints before venturing out into the more general, less constrained parts of the \gls*{acr:scf} landscape where symmetry-broken solutions are abundant and special techniques such as \gls*{acr:noci}\cite{article:Thom2009b, article:Sundstrom2014b} or various projection-based methods\cite{booksec:Mayer1980, article:Scuseria2011a, article:Jimenez-Hoyos2012, article:Ye2019} are needed to restore symmetry and obtain sensible quantum numbers, because symmetry-brokenness can often complicate the interpretation of the \gls*{acr:scf} solutions obtained\cite{article:Small2015, article:Thompson2018, article:Huynh2020}.
\tikzsetexternalprefix{./h4model-impositionofconstraints/tikz/}

\section{\ce{[H4]^{2+}} Model: Imposition of Constraints}
  \label{sec:h4analmodel-impositionofconstraints}

  In order to formulate the analytic equations for the \ce{[H4]^{2+}} model, we begin with a description of the constraints applicable to this system in STO-3G.
  In particular, we discuss the spatial intrinsic constraints that are imposed by the STO-3G basis set in conjunction with the molecular geometry of \ce{[H4]^{2+}} and comment on how they affect the terms in the \gls*{acr:hf} equations.
  We then outline the spin and spatial extrinsic constraints that we shall impose to focus on the numerically located \gls*{acr:uhf} solutions shown in Figure~\ref{fig:h42p-prelim} and also to explore other related local regions.
  This allows us to show that these constraints lead to a family of solutions that are identical in form across different specific extrinsic constraints.

  \subsection{Spatial Intrinsic Constraints}
    \label{subsec:spatialintrinsicconstraints_h42+}

    Table~\ref{table:spatialintrinsicconstraints_h42+} shows the number of non-vanishing independent components of the one- and two-electron integrals for \ce{[H4]^{2+}} in STO-3G with different nuclear arrangements and the corresponding degrees of spatial intrinsic constraints according to Equation~\ref{eq:degreeintrinsicconstraints} and Appendix~\ref{app:nonvanishingeints}.
    Unsurprisingly, as the molecule becomes more symmetric in the sense that the $1s$ \gls*{acr:ao} basis functions become related by more symmetry operations, more constraints are imposed on the electron integrals and the degrees of spatial intrinsic constraints increase accordingly.
    Hence, there are fewer non-zero unique elements in the $\boldsymbol{H}_{\textrm{core}}$ and $\boldsymbol{\Pi}$ tensors by virtue of Equations~\ref{eq:onee} and~\ref{eq:twoe}, resulting in more related terms in the basis-dependent \gls*{acr:hf} equations \ref{eq:hfeqnbasis}.
    In addition, as the \gls*{acr:ao} basis functions are real, all elements of $\boldsymbol{H}_{\textrm{core}}$ and $\boldsymbol{\Pi}$ must also be real and the \gls*{acr:hf} equations \ref{eq:hfeqnbasis} in turn become non-linear polynomial equations over the reals where each term is a monomial of overall degree $1$ or $3$ in the \gls*{acr:mo} coefficients $\boldsymbol{G}$.

    \newlength{\modelsize}
    \setlength{\modelsize}{0.9cm}
    \tikzexternalenable
    \begin{table}[h!]
      \centering
      \caption{Spatial intrinsic constraints for \ce{[H4]^{2+}} in STO-3G.}
      \label{table:spatialintrinsicconstraints_h42+}
      \begin{tabular}{
        >{\centering\arraybackslash} m{2.2cm}
        >{\raggedright\arraybackslash} m{0.6cm}
        S[table-figures-integer=1, table-figures-decimal=0, table-number-alignment=center, table-column-width=0.8cm]
        S[table-figures-integer=2, table-figures-decimal=0, table-number-alignment=center, table-column-width=0.8cm]
        S[table-figures-integer=1, table-figures-decimal=3, table-number-alignment=center, table-column-width=1.3cm]
        S[table-figures-integer=1, table-figures-decimal=3, table-number-alignment=center, table-column-width=1.3cm]
      }
        \toprule
        Shape & $\gls*{struct:gengroup}$ & {$n_1$} & {$n_2$} & {$\eta_1$} & {$\eta_2$}\\
        \midrule
        \parbox[m]{1.8cm}{\useexternalfile{1}{0}{0}{D2}{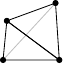}} & $\mathcal{D}_{2}$ & 4 & 19 & 0.600 & 0.655 \\[12pt]
        \parbox[m]{1.8cm}{\useexternalfile{1}{0}{0}{D2h.rec}{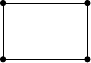}} & $\mathcal{D}_{2h}$ & 4 & 19 & 0.600 & 0.655 \\[12pt]
        \parbox[m]{1.8cm}{\useexternalfile{1}{0}{0}{D4h}{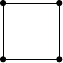}} & $\mathcal{D}_{4h}$ & 3 & 13 & 0.700 & 0.764 \\[12pt]
        \parbox[m]{1.8cm}{\useexternalfile{1}{0}{0}{Td}{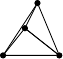}} & $\mathcal{T}_{d}$ & 2 & 7 & 0.800 & 0.873\\
        \bottomrule
      \end{tabular}
    \end{table}
    \tikzexternaldisable

    The Fundamental Theorem of Algebra\cite{book:Fine1997} inspires the holomorphization of conventional \gls*{acr:hf} theory such that in simple cases where the \gls*{acr:hf} equations can be re-parameterized as single univariate polynomial equations, there must exist a constant number of solutions across all molecular geometries\cite{article:Hiscock2014}.
    However, this theorem does not dictate how many of these solutions must be real, nor is it applicable to more complicated problems where multivariate polynomial systems are unavoidable.
    In fact, if the \gls*{acr:hf} equations are considered as polynomial equations in a certain number of unknowns, then B\'{e}zout's theorem imposes an ultimate upper bound on the number of solutions as the product of the polynomial degrees\cite{article:Garcia1980, article:Chen1984, article:Schmid1995, book:Sottile2011}, provided that the number of solutions is finite, but it does not say anything about their reality either.
    This is to be expected since an analysis based solely on the gross algebraic structures of the equations cannot pick up the consequences due to the fine structures within the terms in the polynomials.

    To gain any insight into the connection between symmetry and the reality of \gls*{acr:scf} solutions at all, we must examine the symmetry-induced relations between the monomials in the \gls*{acr:hf} equations.
    The simplicity of the chosen model \ce{[H4]^{2+}} system enables us to achieve this on an analytic level with moderate ease.
    In fact, the values of $n_1$ and $n_2$ shown in Table~\ref{table:spatialintrinsicconstraints_h42+} give the number of one- and two-electron terms in the \gls*{acr:hf} equations after the monomials have been factorized through the common one- and two-electron integrals.
    We believe that restrictions of this nature play a major role in determining whether solutions must be real or can be non-real and we will demonstrate this more carefully in Section~\ref{subsec:consequencesofsymconstraints}.
    But before this can be done, we must introduce some extrinsic constraints to simplify the equations further, as even the high symmetry of $\mathcal{T}_d$ still leaves us with too many unknowns and too complicated equation structures to handle analytically for the purposes of this \ce{[H4]^{2+}} model study.

  \subsection{Spin and Spatial Extrinsic Constraints}
  \label{subsec:spinspatialextrinsicconstraints_h42+}

    To target the \gls*{acr:uhf} solutions shown in Figure~\ref{fig:h42p-prelim}, we can only impose extrinsic spin constraints that are no more restrictive than \gls*{acr:uhf}.
    However, we do not wish to have to deal with any $\hat{s}_z$ symmetry breaking which we feel does not add to our understanding of spatial symmetry constraints in any significant way.
    We shall therefore stay at the \gls*{acr:uhf} level and focus on the $M_S = 0$ solutions.
    Furthermore, we take $\chi_1$ to have $m_s = +\nicefrac{1}{2}$ and $\chi_2$ to have $m_s = -\nicefrac{1}{2}$ without loss of generality.
    Following conventions, we shall use $\alpha$ and $\beta$ to denote the ``spin-up'' and ``spin-down'' components of the electron spin, respectively.

    Spatial extrinsic constraints are more involved to describe since they depend on the underlying point group of the molecule and the symmetry pathway under consideration.
    In particular, any spatial extrinsic constraints we impose must ``respect'' the point-group symmetry of the molecule in the sense that each constraint follows from an equation of the form
    \begin{equation}
      \label{eq:spatialextcons}
      \hat{R} \gls*{wf:hf}[_i] = \sum_j \gls*{wf:hf}[_j] D_{ji}(\hat{R}),
      \quad
      \tr \boldsymbol{D}(\hat{R}) = \chi^{\gls*{struct:rep}_{\gls*{struct:gengroup}}^\mathrm{ir}}(\hat{R})
    \end{equation}
    for a particular $\hat{R} \in \gls*{struct:gengroup}$ where the sum runs over all linearly independent \gls*{acr:hf} determinants that are degenerate and equivalent to $\gls*{wf:hf}[_i]$ by symmetry, $\gls*{struct:rep}_{\gls*{struct:gengroup}}^\mathrm{ir}$ is an \emph{irreducible} representation in the underlying point group $\gls*{struct:gengroup}$ of the molecule, and $\chi^{\gls*{struct:rep}_{\gls*{struct:gengroup}}^\mathrm{ir}}$ its character---the distinction between $\chi$ as a character and $\chi$ as a spin-orbital should be clear from the context.
    In other words, each spatial extrinsic constraint represents a required conserved symmetry element $\hat{R}$ in $\gls*{struct:gengroup}$ (see Ref.~\citenum{article:Huynh2020} for the definition of \emph{symmetry conservation}).

\begin{figure*}[h]
      \centering
      \subfloat[%
        \label{subfig:rectangle}
        Molecular symmetry pathway A.
      ]{
        \centering
        \useexternalfile{1}{0}{0}{D2h.rec-D4h-D2h.rec}{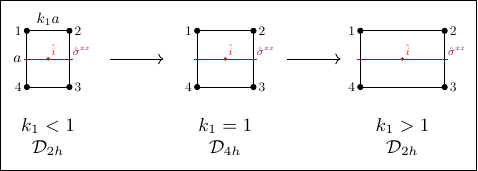}
      }

      \vspace{0.45cm}

      \subfloat[%
        \label{subfig:tetrahedron}
        Molecular symmetry pathway B.
      ]{
        \centering
        \useexternalfile{1}{0}{0}{D4h-D2-Td-D2-D4h}{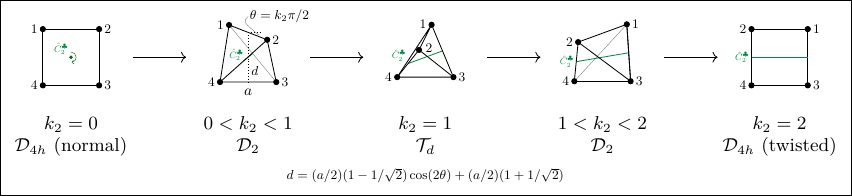}
      }

       \vspace{0.15cm}

       \subfloat{
         \centering
         \useexternalfile{1}{0}{0}{triad}{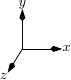}
       }

      \caption{
        Molecular symmetry pathways considered in this study.
        Each pathway is parametrised by $k_i \in \interval[open right]{0}{\infty}, i=1,2$ such that $k_i = 1$ corresponds to the highest-symmetry geometry along that pathway.
        Pathway A is non-periodic, while pathway B is periodic with period $4$.
        The extrinsic constraining elements introduced in Table~\ref{tab:extrinsicconstraints} are also shown here.
      }
      \label{fig:symbreakingpathways}
    \end{figure*}

\begin{figure*}[h]
      \centering
      \subfloat[\label{subfig:paths-axes}]{
        \centering
        \useexternalfile{0.8}{0}{0}{paths0-axes}{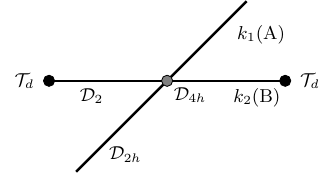}
      }
      \subfloat[\label{subfig:paths-pathwaya}]{
        \centering
        \useexternalfile{0.8}{0}{0}{paths1-pathwaya}{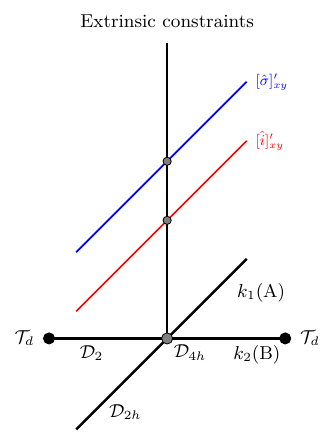}
      }
      \subfloat[\label{subfig:paths-all}]{
        \centering
        \useexternalfile{0.8}{0}{0}{paths2-all}{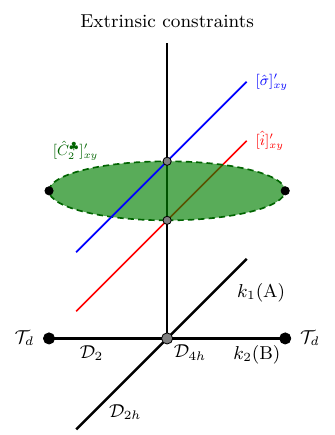}
      }
      \caption{
        Relationship between the molecular symmetry pathways and extrinsic constraints considered in this study.
        In \protect\subref{subfig:paths-axes}, we depict the two-dimensional subspace defined by pathway A (parameterized by $k_1$) and pathway B (parameterized by $k_2$) and mark out the high-symmetry points.
        In \protect\subref{subfig:paths-pathwaya}, we introduce a third axis to represent the extrinsic constraints imposable on the \gls*{acr:scf} solutions, together with the red and blue solid lines representing pathway A subject to the extrinsic constraining spaces $[\hat{i}]'_{xy}$ and $[\hat{\sigma}]'_{xy}$, respectively.
        Finally, in \protect\subref{subfig:paths-all}, we add the green dashed loop representing the periodic pathway B subject to the extrinsic constraining space $[\hat{C}_2^{\clubsuit}]'_{xy}$ chosen to connect the constraining spaces of pathway A.
        In all plots, the high-symmetry points along the molecular symmetry pathways are marked out with black dots ($\mathcal{T}_d$) or grey dots ($\mathcal{D}_{4h}$).
      }
      \label{fig:pathwayvis}
    \end{figure*}

    In Figure~\ref{fig:symbreakingpathways}, we show the two symmetry pathways that we will consider throughout this investigation.
    Each pathway represents a particular direction in which the symmetry of the system can be varied and has been constructed to consist of general low-symmetry segments connecting several high-symmetry special points.
    In addition, the low-symmetry group must be a common subgroup of all the high-symmetry groups so that only the elements of the low-symmetry point group persist throughout the entire pathway and can thus be used to form spatial extrinsic constraints that are well defined at every geometry on the pathway.
    Each pathway is characterized by two parameters: a symmetry factor $k_i$ that controls the variation in symmetry of the system along the pathway, and a scale length $a$ that controls the characteristic distance between neighboring hydrogen atoms.
    Pathway A is also the pathway along which the numerical \gls*{acr:uhf} solutions in Figure~\ref{fig:h42p-prelim} are tracked and shall therefore be the main focus of our discussion.
    We will elaborate later in Section~\ref{subsec:connectingviapathwayb} that, via pathway B, different \emph{extrinsic} constraints along pathway A can be connected.
    The relationship between the two pathways and the high-symmetry points is illustrated in Figure~\ref{subfig:paths-axes}.

    \newlength{\pathstructsize}
    \setlength{\pathstructsize}{0.9cm}
    \tikzexternalenable
    \tikzexternaldisable

    \tikzexternalenable
    \tikzexternaldisable

    We introduce a new notation to facilitate the systematic description of extrinsic constraints.
    If $\psi$ is a generic wavefunction, then $\psi\{\hat{s}, \mu_{s}\}_i[\hat{R}, \chi^{\gls*{struct:rep}_{\gls*{struct:gengroup}}^\mathrm{ir}}(\hat{R})]_j$ denotes that $\psi$ is constrained to be an eigenfunction of a generic spin operator $\hat{s}$ with eigenvalue $\mu_{s}$, and also to conserve symmetry under the spatial operation $\hat{R}$ with character $\chi^{\gls*{struct:rep}_{\gls*{struct:gengroup}}^\mathrm{ir}}(\hat{R})$ as in Equation~\ref{eq:spatialextcons}.
    More than one set of curly (square) brackets can be used to denote multiple spin (spatial) extrinsic constraints, in which case they are distinguished by subscripts such as $i$ and $j$.

    Occasionally, it is desirable to remove one or more constrained conserved symmetries in order to explore larger regions of the \gls*{acr:scf} landscape in the vicinity of the conserved-symmetry constraints.
    However, removing a constraint entirely can sometimes cause the problem to become too general and too intractable analytically.
    Therefore, an alternative would be to \emph{relax} the constraint in a controlled manner so as to keep the problem manageable while still being able to enlarge the \gls*{acr:scf} regions of interest.
    One way this can be done is to allow $\mu_{s}$ to be expectation values of $\hat{s}$ that deviate from its exact eigenvalues, or to let $\tr \boldsymbol{D}(\hat{R})$ in Equation~\ref{eq:spatialextcons} take on values other than exact characters of irreducible representations.
    We then replace $\{\hat{s}, \mu_{s}\}$ and $[\hat{R}, \chi^{\gls*{struct:rep}_{\gls*{struct:gengroup}}^\mathrm{ir}}(\hat{R})]$ with $\{\hat{s}\}'$ and $[\hat{R}]'$ respectively to signify that the \gls*{acr:scf} regions are still being constrained by $\hat{s}$ and $\hat{R}$, but we drop the eigenvalues or character values as they are no longer fixed or meaningful.
    In addition, we use dashes to signify the general symmetry breaking with respect to $\hat{s}$ or $\hat{R}$.

    Using the above notations, we show in Table~\ref{tab:extrinsicconstraints} the \gls*{acr:scf} extrinsic constraints that we will consider for the symmetry pathways A and B in this article.
    The spin extrinsic constraints are strict to ensure that only $M_S = 0$ \gls*{acr:uhf} solutions are obtained, while the spatial extrinsic constraints are loose in the sense that, for each spatial symmetry operation $\hat{R}$ considered ($\hat{R} = \hat{i}, \hat{\sigma}^{xz}, \hat{C}_2^{\clubsuit}$), we define a closed domain of two real parameters $(x, y) \in \mathscr{D} = \interval{-1}{+1} \times \interval{-1}{+1}$ such that each parameter constrains the coefficients of one of the two \glspl*{acr:mo}.
    A visualization of how these extrinsic constraints are related to the symmetry pathways A and B is provided in Figures~\ref{subfig:paths-pathwaya} and~\ref{subfig:paths-all}.

    In each of the \gls*{acr:scf} constraining spaces $[\hat{R}]$ considered, the four corners of $\mathscr{D}$ are special as they correspond to extrinsic constraints that conserve $\hat{R}$-symmetry.
    We will pay particular attention to the solutions that are subject to these constraints since they correspond to true \gls*{acr:scf} stationary points as will be explained in Section~\ref{subsec:symbrokenextrinsicconstraints}.
    We thus give these constraints special shorthand notations as shown in Table~\ref{tab:extrinsicconstraints-special} to facilitate the following discussions.
    While the solutions obtained under these constraints must conserve all the symmetries imposed by the constraints, they are free to break other symmetries without any \textit{a priori} restrictions.
    \newlength{\constrainingsize}
    \setlength{\constrainingsize}{0.6cm}
    \tikzexternalenable
    \begin{table*}
      \centering
      \caption{
        \Glsxtrshort*{acr:scf} extrinsic constraining spaces along pathways A and B.
        Each constraining space is parameterized by two real parameters $(x, y)$ defining a domain $\mathscr{D} = \interval{-1}{+1} \times \interval{-1}{+1}$.
        The constraining elements are shown in color under the \textit{Visualization} column.
        The variations of these constraining elements along their respective pathways are also shown in Figure~\ref{fig:symbreakingpathways}.
        For pathway B, $\hat{C}_2^{\clubsuit}$ denotes the two-fold rotation whose axis coincides with the common perpendicular bisector of the bonds \ce{H^1}---\ce{H^3} and \ce{H^2}---\ce{H^4}, which becomes $\hat{C}_2^z$ at $\mathcal{D}_{4h}$ (normal) and $\hat{C}_2^x$ at $\mathcal{D}_{4h}$ (twisted).
      }
      \label{tab:extrinsicconstraints}
      \renewcommand{\arraystretch}{1.3}
      \resizebox{\textwidth}{!}{%
        \begin{tabular}[t]{
  >{\raggedright\arraybackslash} p{0.9cm}
  >{\raggedright\arraybackslash} p{7.3cm}
  >{\raggedright\arraybackslash} p{4.9cm}
  >{\centering\arraybackslash} p{5.5cm}
}
  \toprule
  Path. & SCF extrinsic constraining space & Coefficient relations & \raggedright\arraybackslash Visualization\\
  \midrule
  A
    & $\begin{multlined}[t]
        \Psi\{\hat{s}_z, 0\}[\hat{i}]'_{xy}
        \\=
          \lvert
            \chi_1\{\hat{s}_z, +\nicefrac{1}{2}\}[\hat{i}]'_{x}
            \;\chi_2\{\hat{s}_z, -\nicefrac{1}{2}\}[\hat{i}]'_{y}
          \rvert
    \end{multlined}$
    & $\begin{alignedat}[t]{3}
        xG_1^{\alpha1,\cdot} &= G_1^{\alpha3,\cdot},\ yG_2^{\beta1,\cdot} &&= G_2^{\beta3,\cdot}\\
        xG_1^{\alpha2,\cdot} &= G_1^{\alpha4,\cdot},\ yG_2^{\beta2,\cdot} &&= G_2^{\beta4,\cdot}
    \end{alignedat}$
    &
    \adjustbox{valign=t}{\useexternalfile{1}{0}{0}{D2h.rec.i.coeffs}{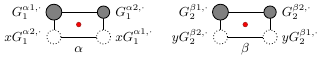}}
  \\
  \addlinespace[8pt]
    & $\begin{multlined}[t]
        \Psi\{\hat{s}_z, 0\}[\hat{\sigma}^{xz}]'_{xy}
        \\=
        \lvert
        \chi_1\{\hat{s}_z, +\nicefrac{1}{2}\}[\hat{\sigma}^{xz}]'_{x}
        \;\chi_2\{\hat{s}_z, -\nicefrac{1}{2}\}[\hat{\sigma}^{xz}]'_{y}
        \rvert
    \end{multlined}$
    & $\begin{alignedat}[t]{3}
        xG_1^{\alpha1,\cdot} &= G_1^{\alpha4,\cdot},\ yG_2^{\beta1,\cdot} &&= G_2^{\beta4,\cdot}\\
        xG_1^{\alpha2,\cdot} &= G_1^{\alpha3,\cdot},\ yG_2^{\beta2,\cdot} &&= G_2^{\beta3,\cdot}
    \end{alignedat}$
    &
    \adjustbox{valign=t}{\useexternalfile{1}{0}{0}{D2h.rec.sigma.coeffs}{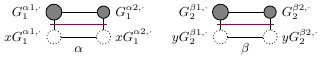}}
  \\
  \addlinespace[16pt]
  B
    & $\begin{multlined}[t]
        \Psi\{\hat{s}_z, 0\}[\hat{C}_2^{\clubsuit}]'_{xy}
        \\=
        \lvert
        \chi_1\{\hat{s}_z, +\nicefrac{1}{2}\}[\hat{C}_2^{\clubsuit}]'_{x}
        \;\chi_2\{\hat{s}_z, -\nicefrac{1}{2}\}[\hat{C}_2^{\clubsuit}]'_{y}
        \rvert
    \end{multlined}$
    & $\begin{alignedat}[t]{3}
        xG_1^{\alpha1,\cdot} &= G_1^{\alpha3,\cdot},\ yG_2^{\beta1,\cdot} &&= G_2^{\beta3,\cdot}\\
        xG_1^{\alpha2,\cdot} &= G_1^{\alpha4,\cdot},\ yG_2^{\beta2,\cdot} &&= G_2^{\beta4,\cdot}
    \end{alignedat}$
    &
    \adjustbox{valign=t}{\useexternalfile{1}{0}{0}{D2.tet.c2z.coeffs}{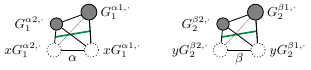}}
  \\
  \bottomrule
\end{tabular}
      }
    \end{table*}
    \tikzexternaldisable

    \begin{table}
      \centering
      \caption{Shorthand notations for special $\hat{R}$-symmetry-conserved extrinsic constraints.}
      \label{tab:extrinsicconstraints-special}
      \ifdefined\twocolumnmode
        \renewcommand{\arraystretch}{1.3}
        \resizebox{\linewidth}{!}{%
          \setlength{\tabcolsep}{6pt}
\begin{tabular}{%
  llcccc
}
  \toprule
  \multicolumn{2}{l}{\multirow{2}{*}{Constraint}} & \multicolumn{4}{c}{$(x, y)$} \\
  \cline{3-6}
  && {\small $(+1, +1)$} & {\small $(-1, +1)$} & {\small $(+1, -1)$} & {\small $(-1, -1)$} \\
  \midrule
  A
    & $\Psi\{\hat{s}_z, 0\}[\hat{i}]'_{xy}$
    & $
    \left\lvert
    \alpha^g \beta^g
    \right\rvert
    $
    & $
    \left\lvert
    \alpha^u \beta^g
    \right\rvert
    $
    & $
    \left\lvert
    \alpha^g \beta^u
    \right\rvert
    $
    & $
    \left\lvert
    \alpha^u \beta^u
    \right\rvert
    $
  \\
  \addlinespace[4pt]
    & $\Psi\{\hat{s}_z, 0\}[\hat{\sigma}^{xz}]'_{xy}$
    & $
    \left\lvert
    \alpha' \beta'
    \right\rvert
    $
    & $
    \left\lvert
    \alpha'' \beta'
    \right\rvert
    $
    & $
    \left\lvert
    \alpha' \beta''
    \right\rvert
    $
    & $
    \left\lvert
    \alpha'' \beta''
    \right\rvert
    $
  \\
  \addlinespace[6pt]
  B
    & $\Psi\{\hat{s}_z, 0\}[\hat{C}_2^{\clubsuit}]'_{xy}$
    & $
    \left\lvert
    \alpha^+ \beta^+
    \right\rvert
    $
    & $
    \left\lvert
    \alpha^- \beta^+
    \right\rvert
    $
    & $
    \left\lvert
    \alpha^+ \beta^-
    \right\rvert
    $
    & $
    \left\lvert
    \alpha^- \beta^-
    \right\rvert
    $
  \\
  \bottomrule
\end{tabular}
        }
      \else
        \setlength{\tabcolsep}{6pt}
\begin{tabular}{%
  llcccc
}
  \toprule
  \multicolumn{2}{l}{\multirow{2}{*}{Constraint}} & \multicolumn{4}{c}{$(x, y)$} \\
  \cline{3-6}
  && {\small $(+1, +1)$} & {\small $(-1, +1)$} & {\small $(+1, -1)$} & {\small $(-1, -1)$} \\
  \midrule
  A
    & $\Psi\{\hat{s}_z, 0\}[\hat{i}]'_{xy}$
    & $
    \left\lvert
    \alpha^g \beta^g
    \right\rvert
    $
    & $
    \left\lvert
    \alpha^u \beta^g
    \right\rvert
    $
    & $
    \left\lvert
    \alpha^g \beta^u
    \right\rvert
    $
    & $
    \left\lvert
    \alpha^u \beta^u
    \right\rvert
    $
  \\
  \addlinespace[4pt]
    & $\Psi\{\hat{s}_z, 0\}[\hat{\sigma}^{xz}]'_{xy}$
    & $
    \left\lvert
    \alpha' \beta'
    \right\rvert
    $
    & $
    \left\lvert
    \alpha'' \beta'
    \right\rvert
    $
    & $
    \left\lvert
    \alpha' \beta''
    \right\rvert
    $
    & $
    \left\lvert
    \alpha'' \beta''
    \right\rvert
    $
  \\
  \addlinespace[6pt]
  B
    & $\Psi\{\hat{s}_z, 0\}[\hat{C}_2^{\clubsuit}]'_{xy}$
    & $
    \left\lvert
    \alpha^+ \beta^+
    \right\rvert
    $
    & $
    \left\lvert
    \alpha^- \beta^+
    \right\rvert
    $
    & $
    \left\lvert
    \alpha^+ \beta^-
    \right\rvert
    $
    & $
    \left\lvert
    \alpha^- \beta^-
    \right\rvert
    $
  \\
  \bottomrule
\end{tabular}
      \fi
    \end{table}

\tikzsetexternalprefix{./h4model-algebraicequations/tikz/}

\section{\ce{[H4]^{2+}} Model: Algebraic Equations}
  \label{sec:h4analmodel-algebraicequations}

  \subsection{Holomorphic Normalization Re-parameterization}
  \label{subsec:normreparam}

    In all cases of extrinsic constraints in Table~\ref{tab:extrinsicconstraints}, it can be shown that the holomorphic normalization of the spin-orbitals $\chi_1$ and $\chi_2$, $\braket{\gls*{bas:spinorb}_i^{*} | \gls*{bas:spinorb}_j} = \delta_{ij}$, translates to%
    \begin{subequations}
      \begin{align}
        2 [
            (G_1^{\alpha1, \cdot})^2 + (G_1^{\alpha2, \cdot})^2
          ] S_1^{-}
        + 4 G_1^{\alpha1, \cdot} G_1^{\alpha2, \cdot} S_2^{-} &= 1,\\
        2 [
            (G_2^{\beta1, \cdot})^2 + (G_2^{\beta2, \cdot})^2
          ] S_1^{+}
        + 4 G_2^{\beta1, \cdot} G_2^{\beta2, \cdot} S_2^{+} &= 1,
      \end{align}%
      \label{eq:generalnormalisation}%
    \end{subequations}%
    where $S_1^{\pm}$ and $S_2^{\pm}$ are functions of the spatial-\gls*{acr:ao}-only overlap integrals $\braket{\varphi_{\mu}^{*} | \varphi_{\nu}}$.
    Since the \glspl*{acr:ao} in STO-3G are real, we do not need to worry about the distinction between the conventional and holomorphic formalisms for the \gls*{acr:ao} integrals and can therefore simply write $\braket{\gls*{bas:spatialbasis}_{\mu} | \gls*{bas:spatialbasis}_{\nu}}$ for $\braket{\gls*{bas:spatialbasis}_{\mu}^{*} | \gls*{bas:spatialbasis}_{\nu}}$.
    The exact functional forms of the coefficients $S_1^{\pm}$ and $S_2^{\pm}$ depend on the precise extrinsic and intrinsic constraints being imposed and are detailed in Section~S-I of the Supplementary Material, but they shall not affect the genericity of the analysis to follow.
    In fact, the holomorphic normalisation constraints in Equation~\ref{eq:generalnormalisation} facilitate the re-parametrisation of the \gls*{acr:mo} coefficients in terms of two angular parameters $\theta_{\alpha}$ and $\theta_{\beta}$ as%
    \begin{subequations}
      \begin{align}
        G_1^{\alpha1, \cdot} &= \sqrt{\frac{\eta^{-}}{2}}
          \left(
            -\frac{\cos\theta_{\alpha}}{\sqrt{1-\gamma^{-}}}
            +\frac{\sin\theta_{\alpha}}{\sqrt{1+\gamma^{-}}}
          \right),
        \\
        G_1^{\alpha2, \cdot} &= \sqrt{\frac{\eta^{-}}{2}}
          \left(
            +\frac{\cos\theta_{\alpha}}{\sqrt{1-\gamma^{-}}}
            +\frac{\sin\theta_{\alpha}}{\sqrt{1+\gamma^{-}}}
          \right),
        \\
        G_2^{\beta1, \cdot} &= \sqrt{\frac{\eta^{+}}{2}}
          \left(
            -\frac{\cos\theta_{\beta}}{\sqrt{1-\gamma^{+}}}
            +\frac{\sin\theta_{\beta}}{\sqrt{1+\gamma^{+}}}
          \right),
        \\
        G_2^{\beta2, \cdot} &= \sqrt{\frac{\eta^{+}}{2}}
          \left(
            +\frac{\cos\theta_{\beta}}{\sqrt{1-\gamma^{+}}}
            +\frac{\sin\theta_{\beta}}{\sqrt{1+\gamma^{+}}}
          \right),
      \end{align}%
      \label{eq:MOcoeffparams}%
    \end{subequations}%
    where
    \begin{equation}
      \gamma^{\pm} = \frac{S_2^{\pm}}{S_1^{\pm}} \qquad \textrm{and} \qquad \eta^{\pm} = \frac{1}{2S_1^{\pm}}
      \label{eq:gammaetacoeffs}
    \end{equation}
    are \gls*{acr:ao}-overlap-dependent coefficients.
    The combination of the extrinsic constraints in Table~\ref{tab:extrinsicconstraints} (four coefficient relations in each case) and the natural constraints provided by the normalization of the \glspl*{acr:mo} (two more coefficient relations) reduces the original eight degrees of freedom to only two which are parameterized by $\theta_{\alpha}$ and $\theta_{\beta}$.
    The problem becomes more tractable since we can now variationally optimize the holomorphic energy expression analytically with relative ease.

  \subsection{Variation of the Holomorphic Energy}
  \label{subsec:holoenergyvariation}

    It turns out that, for all of the extrinsic constraints listed in Table~\ref{tab:extrinsicconstraints} and the associated intrinsic constraints defined by the geometry of the molecule, the one- and two-electron holomorphic energies (signified by tildes) can be written generically as%
    \ifdefined\twocolumnmode
    \begin{subequations}
      \begin{align}
        \tilde{E}_{1} ={}
          & [
            (G_1^{\alpha1, \cdot})^2 + (G_1^{\alpha2, \cdot})^2
          ]\ A_1^{-} \nonumber\\
          & + [
            (G_2^{\beta1, \cdot})^2 + (G_2^{\beta2, \cdot})^2
          ]\ A_1^{+} \nonumber\\
          & + 2G_1^{\alpha1, \cdot}G_1^{\alpha2, \cdot}\ A_2^{-} \nonumber\\
          & + 2G_2^{\beta1, \cdot} G_2^{\beta2, \cdot}\ A_2^{+},\\
        \tilde{E}_{2} ={}
          & [
              (G_1^{\alpha1, \cdot})^2 (G_2^{\beta1, \cdot})^2 + (G_1^{\alpha2, \cdot})^2 (G_2^{\beta1, \cdot})^2
            ]\ A_3 \nonumber\\
          & + 2[
              (G_1^{\alpha1, \cdot})^2 + (G_1^{\alpha2, \cdot})^2
            ]
            G_2^{\beta1, \cdot} G_2^{\beta2, \cdot}\ A_4^{+} \nonumber\\
          & + 2 G_1^{\alpha1, \cdot} G_1^{\alpha2, \cdot}
            [
              (G_2^{\beta1, \cdot})^2 + (G_2^{\beta2, \cdot})^2
            ]\ A_4^{-} \nonumber\\
          & + [
              (G_1^{\alpha1, \cdot})^2 (G_2^{\beta2, \cdot})^2 + (G_1^{\alpha2, \cdot})^2 (G_2^{\beta1, \cdot})^2
            ]\ A_5 \nonumber\\
          & + 4G_1^{\alpha1, \cdot}G_1^{\alpha2, \cdot}G_2^{\beta1, \cdot}G_2^{\beta2, \cdot}\ A_6,
      \end{align}%
      \label{eq:holoenergy}%
    \end{subequations}
  \else
    \begin{subequations}
      \begin{align}
        \tilde{E}_{1} ={}
          & [
            (G_1^{\alpha1, \cdot})^2 + (G_1^{\alpha2, \cdot})^2
          ]\ A_1^{-} +
          [
            (G_2^{\beta1, \cdot})^2 + (G_2^{\beta2, \cdot})^2
          ]\ A_1^{+}
          + 2G_1^{\alpha1, \cdot}G_1^{\alpha2, \cdot}\ A_2^{-}
          + 2G_2^{\beta1, \cdot} G_2^{\beta2, \cdot}\ A_2^{+},\\
        \tilde{E}_{2} ={}
          & [
          (G_1^{\alpha1, \cdot})^2 (G_2^{\beta1, \cdot})^2 + (G_1^{\alpha2, \cdot})^2 (G_2^{\beta1, \cdot})^2
          ]\ A_3  \nonumber\\
          & + 2[
          (G_1^{\alpha1, \cdot})^2 + (G_1^{\alpha2, \cdot})^2
          ]
          G_2^{\beta1, \cdot} G_2^{\beta2, \cdot}\ A_4^{+}
          + 2 G_1^{\alpha1, \cdot} G_1^{\alpha2, \cdot}
          [
          (G_2^{\beta1, \cdot})^2 + (G_2^{\beta2, \cdot})^2
          ]\ A_4^{-} \nonumber\\
          & + [
          (G_1^{\alpha1, \cdot})^2 (G_2^{\beta2, \cdot})^2 + (G_1^{\alpha2, \cdot})^2 (G_2^{\beta1, \cdot})^2
          ]\ A_5
          + 4G_1^{\alpha1, \cdot}G_1^{\alpha2, \cdot}G_2^{\beta1, \cdot}G_2^{\beta2, \cdot}\ A_6,
      \end{align}%
      \label{eq:holoenergy}%
    \end{subequations}%
    \fi
    where the coefficients $A_1^{\pm}$ and $A_2^{\pm}$ are functions of the one-electron integrals $\braket{\varphi_{\mu'} | \hat{h} | \varphi_{\mu}}$, and the coefficients $A_3$, $A_4^{\pm}$, $A_5$, and $A_6$ are functions of the two-electron integrals $\braket{\varphi_{\mu'} \varphi_{\nu'} |\varphi_{\mu}\varphi_{\nu}}$.
    The exact functional forms of these coefficients depend on the precise constraints at hand as shown in Section~S-I of the Supplementary Material and do not affect the following generic analysis, but via these coefficients, the electron integrals govern the nature of the solutions obtained.

    By substituting the parameterization of the \gls*{acr:mo} coefficients in Equation~\ref{eq:MOcoeffparams} into the energy expressions in Equation~\ref{eq:holoenergy}, and then optimizing with respect to the angular parameters $\theta_{\alpha}$ and $\theta_{\beta}$, we obtain a particularly simple system of trigonometric equations which we denote $P\left(\boldsymbol{\theta}; \boldsymbol{B}\right) = 0$:%
    \ifdefined\twocolumnmode
      {\small
        \begin{subequations}
          \label{eq:Psys}
          \begin{align}
            \begin{multlined}[b]
              B_1^{-}  \sin 2\theta_{\alpha}
              + B_2^{--} \sin 2\theta_{\alpha} \cos^2\theta_{\beta}\\
              + B_2^{-+} \sin 2\theta_{\alpha} \sin^2\theta_{\beta}
              + B_3      \cos 2\theta_{\alpha} \sin 2\theta_{\beta} = 0,
            \end{multlined}\\
            \begin{multlined}[b]
              B_1^{+}  \sin 2\theta_{\beta}
              + B_2^{+-} \sin 2\theta_{\beta} \cos^2\theta_{\alpha}\\
              + B_2^{++} \sin 2\theta_{\beta} \sin^2\theta_{\alpha}
              + B_3      \cos 2\theta_{\beta} \sin 2\theta_{\alpha} = 0,
            \end{multlined}%
          \end{align}%
        \end{subequations}%
      }%
    \else
      \begin{subequations}
        \label{eq:Psys}
        \begin{align}
            B_1^{-}  \sin 2\theta_{\alpha}
            + B_2^{--} \sin 2\theta_{\alpha} \cos^2\theta_{\beta}
            + B_2^{-+} \sin 2\theta_{\alpha} \sin^2\theta_{\beta}
            + B_3      \cos 2\theta_{\alpha} \sin 2\theta_{\beta} &= 0,\\
            B_1^{+}  \sin 2\theta_{\beta}
            + B_2^{+-} \sin 2\theta_{\beta} \cos^2\theta_{\alpha}
            + B_2^{++} \sin 2\theta_{\beta} \sin^2\theta_{\alpha}
            + B_3      \cos 2\theta_{\beta} \sin 2\theta_{\alpha} &= 0,%
        \end{align}%
      \end{subequations}%
    \fi
    where $(\theta_\alpha, \theta_\beta) \in \mathbb{C}^2$, and the $B$ coefficients are \textit{definitive} functions of the $\gamma$, $\eta$, and $A$ coefficients:%
    \ifdefined\twocolumnmode
    {
      \scriptsize
      \begin{subequations}
        \begin{align}
          B_1^{\pm}
            &= \frac{2\eta^{\pm}}{1-(\gamma^{\pm})^2}
               \left(
                -\gamma^{\pm}A_1^{\pm} + A_2^{\pm}
               \right),\\
          B_2^{+\pm}
            &= \frac{
                \eta^{-}\eta^{+}
                \left[
                  -\gamma^{+}(A_3 + A_5)
                  + 2(\mp\gamma^{+}A_4^{-} + A_4^{+})
                \right]
               }
               {[1-(\gamma^{+})^2](1 \pm \gamma^{-})}
               \pm 2A_6,\\
          B_2^{-\pm}
            &= \frac{
                \eta^{-}\eta^{+}
                \left[
                  -\gamma^{-}(A_3 + A_5)
                  + 2(\mp\gamma^{-}A_4^{+} + A_4^{-})
                \right]
               }
               {[1-(\gamma^{-})^2](1 \pm \gamma^{+})}
               \pm 2A_6,\\
          B_3
            &= \frac{\eta^{-}\eta^{+}}
                    {\sqrt{1-(\gamma^{-})^2} \sqrt{1-(\gamma^{+})^2}}
               (A_3 - A_5).
        \end{align}%
        \label{eq:Bcoeffs}%
      \end{subequations}%
    }
    \else
      \begin{subequations}
        \begin{align}
          B_1^{\pm}
          &= \frac{2\eta^{\pm}}{1-(\gamma^{\pm})^2}
          \left(
          -\gamma^{\pm}A_1^{\pm} + A_2^{\pm}
          \right),\\
          B_2^{+\pm}
          &= \frac{
            \eta^{-}\eta^{+}
            \left[
            -\gamma^{+}(A_3 + A_5)
            + 2(\mp\gamma^{+}A_4^{-} + A_4^{+})
            \right]
          }
          {[1-(\gamma^{+})^2](1 \pm \gamma^{-})}
          \pm 2A_6,\\
          B_2^{-\pm}
          &= \frac{
            \eta^{-}\eta^{+}
            \left[
            -\gamma^{-}(A_3 + A_5)
            + 2(\mp\gamma^{-}A_4^{+} + A_4^{-})
            \right]
          }
          {[1-(\gamma^{-})^2](1 \pm \gamma^{+})}
          \pm 2A_6,\\
          B_3
          &= \frac{\eta^{-}\eta^{+}}
          {\sqrt{1-(\gamma^{-})^2} \sqrt{1-(\gamma^{+})^2}}
          (A_3 - A_5).
        \end{align}%
        \label{eq:Bcoeffs}%
      \end{subequations}%
    \fi
     By converting $P\left(\boldsymbol{\theta}; \boldsymbol{B}\right) = 0$ to an exponential form using Euler's formula and substituting $z_1=e^{2i\theta_{\alpha}}$ and $z_2=e^{2i\theta_{\beta}}$, we obtain an equivalent polynomial form $\bar{P}\left(\boldsymbol{z}; \boldsymbol{B}\right) = 0$:
    \ifdefined\twocolumnmode
      {\small
        \begin{subequations}
          \label{eq:Pbarsys}
          \begin{align}
            \begin{multlined}[b]
              \left(z_1^2-1\right)
              \left[
                4B_1^{-}z_2
                + B_2^{--}\left(z_2+1\right)^2
                - B_2^{-+}\left(z_2-1\right)^2
              \right]\\
              + 2B_3\left(z_1^2+1\right)\left(z_2^2-1\right) = 0,
            \end{multlined}\\
            \begin{multlined}[b]
              \left(z_2^2-1\right)
              \left[
                4B_1^{+}z_1
                + B_2^{+-}\left(z_1+1\right)^2
                - B_2^{++}\left(z_1-1\right)^2
              \right]\\
              + 2B_3\left(z_1^2-1\right)\left(z_2^2+1\right) = 0.
            \end{multlined}%
          \end{align}%
        \end{subequations}%
      }
    \else
      \begin{subequations}
        \label{eq:Pbarsys}
        \begin{align}
            \left(z_1^2-1\right)
            \left[
              4B_1^{-}z_2
              + B_2^{--}\left(z_2+1\right)^2
              - B_2^{-+}\left(z_2-1\right)^2
            \right]
            + 2B_3\left(z_1^2+1\right)\left(z_2^2-1\right) &= 0,\\
            \left(z_2^2-1\right)
            \left[
              4B_1^{+}z_1
              + B_2^{+-}\left(z_1+1\right)^2
              - B_2^{++}\left(z_1-1\right)^2
            \right]
            + 2B_3\left(z_1^2-1\right)\left(z_2^2+1\right) &= 0.
        \end{align}%
      \end{subequations}%
    \fi
    The system $\bar{P}\left(\boldsymbol{z}; \boldsymbol{B}\right) = 0$ contains the governing equations for the \gls*{acr:scf} solutions in the constraining spaces listed in Table~\ref{tab:extrinsicconstraints}.
    In what follows, we will solve $\bar{P}\left(\boldsymbol{z}; \boldsymbol{B}\right) = 0$ for $z_1$ and $z_2$ analytically to obtain closed-form expressions for the angular parameters $\theta_{\alpha}$ and $\theta_{\beta}$.
\tikzsetexternalprefix{./h4model-solutions/tikz/}

\section{\ce{[H4]^{2+}} Model: Analytic Solutions}
\label{sec:h4analmodel-solutions}

  \subsection{General Forms of Solutions}
  \label{subsec:generalforms}

    Since each equation in $\bar{P}\left(\boldsymbol{z}; \boldsymbol{B}\right) = 0$ has $z_1^2 z_2^2$ as its highest-order term, B\'{e}zout's theorem\cite{article:Chen1984,book:Sottile2011} requires that, if the system $\bar{P}\left(\boldsymbol{z}; \boldsymbol{B}\right) = 0$ has a finite number of solutions, then it can have \emph{at most} 16 solutions.
    However, if we now let $\bar{Q}\left(\boldsymbol{z}; \boldsymbol{B}\right) = 0$ be the corresponding auxiliary system
      \begin{align*}
        \left(2B_3 + B_2^{--} - B_2^{-+}\right)z_1^2z_2^2 &= 0,\\
        \left(2B_3 + B_2^{+-} - B_2^{++}\right)z_1^2z_2^2 &= 0,
      \end{align*}
    which is constructed from the highest-degree terms of the equations in $\bar{P}\left(\boldsymbol{z}; \boldsymbol{B}\right) = 0$,
    then, by Theorem~3.1 of Ref.~\citenum{article:Garcia1980}, that $\bar{Q}\left(\boldsymbol{z}; \boldsymbol{B}\right) = 0$ has non-trivial solutions (\textit{e.g.}, $z_1 \in \mathbb{C}, z_2 = 0$) implies that the system $\bar{P}\left(\boldsymbol{z}; \boldsymbol{B}\right) = 0$ \emph{must} have fewer than \num{16} solutions.
    The upper bound due to B\'{e}zout's theorem is therefore not tight.

    In fact, the system $\bar{P}\left(\boldsymbol{z}; \boldsymbol{B}\right) = 0$ turns out to admit eight solutions as obtained using the symbolic solvers in Mathematica 12.1.\cite{prog:Mathematica}
    And although each solution $(z_1, z_2)$ of $\bar{P}\left(\boldsymbol{z}; \boldsymbol{B}\right) = 0$ yields infinitely many algebraically different solutions for $P\left(\boldsymbol{\theta}; \boldsymbol{B}\right) = 0$ of the form $(\theta_{\alpha} + m\pi, \theta_{\beta} + n\pi)$ for $m, n \in \mathbb{Z}$, they all correspond to \glspl*{acr:mo} that differ from one another by a factor of $\pm 1$ (see Equation~\ref{eq:MOcoeffparams}), and therefore give only a single physically distinct \gls*{acr:scf} determinant.
    It thus suffices to examine only the principal solution $([\theta_{\alpha}], [\theta_{\beta}])$ arising from each $(z_1, z_2)$ pair.
    The analytic forms for these solutions are given in Table~\ref{tab:allsols}.

    \begin{table*}
      \centering
      \caption{
        Analytic \glsxtrshort*{acr:scf} solutions of \ce{[H4]^{2+}} in STO-3G under the extrinsic constraints considered in Table~\ref{tab:extrinsicconstraints}.
        $[\theta_{\alpha}]$ and $[\theta_{\beta}]$ denote the principal solutions of the $P\left(\boldsymbol{\theta}; \boldsymbol{B}\right) = 0$ system and $\Ln z$ denotes the principal natural logarithm of a complex number $z$, \textit{i.e.}, if $z = re^{i\varphi}$, then $\Ln z = \ln r + i\varphi$ for $r \in \mathbb{R}_{+}^{*}$ and $\varphi \in \interval[open left]{-\pi}{\pi}$.
        For solutions $\mathrm{2a}$ and $\mathrm{2b}$, only two out of the four possible $(z_1, z_2)$ combinations are true solutions of $\bar{P}\left(\boldsymbol{z}; \boldsymbol{B}\right) = 0$ while the other two are extraneous---the exact combinations that solve $\bar{P}\left(\boldsymbol{z}; \boldsymbol{B}\right) = 0$ depend on the precise values of the $B$ coefficients.
        When non-real, the angular parameters for the primed and unprimed variants of $\mathrm{2a}$ and $\mathrm{2b}$ are complex-conjugates of each other.
      }
      \label{tab:allsols}
      \ifdefined\twocolumnmode
      \else
        \renewcommand{\arraystretch}{0.7}
      \fi
      \begin{tabular}{
  c
  >{\centering\arraybackslash}p{3cm}
  >{\centering\arraybackslash}p{3cm}
  >{\centering\arraybackslash}p{3.5cm}
  >{\centering\arraybackslash}p{3.5cm}
}
  \toprule
  Solution     & $z_1$ & $z_2$ & $[\theta_{\alpha}]$ & $[\theta_{\beta}]$ \\
  \midrule
  $\mathrm{1a}\hphantom{'}$ & $+1$  & $+1$  & $0$                 & $0$            \\
  $\mathrm{1b}\hphantom{'}$ & $+1$  & $-1$  & $0$                 & $\pi/2$        \\
  $\mathrm{1c}\hphantom{'}$ & $-1$  & $+1$  & $\pi/2$             & $0$            \\
  $\mathrm{1d}\hphantom{'}$ & $-1$  & $-1$  & $\pi/2$             & $\pi/2$        \\
  \addlinespace[6pt]
  $\mathrm{2a}\hphantom{'}$ & $\left(u^{--} + iv^{-+}\right)^2$ & \multirow{2}{*}{$\left(u^{+-} \pm iv^{++}\right)^2$} & $-i\Ln\left(u^{--} + iv^{-+}\right)$ & \multirow{2}{*}{$-i\Ln\left(u^{+-} \pm iv^{++}\right)$}\\
  $\mathrm{2a}'$ & $\left(u^{--} - iv^{-+}\right)^2$ & & $-i\Ln\left(u^{--} - iv^{-+}\right)$ &\\
  \addlinespace[6pt]
  $\mathrm{2b}\hphantom{'}$ & $\left(u^{-+} + iv^{--}\right)^2$ & \multirow{2}{*}{$\left(u^{++} \pm iv^{+-}\right)^2$} & $-i\Ln\left(u^{-+} + iv^{--}\right)$ & \multirow{2}{*}{$-i\Ln\left(u^{++} \pm iv^{+-}\right)$}\\
  $\mathrm{2b}'$ & $\left(u^{-+} - iv^{--}\right)^2$ & & $-i\Ln\left(u^{-+} - iv^{--}\right)$ &\\
  \bottomrule
\end{tabular}
{\footnotesize
\begin{alignat*}{2}
  u^{-\mp} &= \frac{\sqrt{D_1^{-+} \mp D_2^{-}\sqrt{\Delta}}}{\sqrt{D_1^{--} + D_1^{-+}}}
  \qquad
  v^{-\pm} &&= \frac{\sqrt{D_1^{--} \pm D_2^{-}\sqrt{\Delta}}}{\sqrt{D_1^{--} + D_1^{-+}}}\\[6pt]
  u^{+\mp} &= \frac{\sqrt{D_1^{++} \mp D_2^{+}\sqrt{\Delta}}}{\sqrt{D_1^{+-} + D_1^{++}}}
  \qquad
  v^{+\pm} &&= \frac{\sqrt{D_1^{+-} \pm D_2^{+}\sqrt{\Delta}}}{\sqrt{D_1^{+-} + D_1^{++}}}
\end{alignat*}
\begin{align*}
  D_1^{-\pm}
    &=
      \begin{aligned}[t]
         4B_3^4
         &\pm 2(B_1^{-} + B_2^{--})
              (B_1^{-} + B_2^{-+})
              (B_1^{+} + B_2^{+\pm})
              (B_2^{+-} - B_2^{++})\\
         &+ B_3^2 \left[
            (2B_1^{-} + B_2^{--} + B_2^{-+})^2
            \pm (B_2^{--} - B_2^{-+})
                (2B_1^{+} - B_2^{+\mp} +3B_2^{+\pm})
         \right]
      \end{aligned} \\[5pt]
  D_1^{+\pm}
    &=
      \begin{aligned}[t]
         4B_3^4
         &\pm 2(B_1^{+} + B_2^{++})
              (B_1^{+} + B_2^{+-})
              (B_1^{-} + B_2^{-\pm})
              (B_2^{--} - B_2^{-+})\\
         &+ B_3^2 \left[
              (2B_1^{+} + B_2^{+-} + B_2^{++})^2
              \pm (B_2^{+-} - B_2^{++})
              (2B_1^{-} - B_2^{-\mp} +3B_2^{-\pm})
         \right]
      \end{aligned} \\[5pt]
  D_2^{\pm}
    &= B_3 (2B_1^{\pm} + B_2^{\pm-} + B_2^{\pm+})\\[7pt]
  \Delta
    &=
    \begin{aligned}[t]
      4B_3^4
       &+ 4(B_1^{-} + B_2^{--})
          (B_1^{-} + B_2^{-+})
          (B_1^{+} + B_2^{+-})
          (B_1^{+} + B_2^{++})\\
       &+ B_3^2 \left[
            (2B_1^{-} + B_2^{--} + B_2^{-+})^2
          + (2B_1^{+} + B_2^{+-} + B_2^{++})^2
          - 2(B_2^{--} - B_2^{-+})(B_2^{+-} - B_2^{++})
       \right]
    \end{aligned}
\end{align*}
}
    \end{table*}

    \paragraph{Persistently Real Solutions.}
      $\bar{P}\left(\boldsymbol{z}; \boldsymbol{B}\right) = 0$ admits four obvious solutions (labeled $\mathrm{1a}$--$\mathrm{d}$ in Table~\ref{tab:allsols}) which give rise to $\theta_{\alpha}$ and $\theta_{\beta}$ that are real and independent of the $B$ coefficients, so long as the functional forms of $P\left(\boldsymbol{\theta}; \boldsymbol{B}\right) = 0$ and $\bar{P}\left(\boldsymbol{z}; \boldsymbol{B}\right) = 0$, which are determined by the various intrinsic and extrinsic constraints under consideration (Tables~\ref{table:spatialintrinsicconstraints_h42+} and~\ref{tab:extrinsicconstraints}), remain unchanged.
      In other words, when the molecular symmetry of \ce{[H4]^{2+}} has been fixed and appropriate requirements for symmetry conservation have been imposed on the \glspl*{acr:mo}, solutions $\mathrm{1a}$--$\mathrm{d}$ will always exist in the real domain of the coefficients, regardless of the \ce{H}---\ce{H} bond lengths, and must therefore always be locatable by conventional \gls*{acr:scf} \gls*{acr:hf} methods.
      In fact, they are controlled entirely by the \gls*{acr:ao}-overlap integrals via the $\gamma$ and $\eta$ coefficients in Equation~\ref{eq:MOcoeffparams} while the core-Hamiltonian and the two-electron \gls*{acr:ao} integrals have no effects on them.

    \paragraph{Transiently Real Solutions.}
      $\bar{P}\left(\boldsymbol{z}; \boldsymbol{B}\right) = 0$ admits four more solutions (labeled $\mathrm{2a}$, $\mathrm{2a}'$, $\mathrm{2b}$, and $\mathrm{2b}'$ in Table~\ref{tab:allsols}) which give rise to generally complex $\theta_{\alpha}$ and $\theta_{\beta}$.
      From the general form $[\theta_{\sigma}] = -i\Ln(u_{\sigma}\pm iv_{\sigma})$ for the principal values of the angular parameters, if we write
      \begin{equation}
        u_{\sigma}\pm iv_{\sigma} = r_{\sigma}e^{i\varphi_{\sigma}}, \quad r_{\sigma} \in \mathbb{R}_{+}^{*}, \quad \varphi_{\sigma} \in \interval[open left]{-\pi}{\pi},
        \label{eq:uvrecast}
      \end{equation}
      then
      \begin{equation}
        [\theta_{\sigma}] = \varphi_{\sigma} -i\ln r_{\sigma},
        \label{eq:thetarecast}
      \end{equation}
      which implies that $\theta_{\sigma}$ is real if and only if the quantity
      \ifdefined\twocolumnmode
        \begin{align}
          \rho_{\sigma}
          &\equiv \Im [\theta_{\sigma}] = -\ln r_{\sigma}\nonumber\\
          &= -\ln \sqrt{\left(\Re u_{\sigma} \mp \Im v_{\sigma}\right)^2 +
                       \left(\Im u_{\sigma} \pm \Re v_{\sigma}\right)^2}
          \label{eq:indicator}
        \end{align}
      \else
        \begin{equation}
          \rho_{\sigma}
          \equiv \Im [\theta_{\sigma}]
          = -\ln \sqrt{\left(\Re u_{\sigma} \mp \Im v_{\sigma}\right)^2 +
                       \left(\Im u_{\sigma} \pm \Re v_{\sigma}\right)^2}
          \label{eq:indicator}
        \end{equation}
      \fi
      vanishes.
      Here, $\sigma \in \{\alpha, \beta\}$ and $u_{\sigma}, v_{\sigma}$ refer to the appropriate variants of $u^{\pm\pm}$ and $v^{\pm\pm}$ that correspond to the solution of interest as given in Table~\ref{tab:allsols}.
      The relation between $\theta_{\alpha}$ and $\theta_{\beta}$ imposed by $P\left(\boldsymbol{\theta}; \boldsymbol{B}\right) = 0$ in Equation~\ref{eq:Psys} requires that if $\theta_{\alpha}$ is real, then so is $\theta_{\beta}$, and \textit{vice versa}, which enables us to define $\rho \equiv \rho_{\alpha}$ as a reality indicator for solutions $\mathrm{2a}$ and $\mathrm{2b}$.

\begin{figure*}
        \centering
        \ifdefined\twocolumnmode
          \useexternalfile{0.90}{34.92221pt}{7.54836pt}{regimeplot.D2h.rec.i.single.all}{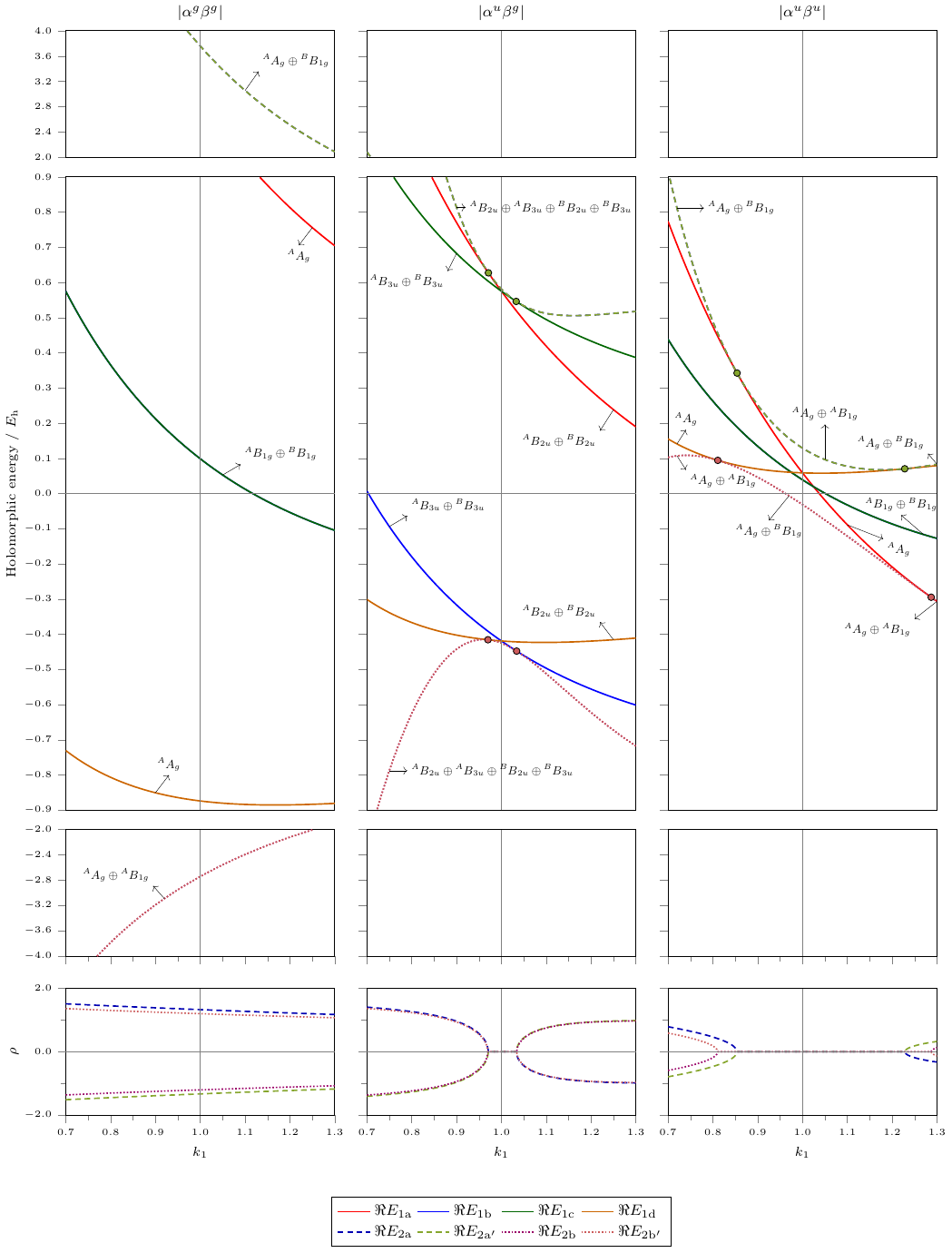}
        \else
          \useexternalfile{0.80}{34.92221pt}{7.54836pt}{regimeplot.D2h.rec.i.single.all}{figure4}
        \fi
        \caption{
          Holomorphic energy, symmetry, and reality indicator of the analytic solutions for \ce{[H4]^{2+}} in STO-3G for the special $\hat{i}$-symmetry-conserved extrinsic constraints in the vicinity of $\mathcal{D}_{4h}$ along molecular symmetry pathway A at $a = \SI{1.058350}{\angstrom}$.
          The holomorphic energy is real in all cases.
          The exact locations where solutions $\mathrm{2a}$ and $\mathrm{2b}$ undergo real--non-real transition are indicated by colored dots on the energy curves.
          The irreducible corepresentations in $\mathcal{T} \otimes \mathcal{D}_{2h}$ are denoted by $\prescript{\gamma}{}{\Gamma}$ where $\gamma$ is an irreducible corepresentation in $\mathcal{T}$ and $\Gamma$ an irreducible representation in $\mathcal{D}_{2h}$.
        }
        \label{fig:pathway-a-energy-i}
      \end{figure*}

      It turns out that all of the numerically located $M_S = 0$ conventional \gls*{acr:uhf} solutions shown in Figure~\ref{fig:h42p-prelim} conserve $\hat{i}$-symmetry along molecular symmetry pathway A.
      As such, they can all be captured by the analytic solutions obtained under the special $\hat{i}$-symmetry-conserved extrinsic constraints $\left\lvert\alpha^g \beta^g\right\rvert$, $\left\lvert\alpha^u \beta^g\right\rvert$, and $\left\lvert\alpha^u \beta^u\right\rvert$.
      Table~\ref{tab:analyticalnumericalidentification} maps the analytic solutions to the numerical solutions using the labels in Figure~\ref{fig:h42p-prelim}.
      We therefore focus on the analytic solutions for these three extrinsic  constraints in the following discussion.
      The holomorphic energies of these solutions together with their reality indicators are plotted in Figure~\ref{fig:pathway-a-energy-i}, the corresponding variations of their spin-orbitals are shown in Table~\ref{tab:pathway-a-orbitals-i}, and the accompanying animations in the included video (see Section~S-II in the Supplementary Material).

      \begin{table}
        \centering
        \caption{
          Identification of analytic solutions with numerical \glsxtrshort*{acr:uhf} solutions along molecular pathway A at $a = \SI{1.058350}{\angstrom}$.
          Repeated labels indicate solutions related by spatial symmetry.
        }
        \label{tab:analyticalnumericalidentification}
        \ifdefined\twocolumnmode
        \else
          \renewcommand{\arraystretch}{0.7}
        \fi
        \begin{tabular}{
            >{\raggedright\arraybackslash} b{1.0cm}
            >{\raggedright\arraybackslash} b{1.5cm}
            >{\raggedright\arraybackslash} b{1.5cm}
            >{\raggedright\arraybackslash} b{1.5cm}
          }
          \toprule
          Sol.
            & $\left\lvert \alpha^g \beta^g \right\rvert$
            & $\left\lvert \alpha^u \beta^g \right\rvert$
            & $\left\lvert \alpha^u \beta^u \right\rvert$ \\
          \midrule
          $\mathrm{1a}$
            & $\mathrm{E}$
            & $\mathrm{D}$
            & $\mathrm{C}''$ \\
          $\mathrm{1b}$
            & ${\mathrm{C}^4}'$
            & $\mathrm{B}'$
            & $\mathrm{C}'$ \\
          $\mathrm{1c}$
            & ${\mathrm{C}^4}'$
            & $\mathrm{D}'$
            & $\mathrm{C}'$ \\
          $\mathrm{1d}$
            & $\mathrm{A}$
            & $\mathrm{B}''$
            & $\mathrm{C}'''$ \\
          \addlinespace[6pt]
          $\mathrm{2a}$
            & -
            & $\mathrm{D}''$
            & ${\mathrm{C}^5}'$ \\
          $\mathrm{2a}'$
            & -
            & $\mathrm{D}''$
            & ${\mathrm{C}^5}'$ \\
          $\mathrm{2b}$
            & -
            & $\mathrm{B}$
            & $\mathrm{C}$ \\
          $\mathrm{2b}'$
            & -
            & $\mathrm{B}$
            & $\mathrm{C}$ \\
          \bottomrule
        \end{tabular}
      \end{table}

      \tikzsetexternalprefix{./h4model-solutions/tikz/energy/}
      \tikzexternalenable
      \tikzexternaldisable

  \subsection{Consequences of Symmetry Constraints}
  \label{subsec:consequencesofsymconstraints}
    \subsubsection{Strong Reality Requirements}
    \label{subsubsec:strongreality}
      The dependence of the reality indicator $\rho$ on the $B$ coefficients, and hence the electron integrals, via $u_{\sigma}$ and $v_{\sigma}$ (Equation~\ref{eq:indicator}) prompts the questions of whether there exist conditions on these coefficients such that $\rho$ is forced to vanish identically, and how these conditions are related to the various constraints imposed by symmetry on the system.
      General answers to these questions are challenging to obtain without a more in-depth analysis of the algebraic complexity of the \gls*{acr:hf} equations which we do not intend to carry out in the current study.
      Instead, we merely wish to demonstrate the existence of several such conditions that arise from the interplay between the intrinsic and extrinsic constraints exhibited by the model \ce{[H4]^{2+}} system so as to provide some insight into the attributes of the transiently real solutions.

      \begin{table}
        \centering
        \caption{
          Vanishing electron-integral coefficients along pathway A.
        }
        \label{tab:vanishingcoeffs-i}
        \ifdefined\twocolumnmode
          \renewcommand{\arraystretch}{1.5}
        \else
        \fi
        \begin{tabular}{
            >{\raggedright\arraybackslash} b{2.0cm}
            >{\raggedright\arraybackslash} b{1.5cm}
            >{\raggedright\arraybackslash} b{3.0cm}
          }
          \toprule
          Extrinsic constraints & \makecell[bl]{$\mathcal{D}_{2h}$\\ $(k_1 \ne 1)$} & \makecell[bl]{$\mathcal{D}_{4h}$\\ $(k_1 = 1)$} \\
          \midrule
          $\left\lvert \alpha^g \beta^g \right\rvert$
            & -
            & - \\
           $\left\lvert \alpha^u \beta^g \right\rvert$
            & -
            & \makecell[tl]{
              $S_2^{-} \quad A_2^{-} \quad A_4^{-} \quad A_6$ \\
              $ \gamma^{-} \quad B_1^{-} \quad B_2^{-\pm}$} \\
          $\left\lvert \alpha^u \beta^u \right\rvert$
            & -
            & \makecell[tl]{
              $S_2^{\pm} \quad A_2^{\pm} \quad A_4^{\pm}$ \\
              $ \gamma^{\pm} \quad B_1^{\pm}$} \\
          \bottomrule
        \end{tabular}
      \end{table}

      In Table~\ref{tab:vanishingcoeffs-i}, we list the electron-integral coefficients $S$ and $A$ that are forced to vanish identically by particular combinations of spatial intrinsic constraints ($\mathcal{D}_{4h}$) and spin and spatial extrinsic constraints ($\left\lvert \alpha^u \beta^g \right\rvert$ and $\left\lvert \alpha^u \beta^u \right\rvert$) along pathway A (\textit{cf.} the explicit functional forms for these coefficients in Section~S-I of the Supplementary Material).
      Consequently, by Equations~\ref{eq:gammaetacoeffs} and~\ref{eq:Bcoeffs}, several related $\gamma$ and $B$ coefficients must vanish, as also listed in Table~\ref{tab:vanishingcoeffs-i}.
      This simplifies $u_{\sigma}$ and $v_{\sigma}$ via the various expressions for the $D$ and $\Delta$ coefficients in Table~\ref{tab:allsols} and eventually annihilates $\rho$ for both solutions $\mathrm{2a}$ and $\mathrm{2b}$, thus mandating these generally complex solutions to be real.
      In fact, the indicator plots for the extrinsic constraints $\left\lvert \alpha^u \beta^g \right\rvert$ and $\left\lvert \alpha^u \beta^u \right\rvert$ in Figure~\ref{fig:pathway-a-energy-i} show clearly the vanishing of $\rho$ at $\mathcal{D}_{4h}$.
      This explains the ability to numerically locate the transiently real solutions $\mathrm{B}$, $\mathrm{C}$, ${\mathrm{C}^5}'$, and $\mathrm{D}''$ in the vicinity of $\mathrm{D}_{4h}$ using conventional \gls*{acr:hf}, but not further away.
      This also reveals that there are four more solutions---namely, the transiently real solutions of $\left\lvert \alpha^g \beta^g \right\rvert$---that cannot be found numerically using conventional \gls*{acr:hf} for the particular length scale $a = \SI{1.058350}{\angstrom}$ even at $\mathrm{D}_{4h}$ since the combined extrinsic constraint of $\left\lvert \alpha^g \beta^g \right\rvert$ and intrinsic constraint of $\mathcal{D}_{4h}$ are not sufficient to force them to be real.

\begin{figure*}
        \centering
        \useexternalfile{0.9}{29.85779pt}{45.02512pt}{regimeplot.surf.D2h.rec.i.all.all}{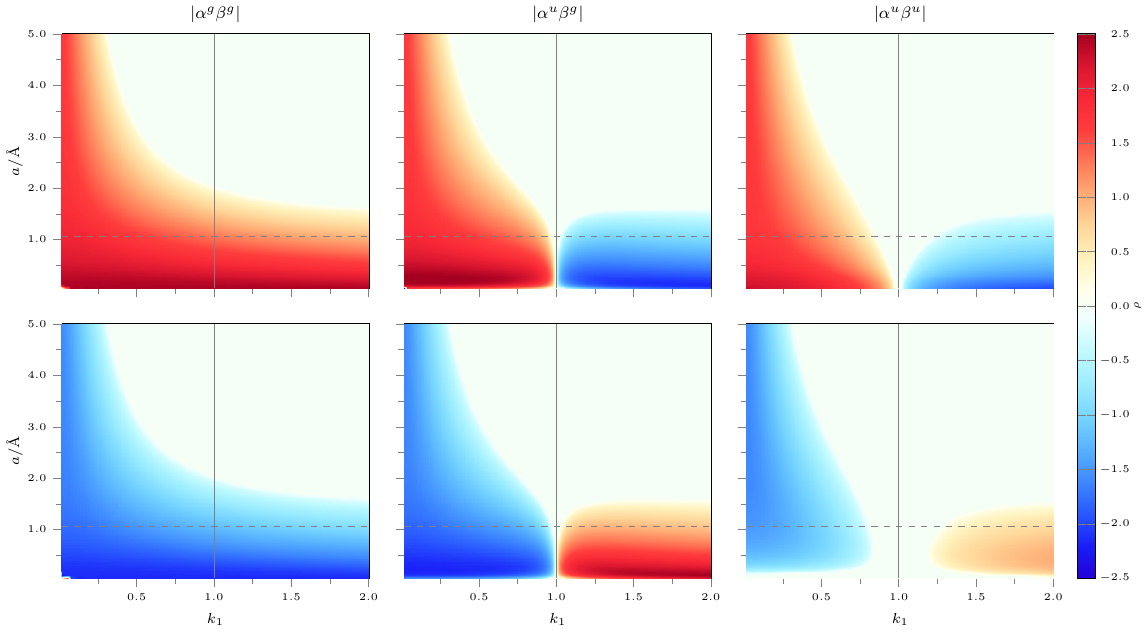}
        \caption{
          Reality indicator of solutions $\mathrm{2a}$ and $\mathrm{2b}$ along pathway A at different scale lengths $a$ for the special $\hat{i}$-symmetry-conserved extrinsic constraints.
          In each case, the reality indicator for solutions $\mathrm{2a}$ is shown in the top panel while that for solutions $\mathrm{2b}$ in the bottom panel.
          The vertical line at $k_1 = 1$ indicates $\mathcal{D}_{4h}$ symmetry, and the dashed horizontal line indicates $a = \SI{1.058350}{\angstrom}$.
          The real regimes are white whereas the non-real regimes are colored.
        }
        \label{fig:pathway-a-realityregimes-i}
      \end{figure*}

      The required vanishing of $\rho$ at $\mathcal{D}_{4h}$ for $\left\lvert \alpha^u \beta^g \right\rvert$ and $\left\lvert \alpha^u \beta^u \right\rvert$ depends only on the combinations of intrinsic and extrinsic symmetry constraints but not at all on the actual value of the scale length $a$.
      We thus consider such reality requirements to be \emph{strong}.
      This is illustrated in Figure~\ref{fig:pathway-a-realityregimes-i} where the reality indicator $\rho$ for solutions $\mathrm{2a}$ and $\mathrm{2b}$ is plotted over both geometrical parameters $a$ and $k_1$.
      These plots show that the aforementioned required reality for $\left\lvert \alpha^u \beta^g \right\rvert$ and $\left\lvert \alpha^u \beta^u \right\rvert$ at $\mathcal{D}_{4h}$ holds for all values of $a$.

      \tikzsetexternalprefix{./h4model-solutions/tikz/realityregimes/}
      \tikzexternalenable
      \tikzexternaldisable

    \subsubsection{Weak Reality Requirements}

      Figure~\ref{fig:pathway-a-realityregimes-i} also reveals the existence of ``seas'' of real solutions (white regimes) surrounding ``islands'' of non-real holomorphic solutions (colored regimes).
      In other words, these features show that, even at configurations where the electron-integral coefficients are not forced to vanish identically by symmetry as indicated in Table~\ref{tab:vanishingcoeffs-i}, the reality indicator $\rho$ can still vanish and the corresponding transiently real solutions are still required to be real-valued.
      This turns out to be a consequence of the fine balances between the electron-integral coefficients.
      An algebraic consideration as detailed in Section~S-III of the Supplementary Material shows that, under the constraints listed in Table~\ref{tab:extrinsicconstraints}, a more general condition for $\rho$ to vanish is
      \begin{subequations}
        \begin{equation}
          \frac{D_{1}^{-+} - D_{2}^{-}\sqrt{\Delta}}{D_{1}^{--} + D_{1}^{-+}} \ge 0 \quad \textrm{and} \quad
          \frac{D_{1}^{--} + D_{2}^{-}\sqrt{\Delta}}{D_{1}^{--} + D_{1}^{-+}} \ge 0
        \end{equation}%
      for solutions $\mathrm{2a}$, and
        \begin{equation}
          \frac{D_{1}^{-+} + D_{2}^{-}\sqrt{\Delta}}{D_{1}^{+-} + D_{1}^{++}} \ge 0 \quad \textrm{and} \quad
          \frac{D_{1}^{--} - D_{2}^{-}\sqrt{\Delta}}{D_{1}^{+-} + D_{1}^{++}} \ge 0
        \end{equation}%
        \label{eq:weakrealityineqs}%
      \end{subequations}%
      for solutions $\mathrm{2b}$.
      These inequalities implicitly define the regions of reality over the parameter space of interest, and they are now dependent on the scale length $a$ via the electron-integral coefficients, even though their forms are still fixed by the imposed constraints.
      For this reason, we consider the resulting reality requirements to be \emph{weak}.

    \subsubsection{Coalescence Boundaries}
    \label{subsubsec:coalescenceboundaries}

      When the conditions in Inequalities~\ref{eq:weakrealityineqs} cease to hold, the corresponding transiently real solutions become non-real.
      If the $D$ and $\Delta$ coefficients vary smoothly, then the onset of this transition occurs when
      \begin{subequations}
        \begin{equation}
          D_{1}^{-+} - D_{2}^{-}\sqrt{\Delta} = 0
          \quad \textrm{or} \quad
          D_{1}^{--} + D_{2}^{-}\sqrt{\Delta} = 0
        \end{equation}%
      for solutions $\mathrm{2a}$, and
        \begin{equation}
          D_{1}^{-+} + D_{2}^{-}\sqrt{\Delta} = 0
          \quad \textrm{or} \quad
          D_{1}^{--} - D_{2}^{-}\sqrt{\Delta} = 0
        \end{equation}%
        \label{eq:boundariesDcoeffs}%
      \end{subequations}%
      for solutions $\mathrm{2b}$.
      These equations give implicit descriptions of the locations of the \emph{coalescence boundaries} across which real conventional \gls*{acr:hf} solutions that we classify in this article as transiently real are commonly known to coalesce and disappear as they become holomorphically non-real.

      When the forms of the transiently real solutions $\mathrm{2a}$ and $\mathrm{2b}$ (Table~\ref{tab:allsols}) are subject to the conditions in Equation~\ref{eq:boundariesDcoeffs}, the primed and unprimed variants become identical to each other and also to one of the persistently real solutions.
      The coalescence boundaries in all cases exhibit a triple degeneracy (ignoring any additional degeneracies due to time-reversal symmetry) where each pair of corresponding transiently real solutions are required to coalesce with each other and with one of the persistently real solutions as they transition between the real and non-real regimes.
      Such coalescence points are also marked out in Figure~\ref{fig:pathway-a-energy-i} and can be seen to form one-dimensional boundaries over the $a$-$k_1$ plane in Figure~\ref{fig:pathway-a-realityregimes-i}.
\tikzsetexternalprefix{./h4model-symmetry/tikz/}

\section{\ce{[H4]^{2+}} Model: Solution Symmetry}
\label{sec:h4analmodel-symmetry}

  \tikzsetexternalprefix{./h4model-solutions/tikz/orbs/}
  \newlength{\orbsize}
  \setlength{\orbsize}{0.65cm}
  \tikzexternalenable
  \begin{table*}
    \centering
    \caption{
      Spin-orbitals of special $\hat{i}$-symmetry-conserved solutions at three representative points along molecular pathway A.
      Each circle represents the coefficient of a $1s$ \glsxtrshort*{acr:ao}: the circle's area is proportional to the magnitude of the coefficient, and the angular position of the dot on the circumference, as well as the hue of the fill color, indicates its phase.
      Circles corresponding to real coefficients have thicker boundaries.
      Shown beneath the spin-orbitals are the reality indicator $\rho$ and the symmetry classifications in $\mathcal{T} \otimes \mathcal{D}_{2h}$.
      The irreducible corepresentations in $\mathcal{T} \otimes \mathcal{D}_{2h}$ are denoted by $\prescript{\gamma}{}{\Gamma}$ where $\gamma$ is an irreducible corepresentation in $\mathcal{T}$ and $\Gamma$ an irreducible representation in $\mathcal{D}_{2h}$.
      Only one set of symmetry symbols is given for brevity if the solution symmetry is the same at all three $k_1$ values.
      See also the accompanied animations (Section S-II of the Supplementary Material) that show the variations of these spin-orbitals along pathway A.
    }
    \label{tab:pathway-a-orbitals-i}
    \ifdefined\twocolumnmode
      \renewcommand{\arraystretch}{1.3}
    \else
      \renewcommand{\arraystretch}{0.65}
    \fi
    \resizebox{\textwidth}{!}{%
      \begin{tabular}[t]{
  >{\raggedright\arraybackslash} m{0.5cm}
  *{3}{>{\centering\arraybackslash} m{2.0cm}}|
  *{3}{>{\centering\arraybackslash} m{2.0cm}}|
  *{3}{>{\centering\arraybackslash} m{2.0cm}}
}
  \toprule
  \multirow{2}{*}{Sol.} & \multicolumn{3}{c|}{$\lvert\alpha^g\beta^g\rvert$} & \multicolumn{3}{c|}{$\lvert\alpha^u\beta^g\rvert$} & \multicolumn{3}{c}{$\lvert\alpha^u\beta^u\rvert$}\\
  & $k_1 = 0.70$ & $k_1 = 1.00$ & $k_1 = 1.30$ & $k_1 = 0.70$ & $k_1 = 1.00$ & $k_1 = 1.30$ & $k_1 = 0.70$ & $k_1 = 1.00$ & $k_1 = 1.30$\\
  \midrule
  $\mathrm{1a}$
    & \parbox[m]{2.0cm}{\useexternalfile{0.67}{0}{0}{i.gg.k0.70.1a}{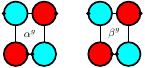}}
    & \parbox[m]{2.0cm}{\useexternalfile{0.67}{0}{0}{i.gg.k1.00.1a}{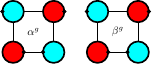}}
    & \parbox[m]{2.0cm}{\useexternalfile{0.67}{0}{0}{i.gg.k1.30.1a}{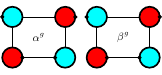}}
    & \parbox[m]{2.0cm}{\useexternalfile{0.67}{0}{0}{i.ug.k0.70.1a}{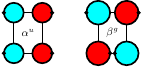}}
    & \parbox[m]{2.0cm}{\useexternalfile{0.67}{0}{0}{i.ug.k1.00.1a}{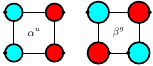}}
    & \parbox[m]{2.0cm}{\useexternalfile{0.67}{0}{0}{i.ug.k1.30.1a}{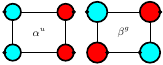}}
    & \parbox[m]{2.0cm}{\useexternalfile{0.67}{0}{0}{i.uu.k0.70.1a}{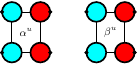}}
    & \parbox[m]{2.0cm}{\useexternalfile{0.67}{0}{0}{i.uu.k1.00.1a}{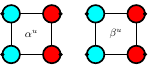}}
    & \parbox[m]{2.0cm}{\useexternalfile{0.67}{0}{0}{i.uu.k1.30.1a}{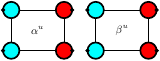}} \\
    & \multicolumn{3}{c|}{\scalebox{0.75}{$\prescript{A}{}{A}_{g}$}}
    & \multicolumn{3}{c|}{\scalebox{0.75}{$\prescript{A}{}{B}_{2u} \oplus \prescript{B}{}{B}_{2u}$}}
    & \multicolumn{3}{c}{\scalebox{0.75}{$\prescript{A}{}{A}_{g} $}} \\[8pt]
  $\mathrm{1b}$
    & \parbox[m]{2.0cm}{\useexternalfile{0.67}{0}{0}{i.gg.k0.70.1b}{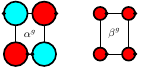}}
    & \parbox[m]{2.0cm}{\useexternalfile{0.67}{0}{0}{i.gg.k1.00.1b}{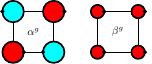}}
    & \parbox[m]{2.0cm}{\useexternalfile{0.67}{0}{0}{i.gg.k1.30.1b}{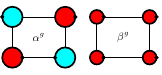}}
    & \parbox[m]{2.0cm}{\useexternalfile{0.67}{0}{0}{i.ug.k0.70.1b}{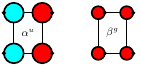}}
    & \parbox[m]{2.0cm}{\useexternalfile{0.67}{0}{0}{i.ug.k1.00.1b}{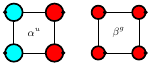}}
    & \parbox[m]{2.0cm}{\useexternalfile{0.67}{0}{0}{i.ug.k1.30.1b}{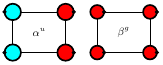}}
    & \parbox[m]{2.0cm}{\useexternalfile{0.67}{0}{0}{i.uu.k0.70.1b}{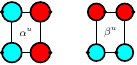}}
    & \parbox[m]{2.0cm}{\useexternalfile{0.67}{0}{0}{i.uu.k1.00.1b}{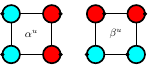}}
    & \parbox[m]{2.0cm}{\useexternalfile{0.67}{0}{0}{i.uu.k1.30.1b}{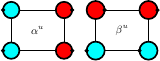}} \\
    & \multicolumn{3}{c|}{\scalebox{0.75}{$\prescript{A}{}{B}_{1g} \oplus \prescript{B}{}{B}_{1g}$}}
    & \multicolumn{3}{c|}{\scalebox{0.75}{$\prescript{A}{}{B}_{3u} \oplus \prescript{B}{}{B}_{3u}$}}
    & \multicolumn{3}{c}{\scalebox{0.75}{$\prescript{A}{}{B}_{1g} \oplus \prescript{B}{}{B}_{1g}$}} \\[8pt]
  $\mathrm{1c}$
    & \parbox[m]{2.0cm}{\useexternalfile{0.67}{0}{0}{i.gg.k0.70.1c}{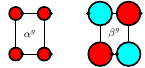}}
    & \parbox[m]{2.0cm}{\useexternalfile{0.67}{0}{0}{i.gg.k1.00.1c}{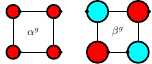}}
    & \parbox[m]{2.0cm}{\useexternalfile{0.67}{0}{0}{i.gg.k1.30.1c}{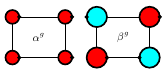}}
    & \parbox[m]{2.0cm}{\useexternalfile{0.67}{0}{0}{i.ug.k0.70.1c}{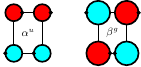}}
    & \parbox[m]{2.0cm}{\useexternalfile{0.67}{0}{0}{i.ug.k1.00.1c}{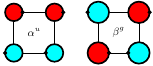}}
    & \parbox[m]{2.0cm}{\useexternalfile{0.67}{0}{0}{i.ug.k1.30.1c}{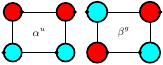}}
    & \parbox[m]{2.0cm}{\useexternalfile{0.67}{0}{0}{i.uu.k0.70.1c}{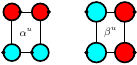}}
    & \parbox[m]{2.0cm}{\useexternalfile{0.67}{0}{0}{i.uu.k1.00.1c}{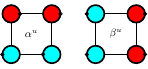}}
    & \parbox[m]{2.0cm}{\useexternalfile{0.67}{0}{0}{i.uu.k1.30.1c}{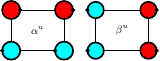}} \\
    & \multicolumn{3}{c|}{\scalebox{0.75}{$\prescript{A}{}{B}_{1g} \oplus \prescript{B}{}{B}_{1g}$}}
    & \multicolumn{3}{c|}{\scalebox{0.75}{$\prescript{A}{}{B}_{3u} \oplus \prescript{B}{}{B}_{3u}$}}
    & \multicolumn{3}{c}{\scalebox{0.75}{$\prescript{A}{}{B}_{1g} \oplus \prescript{B}{}{B}_{1g}$}} \\[8pt]
  $\mathrm{1d}$
    & \parbox[m]{2.0cm}{\useexternalfile{0.67}{0}{0}{i.gg.k0.70.1d}{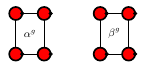}}
    & \parbox[m]{2.0cm}{\useexternalfile{0.67}{0}{0}{i.gg.k1.00.1d}{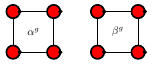}}
    & \parbox[m]{2.0cm}{\useexternalfile{0.67}{0}{0}{i.gg.k1.30.1d}{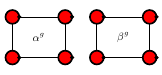}}
    & \parbox[m]{2.0cm}{\useexternalfile{0.67}{0}{0}{i.ug.k0.70.1d}{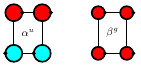}}
    & \parbox[m]{2.0cm}{\useexternalfile{0.67}{0}{0}{i.ug.k1.00.1d}{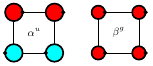}}
    & \parbox[m]{2.0cm}{\useexternalfile{0.67}{0}{0}{i.ug.k1.30.1d}{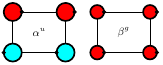}}
    & \parbox[m]{2.0cm}{\useexternalfile{0.67}{0}{0}{i.uu.k0.70.1d}{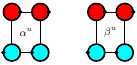}}
    & \parbox[m]{2.0cm}{\useexternalfile{0.67}{0}{0}{i.uu.k1.00.1d}{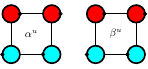}}
    & \parbox[m]{2.0cm}{\useexternalfile{0.67}{0}{0}{i.uu.k1.30.1d}{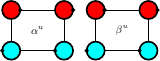}} \\
    & \multicolumn{3}{c|}{\scalebox{0.75}{$\prescript{A}{}{A}_{g}$}}
    & \multicolumn{3}{c|}{\scalebox{0.75}{$\prescript{A}{}{B}_{2u} \oplus \prescript{B}{}{B}_{2u}$}}
    & \multicolumn{3}{c}{\scalebox{0.75}{$\prescript{A}{}{A}_{g}$}} \\[12pt]
  $\mathrm{2a}$
    & \parbox[m]{2.0cm}{\useexternalfile{0.67}{0}{0}{i.gg.k0.70.2a}{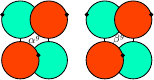}}
    & \parbox[m]{2.0cm}{\useexternalfile{0.67}{0}{0}{i.gg.k1.00.2a}{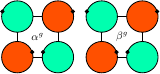}}
    & \parbox[m]{2.0cm}{\useexternalfile{0.67}{0}{0}{i.gg.k1.30.2a}{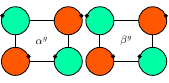}}
    & \parbox[m]{2.0cm}{\useexternalfile{0.67}{0}{0}{i.ug.k0.70.2a}{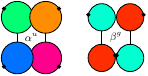}}
    & \parbox[m]{2.0cm}{\useexternalfile{0.67}{0}{0}{i.ug.k1.00.2a}{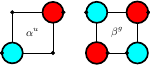}}
    & \parbox[m]{2.0cm}{\useexternalfile{0.67}{0}{0}{i.ug.k1.30.2a}{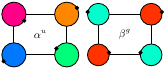}}
    & \parbox[m]{2.0cm}{\useexternalfile{0.67}{0}{0}{i.uu.k0.70.2a}{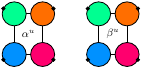}}
    & \parbox[m]{2.0cm}{\useexternalfile{0.67}{0}{0}{i.uu.k1.00.2a}{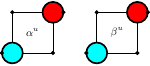}}
    & \parbox[m]{2.0cm}{\useexternalfile{0.67}{0}{0}{i.uu.k1.30.2a}{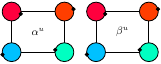}} \\
    & \scalebox{0.75}{$\rho = +1.514$}
    & \scalebox{0.75}{$\rho = +1.331$}
    & \scalebox{0.75}{$\rho = +1.177$}
    & \scalebox{0.75}{$\rho = +1.409$}
    & \scalebox{0.75}{$\rho = +0.000$}
    & \scalebox{0.75}{$\rho = -0.984$}
    & \scalebox{0.75}{$\rho = +0.791$}
    & \scalebox{0.75}{$\rho = +0.000$}
    & \scalebox{0.75}{$\rho = -0.326$} \\[-2pt]
    & \multicolumn{3}{c|}{\scalebox{0.75}{$\prescript{A}{}{A}_{g} \oplus \prescript{B}{}{B}_{1g}$}}
    & \multicolumn{3}{c|}{\scalebox{0.75}{$\prescript{A}{}{B}_{2u} \oplus \prescript{A}{}{B}_{3u} \oplus \prescript{B}{}{B}_{2u} \oplus \prescript{B}{}{B}_{3u}\hphantom{\,\oplus}$}}
    & \scalebox{0.75}{$\prescript{A}{}{A}_{g} \oplus \prescript{B}{}{B}_{1g}$}
    & \scalebox{0.75}{$\prescript{A}{}{A}_{g} \oplus \prescript{A}{}{B}_{1g}$}
    & \scalebox{0.75}{$\prescript{A}{}{A}_{g} \oplus \prescript{B}{}{B}_{1g}$} \\[8pt]
  $\mathrm{2a}'$
    & \parbox[m]{2.0cm}{\useexternalfile{0.67}{0}{0}{i.gg.k0.70.2ap}{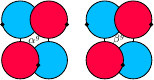}}
    & \parbox[m]{2.0cm}{\useexternalfile{0.67}{0}{0}{i.gg.k1.00.2ap}{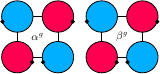}}
    & \parbox[m]{2.0cm}{\useexternalfile{0.67}{0}{0}{i.gg.k1.30.2ap}{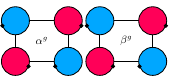}}
    & \parbox[m]{2.0cm}{\useexternalfile{0.67}{0}{0}{i.ug.k0.70.2ap}{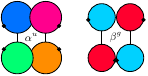}}
    & \parbox[m]{2.0cm}{\useexternalfile{0.67}{0}{0}{i.ug.k1.00.2ap}{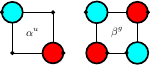}}
    & \parbox[m]{2.0cm}{\useexternalfile{0.67}{0}{0}{i.ug.k1.30.2ap}{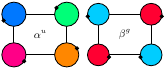}}
    & \parbox[m]{2.0cm}{\useexternalfile{0.67}{0}{0}{i.uu.k0.70.2ap}{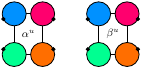}}
    & \parbox[m]{2.0cm}{\useexternalfile{0.67}{0}{0}{i.uu.k1.00.2ap}{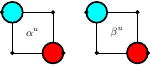}}
    & \parbox[m]{2.0cm}{\useexternalfile{0.67}{0}{0}{i.uu.k1.30.2ap}{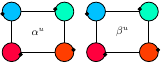}} \\
    & \scalebox{0.75}{$\rho = -1.514$}
    & \scalebox{0.75}{$\rho = -1.331$}
    & \scalebox{0.75}{$\rho = -1.177$}
    & \scalebox{0.75}{$\rho = -1.409$}
    & \scalebox{0.75}{$\rho = +0.000$}
    & \scalebox{0.75}{$\rho = +0.984$}
    & \scalebox{0.75}{$\rho = -0.791$}
    & \scalebox{0.75}{$\rho = +0.000$}
    & \scalebox{0.75}{$\rho = +0.326$} \\[-2pt]
    & \multicolumn{3}{c|}{\scalebox{0.75}{$\prescript{A}{}{A}_{g} \oplus \prescript{B}{}{B}_{1g}$}}
    & \multicolumn{3}{c|}{\scalebox{0.75}{$\prescript{A}{}{B}_{2u} \oplus \prescript{A}{}{B}_{3u} \oplus \prescript{B}{}{B}_{2u} \oplus \prescript{B}{}{B}_{3u}\hphantom{\,\oplus}$}}
    & \scalebox{0.75}{$\prescript{A}{}{A}_{g} \oplus \prescript{B}{}{B}_{1g}$}
    & \scalebox{0.75}{$\prescript{A}{}{A}_{g} \oplus \prescript{A}{}{B}_{1g}$}
    & \scalebox{0.75}{$\prescript{A}{}{A}_{g} \oplus \prescript{B}{}{B}_{1g}$} \\[8pt]
  $\mathrm{2b}$
    & \parbox[m]{2.0cm}{\useexternalfile{0.67}{0}{0}{i.gg.k0.70.2b}{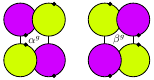}}
    & \parbox[m]{2.0cm}{\useexternalfile{0.67}{0}{0}{i.gg.k1.00.2b}{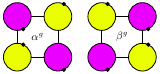}}
    & \parbox[m]{2.0cm}{\useexternalfile{0.67}{0}{0}{i.gg.k1.30.2b}{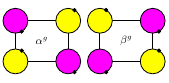}}
    & \parbox[m]{2.0cm}{\useexternalfile{0.67}{0}{0}{i.ug.k0.70.2b}{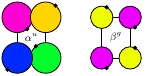}}
    & \parbox[m]{2.0cm}{\useexternalfile{0.67}{0}{0}{i.ug.k1.00.2b}{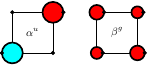}}
    & \parbox[m]{2.0cm}{\useexternalfile{0.67}{0}{0}{i.ug.k1.30.2b}{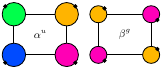}}
    & \parbox[m]{2.0cm}{\useexternalfile{0.67}{0}{0}{i.uu.k0.70.2b}{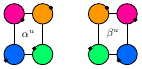}}
    & \parbox[m]{2.0cm}{\useexternalfile{0.67}{0}{0}{i.uu.k1.00.2b}{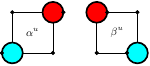}}
    & \parbox[m]{2.0cm}{\useexternalfile{0.67}{0}{0}{i.uu.k1.30.2b}{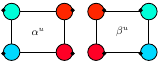}} \\
    & \scalebox{0.75}{$\rho = -1.364$}
    & \scalebox{0.75}{$\rho = -1.202$}
    & \scalebox{0.75}{$\rho = -1.075$}
    & \scalebox{0.75}{$\rho = -1.368$}
    & \scalebox{0.75}{$\rho = +0.000$}
    & \scalebox{0.75}{$\rho = +0.972$}
    & \scalebox{0.75}{$\rho = -0.589$}
    & \scalebox{0.75}{$\rho = +0.000$}
    & \scalebox{0.75}{$\rho = +0.131$} \\[-2pt]
    & \multicolumn{3}{c|}{\scalebox{0.75}{$\prescript{A}{}{A}_{g} \oplus \prescript{A}{}{B}_{1g}$}}
    & \multicolumn{3}{c|}{\scalebox{0.75}{$\prescript{A}{}{B}_{2u} \oplus \prescript{A}{}{B}_{3u} \oplus \prescript{B}{}{B}_{2u} \oplus \prescript{B}{}{B}_{3u}$}}
    & \scalebox{0.75}{$\prescript{A}{}{A}_{g} \oplus \prescript{A}{}{B}_{1g}$}
    & \scalebox{0.75}{$\prescript{A}{}{A}_{g} \oplus \prescript{B}{}{B}_{1g}$}
    & \scalebox{0.75}{$\prescript{A}{}{A}_{g} \oplus \prescript{A}{}{B}_{1g}$} \\[8pt]
  $\mathrm{2b}'$
    & \parbox[m]{2.0cm}{\useexternalfile{0.67}{0}{0}{i.gg.k0.70.2bp}{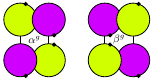}}
    & \parbox[m]{2.0cm}{\useexternalfile{0.67}{0}{0}{i.gg.k1.00.2bp}{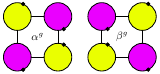}}
    & \parbox[m]{2.0cm}{\useexternalfile{0.67}{0}{0}{i.gg.k1.30.2bp}{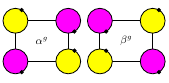}}
    & \parbox[m]{2.0cm}{\useexternalfile{0.67}{0}{0}{i.ug.k0.70.2bp}{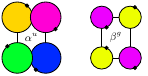}}
    & \parbox[m]{2.0cm}{\useexternalfile{0.67}{0}{0}{i.ug.k1.00.2bp}{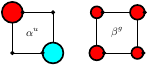}}
    & \parbox[m]{2.0cm}{\useexternalfile{0.67}{0}{0}{i.ug.k1.30.2bp}{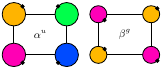}}
    & \parbox[m]{2.0cm}{\useexternalfile{0.67}{0}{0}{i.uu.k0.70.2bp}{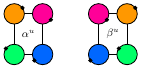}}
    & \parbox[m]{2.0cm}{\useexternalfile{0.67}{0}{0}{i.uu.k1.00.2bp}{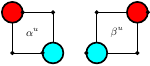}}
    & \parbox[m]{2.0cm}{\useexternalfile{0.67}{0}{0}{i.uu.k1.30.2bp}{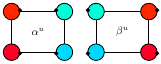}} \\
    & \scalebox{0.75}{$\rho = +1.364$}
    & \scalebox{0.75}{$\rho = +1.202$}
    & \scalebox{0.75}{$\rho = +1.075$}
    & \scalebox{0.75}{$\rho = +1.368$}
    & \scalebox{0.75}{$\rho = +0.000$}
    & \scalebox{0.75}{$\rho = -0.972$}
    & \scalebox{0.75}{$\rho = +0.589$}
    & \scalebox{0.75}{$\rho = +0.000$}
    & \scalebox{0.75}{$\rho = -0.131$} \\[-2pt]
    & \multicolumn{3}{c|}{\scalebox{0.75}{$\prescript{A}{}{A}_{g} \oplus \prescript{A}{}{B}_{1g}$}}
    & \multicolumn{3}{c|}{\scalebox{0.75}{$\prescript{A}{}{B}_{2u} \oplus \prescript{A}{}{B}_{3u} \oplus \prescript{B}{}{B}_{2u} \oplus \prescript{B}{}{B}_{3u}$}}
    & \scalebox{0.75}{$\prescript{A}{}{A}_{g} \oplus \prescript{A}{}{B}_{1g}$}
    & \scalebox{0.75}{$\prescript{A}{}{A}_{g} \oplus \prescript{B}{}{B}_{1g}$}
    & \scalebox{0.75}{$\prescript{A}{}{A}_{g} \oplus \prescript{A}{}{B}_{1g}$} \\
  \bottomrule
\end{tabular}
    }

    \vspace{10pt}
    \ifdefined\usecentralfigs
      \includegraphics[scale=0.7]{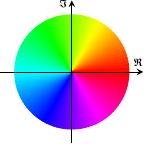}
    \else
      \includegraphics[scale=0.7]{h4model-symmetry/tikz/tikzoutput/colorwheel}
    \fi
  \end{table*}
  \tikzexternaldisable

  \subsection{Complex-Conjugation Symmetry}
  \label{subsec:conjsym}

    \subsubsection{Real Regimes}

      The solutions of $\bar{P}\left(\boldsymbol{z}; \boldsymbol{B}\right) = 0$ exhibit special behaviors under complex conjugation due to the forms they adopt.
      Trivially, the persistently real solutions $\mathrm{1a}$--$\mathrm{1d}$ are invariant under the action of $\gls*{op:holoconj}$ (defined in Equation~\ref{eq:kappaPsi}), so that their conventional and holomorphic energies coincide and must both be real:
      \begin{subequations}
        \begin{align}
            E[\gls*{wf:hf}]
            &= \frac{\braket{\gls*{wf:hf} | \gls*{op:hamil} | \gls*{wf:hf}}}{\braket{\gls*{wf:hf} | \gls*{wf:hf}}}
            = \frac{\braket{\gls*{op:holoconj}\gls*{wf:hf} | \gls*{op:hamil} | \gls*{op:holoconj}\gls*{wf:hf}}^*}{\braket{\gls*{op:holoconj}\gls*{wf:hf} | \gls*{op:holoconj}\gls*{wf:hf}}^*} \nonumber\\
            &= \frac{\braket{\gls*{wf:hf} | \gls*{op:hamil} | \gls*{wf:hf}}^*}{\braket{\gls*{wf:hf} | \gls*{wf:hf}}^*}
            = E^*[\gls*{wf:hf}],\\
          \tilde{E}[\gls*{wf:hf}]
            &= \frac{\braket{\gls*{op:holoconj}\gls*{wf:hf} | \gls*{op:hamil} | \gls*{wf:hf}}}{\braket{\gls*{op:holoconj}\gls*{wf:hf} | \gls*{wf:hf}}}
            = \frac{\braket{\gls*{wf:hf} | \gls*{op:hamil} | \gls*{wf:hf}}}{\braket{\gls*{wf:hf} | \gls*{wf:hf}}}
            = E[\gls*{wf:hf}],
        \end{align}%
        \label{eq:realconvholoenergy}%
      \end{subequations}%
      where we have used the antiunitarity of $\gls*{op:holoconj}$ and its commutativity with $\gls*{op:hamil}$ for the second equality in the first line.

      The transiently real solutions, however, behave less straightforwardly.
      Let us take $\gls*{wf:hf}$ to be any one of them.
      In the regimes where the reality indicator $\rho$ vanishes for this solution, its angular parameters $\theta_{\sigma}$ are real, its spin-orbitals $\chi_i$ must be real-valued, and $\gls*{wf:hf}$ itself must once again be invariant under $\gls*{op:holoconj}$.
      The conditions of Equation~\ref{eq:realconvholoenergy} thus apply and the same conclusion can be drawn for the energies of this solution in these regimes of reality.

    \subsubsection{Non-Real Regimes}
      \label{subsubsec:nonrealregimes}
      \paragraph{Energy Reality.}
        Outside of the real regimes, $\rho$ is non-zero, $\theta_{\sigma}$ is complex, and generally there is not much to say about the behavior of $\gls*{wf:hf}$ under $\gls*{op:holoconj}$.
        However, it turns out that the real part of $[\theta_{\sigma}]$ is \emph{not} arbitrary, and this dictates how $\gls*{op:holoconj}$ affects $\gls*{wf:hf}$.
        From Equations~\ref{eq:uvrecast}~and~\ref{eq:thetarecast}, we obtain
        \begin{equation}
          \Re[\theta_{\sigma}]
          = \varphi_{\sigma} = \Arg(u_{\sigma} \pm iv_{\sigma})
        \end{equation}
        where $\Arg$ is the principal argument function.
        As explained in Section~S-III of the Supplementary Material, under the conditions in this work, $u_{\sigma}$ and $v_{\sigma}$ are either purely real or purely imaginary, and hence, for $\rho$ to be non-zero, one of them must be real and the other one imaginary.
        Consequently, $u_{\sigma} \pm iv_{\sigma}$ must also be either purely real or purely imaginary, so $\varphi_{\sigma}$ can only take values of $0$ or $\pm\pi/2$, and by virtue of Equation~\ref{eq:MOcoeffparams}, we deduce that
        \begin{equation*}
          \begin{cases*}
            (G_i^{\sigma1,\cdot})^* = -G_i^{\sigma2,\cdot} & if $[\theta_{\sigma}] = i\rho_{\sigma}$,\\
            (G_i^{\sigma1,\cdot})^* = +G_i^{\sigma2,\cdot} & if $[\theta_{\sigma}] = \pm\frac{\pi}{2} + i\rho_{\sigma}$.
          \end{cases*}
        \end{equation*}
        Because of these relations, the effect of $\gls*{op:holoconj}$ on $\gls*{wf:hf}$ as given by Equations~\ref{eq:kappaPsi}~and~\ref{eq:chiconj} is thus
        \begin{equation}
          \gls*{op:holoconj}\gls*{wf:hf} = \pm\hat{R}\gls*{wf:hf}
          \label{eq:timerevunitaryop}
        \end{equation}
        where $\hat{R}$ interchanges the \gls*{acr:mo} coefficients on \ce{H^1} with those on \ce{H^2} and those on \ce{H^3} with those on \ce{H^4} for every spin-orbital $\chi_i$ (see Table~\ref{tab:extrinsicconstraints}).
        The operator $\hat{R}$ is therefore a unitary spatial-symmetry operation of the system.
        As $\gls*{op:holoconj}$ is involutory, \textit{i.e.}, $\gls*{op:holoconj}^2 = \id$\cite{book:Wigner1959}, it follows that $\hat{R}$ must be too.
        For all of the constraints in this work, we can identify $\hat{R}$ with the $\hat{C}_2^{y}$ rotation whose axis forms the common perpendicular bisector of the \ce{H^1}---\ce{H^2} and \ce{H^3}---\ce{H^4} bonds.
        This relation can be better appreciated by inspecting the forms of the spin-orbitals for the transiently real solutions with $\rho \ne 0$ in Table~\ref{tab:pathway-a-orbitals-i}.
        We then deduce that the conventional energy is still real:
        \begin{subequations}
          \ifdefined\twocolumnmode
            \begin{align}
              E[\gls*{wf:hf}]
                &= \frac{\braket{\gls*{wf:hf} | \gls*{op:hamil} | \gls*{wf:hf}}}{\braket{\gls*{wf:hf} | \gls*{wf:hf}}}
                = \frac{\braket{\gls*{op:holoconj}\gls*{wf:hf} | \gls*{op:hamil} | \gls*{op:holoconj}\gls*{wf:hf}}^*}{\braket{\gls*{op:holoconj}\gls*{wf:hf} | \gls*{op:holoconj}\gls*{wf:hf}}^*} \nonumber\\
                &= \frac{\braket{\hat{R}\gls*{wf:hf} | \gls*{op:hamil} | \hat{R}\gls*{wf:hf}}^*}{\braket{\hat{R}\gls*{wf:hf} | \hat{R}\gls*{wf:hf}}^*}
                = \frac{\braket{\gls*{wf:hf} | \gls*{op:hamil} | \gls*{wf:hf}}^*}{\braket{\gls*{wf:hf} | \gls*{wf:hf}}^*} \nonumber\\
                &= E^*[\gls*{wf:hf}],
              \label{eq:realconvEcomplexregion}
            \end{align}
          \else
            \begin{align}
              E[\gls*{wf:hf}]
                &= \frac{\braket{\gls*{wf:hf} | \gls*{op:hamil} | \gls*{wf:hf}}}{\braket{\gls*{wf:hf} | \gls*{wf:hf}}}
                = \frac{\braket{\gls*{op:holoconj}\gls*{wf:hf} | \gls*{op:hamil} | \gls*{op:holoconj}\gls*{wf:hf}}^*}{\braket{\gls*{op:holoconj}\gls*{wf:hf} | \gls*{op:holoconj}\gls*{wf:hf}}^*} \nonumber\\
                &= \frac{\braket{\hat{R}\gls*{wf:hf} | \gls*{op:hamil} | \hat{R}\gls*{wf:hf}}^*}{\braket{\hat{R}\gls*{wf:hf} | \hat{R}\gls*{wf:hf}}^*}
                = \frac{\braket{\gls*{wf:hf} | \gls*{op:hamil} | \gls*{wf:hf}}^*}{\braket{\gls*{wf:hf} | \gls*{wf:hf}}^*} = E^*[\gls*{wf:hf}],
              \label{eq:realconvEcomplexregion}
            \end{align}
          \fi
          but the holomorphic energy is no longer identical to the conventional energy:
          \ifdefined\twocolumnmode
            \begin{align}
              \tilde{E}[\gls*{wf:hf}]
                &= \frac{\braket{\gls*{op:holoconj}\gls*{wf:hf} | \gls*{op:hamil} | \gls*{wf:hf}}}{\braket{\gls*{op:holoconj}\gls*{wf:hf} | \gls*{wf:hf}}}
                = \frac{\braket{\hat{R}\gls*{wf:hf} | \gls*{op:hamil} | \gls*{wf:hf}}}{\braket{\hat{R}\gls*{wf:hf} | \gls*{wf:hf}}} \nonumber\\
                &\ne E[\gls*{wf:hf}].
            \end{align}
          \else
            \begin{align}
              \tilde{E}[\gls*{wf:hf}]
                &= \frac{\braket{\gls*{op:holoconj}\gls*{wf:hf} | \gls*{op:hamil} | \gls*{wf:hf}}}{\braket{\gls*{op:holoconj}\gls*{wf:hf} | \gls*{wf:hf}}}
                = \frac{\braket{\hat{R}\gls*{wf:hf} | \gls*{op:hamil} | \gls*{wf:hf}}}{\braket{\hat{R}\gls*{wf:hf} | \gls*{wf:hf}}} \ne E[\gls*{wf:hf}].
            \end{align}
          \fi
          In addition, making use of the involutority and unitarity of $\hat{R}$, we obtain
          \ifdefined\twocolumnmode
            \begin{align}
              \tilde{E}[\gls*{wf:hf}]
                &= \frac{\braket{\gls*{op:holoconj}\gls*{wf:hf} | \gls*{op:hamil} | \gls*{wf:hf}}}{\braket{\gls*{op:holoconj}\gls*{wf:hf} | \gls*{wf:hf}}}
                = \frac{\braket{\gls*{wf:hf} | \gls*{op:hamil} | \gls*{op:holoconj}\gls*{wf:hf}}^*}{\braket{\gls*{wf:hf} | \gls*{op:holoconj}\gls*{wf:hf}}^*} \nonumber\\
                &= \frac{\braket{\gls*{wf:hf} | \gls*{op:hamil} | \hat{R}\gls*{wf:hf}}^*}{\braket{\gls*{wf:hf} | \hat{R}\gls*{wf:hf}}^*}
                = \frac{\braket{\hat{R}\gls*{wf:hf} | \gls*{op:hamil} | \gls*{wf:hf}}^*}{\braket{\hat{R}\gls*{wf:hf} | \gls*{wf:hf}}^*} \nonumber\\
                &= \tilde{E}^*[\gls*{wf:hf}],
                \label{eq:realholoEcomplexregion}
            \end{align}%
          \else
            \begin{align}
              \tilde{E}[\gls*{wf:hf}]
                &= \frac{\braket{\gls*{op:holoconj}\gls*{wf:hf} | \gls*{op:hamil} | \gls*{wf:hf}}}{\braket{\gls*{op:holoconj}\gls*{wf:hf} | \gls*{wf:hf}}}
                = \frac{\braket{\gls*{wf:hf} | \gls*{op:hamil} | \gls*{op:holoconj}\gls*{wf:hf}}^*}{\braket{\gls*{wf:hf} | \gls*{op:holoconj}\gls*{wf:hf}}^*} \nonumber\\
                &= \frac{\braket{\gls*{wf:hf} | \gls*{op:hamil} | \hat{R}\gls*{wf:hf}}^*}{\braket{\gls*{wf:hf} | \hat{R}\gls*{wf:hf}}^*}
                = \frac{\braket{\hat{R}\gls*{wf:hf} | \gls*{op:hamil} | \gls*{wf:hf}}^*}{\braket{\hat{R}\gls*{wf:hf} | \gls*{wf:hf}}^*} = \tilde{E}^*[\gls*{wf:hf}],
                \label{eq:realholoEcomplexregion}
            \end{align}%
          \fi
          \label{eq:complconvholoenergy}%
        \end{subequations}%
        which shows that the holomorphic energy must also be real, despite the non-real coefficients.
        Therefore, there is no need to show the imaginary parts of the holomorphic energy for the transiently real solutions in Figure~\ref{fig:pathway-a-energy-i}.

        The required reality exhibited by the energy functionals in non-real-coefficient regimes is due specifically to Equation~\ref{eq:timerevunitaryop} which relates the action of an antiunitary operator $\gls*{op:holoconj}$ to that of an involutory unitary spatial-symmetry operator $\hat{R}$.
        In fact, more generally, for any determinant $\gls*{wf:hf}$ with complex \gls*{acr:mo} coefficients, if we insist that the energies be real, then
        \ifdefined\twocolumnmode
          \begin{align*}
            E[\gls*{wf:hf}] &= E^*[\gls*{wf:hf}]\\
            \Rightarrow
            \frac{\braket{\gls*{wf:hf} | \gls*{op:hamil} | \gls*{wf:hf}}}{\braket{\gls*{wf:hf} | \gls*{wf:hf}}} &= \frac{\braket{\gls*{op:holoconj}\gls*{wf:hf} | \gls*{op:hamil} | \gls*{op:holoconj}\gls*{wf:hf}}}{\braket{\gls*{op:holoconj}\gls*{wf:hf} | \gls*{op:holoconj}\gls*{wf:hf}}}
          \end{align*}
        \else
          \begin{equation*}
            E[\gls*{wf:hf}] = E^*[\gls*{wf:hf}]
            \Rightarrow
            \frac{\braket{\gls*{wf:hf} | \gls*{op:hamil} | \gls*{wf:hf}}}{\braket{\gls*{wf:hf} | \gls*{wf:hf}}} = \frac{\braket{\gls*{op:holoconj}\gls*{wf:hf} | \gls*{op:hamil} | \gls*{op:holoconj}\gls*{wf:hf}}}{\braket{\gls*{op:holoconj}\gls*{wf:hf} | \gls*{op:holoconj}\gls*{wf:hf}}}
          \end{equation*}
        \fi
        and
        \ifdefined\twocolumnmode
          \begin{align*}
            \tilde{E}[\gls*{wf:hf}] &= \tilde{E}^*[\gls*{wf:hf}]\\
            \Rightarrow
            \frac{\braket{\gls*{op:holoconj}\gls*{wf:hf} | \gls*{op:hamil} | \gls*{wf:hf}}}{\braket{\gls*{op:holoconj}\gls*{wf:hf} | \gls*{wf:hf}}} &= \frac{\braket{\gls*{wf:hf} | \gls*{op:hamil} | \gls*{op:holoconj}\gls*{wf:hf}}}{\braket{\gls*{wf:hf} | \gls*{op:holoconj}\gls*{wf:hf}}},
          \end{align*}
        \else
          \begin{equation*}
            \tilde{E}[\gls*{wf:hf}] = \tilde{E}^*[\gls*{wf:hf}]
            \Rightarrow
            \frac{\braket{\gls*{op:holoconj}\gls*{wf:hf} | \gls*{op:hamil} | \gls*{wf:hf}}}{\braket{\gls*{op:holoconj}\gls*{wf:hf} | \gls*{wf:hf}}} = \frac{\braket{\gls*{wf:hf} | \gls*{op:hamil} | \gls*{op:holoconj}\gls*{wf:hf}}}{\braket{\gls*{wf:hf} | \gls*{op:holoconj}\gls*{wf:hf}}},
          \end{equation*}
        \fi
        both of which imply that the action of $\gls*{op:holoconj}$ on $\gls*{wf:hf}$ must be identifiable with an involutory unitary symmetry operation $\hat{U}$,
        \begin{equation}
          \gls*{op:holoconj}\gls*{wf:hf} = \pm \hat{U}\gls*{wf:hf},
        \end{equation}
        where $\hat{U}$ need not be a spatial symmetry operation, as long as $\hat{U}$ acts on the same wavefunction space as $\gls*{op:holoconj}$ and $\gls*{op:hamil}$ do and $\hat{U}$ commutes with $\gls*{op:hamil}$.

      \paragraph{Transiently Real Pairs.}
        The reality of $\tilde{E}[\gls*{wf:hf}]$ means that, if $\tilde{E}[\gls*{wf:hf}]$ is a stationary point in the \gls*{acr:scf} landscape with a particular set of extrinsic constraints, then so is $\tilde{E}[\gls*{op:holoconj}\gls*{wf:hf}] = \tilde{E}^*[\gls*{wf:hf}] = \tilde{E}[\gls*{wf:hf}]$.
        Thus, in non-real regimes, if $\gls*{wf:hf}$ is one of the transiently real solutions, then $\gls*{op:holoconj}\gls*{wf:hf}$ must be a \emph{different} transiently real solution, which means that the transiently real solutions must occur in complex-conjugate pairs when they are non-real.
        This is not at all surprising given the algebraic form of $P\left(\boldsymbol{\theta}; \boldsymbol{B}\right) = 0$ in Equation~\ref{eq:Psys} in which all of the $B$ coefficients are real.
        In fact, from Table~\ref{tab:allsols} and from the fact that $u_{\sigma}$ and $v_{\sigma}$ are either purely real or purely imaginary, it can be shown that $\rho = -\rho'$, where $\rho$ and $\rho'$ are the reality indicators, which are also the imaginary parts, of the unprimed and primed $\mathrm{2a}$ and $\mathrm{2b}$ solutions, respectively.

        Due to the identification of $\gls*{op:holoconj}$ with $\hat{C}_2^{y}$ in Equation~\ref{eq:timerevunitaryop}, the primed and unprimed solutions are also related by the $\hat{C}_2^{y}$ spatial symmetry in the non-real regimes.
        But unlike the antilinear $\gls*{op:holoconj}$, $\hat{C}_2^{y}$ is linear and we expect its action on $\gls*{wf:hf}$ to be independent of whether $\gls*{wf:hf}$ is real or non-real.
        The $\hat{C}_2^{y}$-relation between the primed and unprimed solutions therefore persists through all regimes.
        Consider for example the forms of the $\mathrm{2a}$ and $\mathrm{2a}'$ solutions in the $\lvert\alpha^u\beta^g\rvert$ constraint shown in Table~\ref{tab:pathway-a-orbitals-i}.
        The $\hat{C}_2^{y}$-relation between $\mathrm{2a}$ and $\mathrm{2a}'$ can be seen at all three values of $k_1$ plotted, regardless of whether these solutions are real or not.
        The $\gls*{op:holoconj}$-relation between them, however, is absent at $k_1 = 1.00$ where these two solutions are real.

  \subsection{Time-Reversal Symmetry}
  \label{subsec:timerevsym}

    Generally, on the domain $\mathscr{D}$ within each \emph{spatial} extrinsic constraining space (Table~\ref{tab:extrinsicconstraints}), if $x \ne y$, the spatial parts of $\chi_1$ and $\chi_2$ must always be linearly independent.
    In other words, if $\hat{q}$ is a one-particle spin rotation through an angle of $\pi$ about the $y$-axis such that
    \begin{equation*}
      \hat{q}
      \begin{pmatrix}
        \alpha & \beta
      \end{pmatrix}
      =
      \begin{pmatrix}
        \alpha & \beta
      \end{pmatrix}
      \begin{pmatrix*}[r]
        0 & -1\\
        1 &  0
      \end{pmatrix*},
    \end{equation*}
    then $\hat{q}\chi_1$ and $\chi_2$ are linearly independent, as are $\chi_1$ and $\hat{q}\chi_2$.
    Hence, for any of the solutions $\gls*{wf:hf}$ obtained under an extrinsic constraint with $x \ne y$ such as $\lvert\alpha^u \beta^g\rvert$, if we define the total spin rotation
    \begin{equation}
      \label{eq:totspinrot}
      \gls*{op:piyspinrot} \equiv \hat{q}_1\hat{q}_2
    \end{equation}
    where $\hat{q}_i$ acts on the $i$\textsuperscript{th} particle, then $\gls*{wf:hf}$ and $\gls*{op:piyspinrot}\gls*{wf:hf}$ must be linearly independent.
    On the other hand, when $x = y$, such as in the $\lvert\alpha^g \beta^g\rvert$ and $\lvert\alpha^u \beta^u\rvert$ cases, the linear independence between $\gls*{wf:hf}$ and $\gls*{op:piyspinrot}\gls*{wf:hf}$ is no longer always guaranteed.
    We will distinguish between two kinds of solutions: in the first kind, the spatial parts of $\chi_1$ and $\chi_2$ are identical up to a phase factor such that $\hat{q}\chi_1 = \pm \chi_2$ and hence $\gls*{op:piyspinrot}\gls*{wf:hf} = \pm\gls*{wf:hf}$; and in the second kind, $\hat{q}\chi_1 \ne \pm \chi_2$ so that $\gls*{op:piyspinrot}\gls*{wf:hf}$ and $\gls*{wf:hf}$ are linearly independent.
    For brevity, we shall refer to solutions of the first kind as \emph{\glsxtrshort*{acr:rhf}-like solutions} and those of the second kind \emph{non-\glsxtrshort*{acr:rhf} solutions}.
    Note that, despite these names, all of the solutions discussed in this work are still \gls*{acr:uhf} in nature due to the spin extrinsic constraint under which they were obtained (see Section~\ref{subsec:spinspatialextrinsicconstraints_h42+}).

    Let us now consider the antiunitary time-reversal operator \cite{book:Wigner1959} for \ce{[H4]^{2+}}:
    \begin{equation}
      \gls*{op:timerev} = \gls*{op:piyspinrot}\gls*{op:holoconj} = \gls*{op:holoconj}\gls*{op:piyspinrot},
    \end{equation}
    where $\gls*{op:holoconj}$ is given by Equations~\ref{eq:kappaPsi}~and~\ref{eq:chiconj}.
    Appendix~\ref{app:timerevgroup} gives the character table and irreducible corepresentations\cite{book:Wigner1959,article:Cracknell1966,article:Newmarch1981,article:Bradley1966} for the time-reversal group $\mathcal{T}$ generated by $\gls*{op:timerev}$.
    For the persistently real solutions and the transiently real solutions in the real regimes, $\gls*{op:holoconj}$ is simply the identity, and therefore the effect of time reversal on these solutions boils down to their behaviors under $\gls*{op:piyspinrot}$ as discussed in the previous paragraph.
    Specifically, for \gls*{acr:rhf}-like solutions,
    \begin{equation*}
      \gls*{op:piyspinrot}\gls*{wf:hf} = \pm\gls*{wf:hf} \Rightarrow \gls*{op:timerev}\gls*{wf:hf} = \pm\gls*{wf:hf},
    \end{equation*}
    so that $\gls*{wf:hf}$ conserves time-reversal symmetry in the sense that it spans a one-dimensional irreducible corepresentation in $\mathcal{T}$.
    Conversely, for non-\gls*{acr:rhf} solutions, $\gls*{wf:hf}$ and $\gls*{op:timerev}\gls*{wf:hf}$ are linearly independent and together they span a two-dimensional reducible corepresentation in $\mathcal{T}$; we thus say that $\gls*{wf:hf}$ is now time-reversal symmetry-broken.

    More specifically, we note that, whenever the two spin-orbitals $\chi_1$ and $\chi_2$ are subject to the same spatial extrinsic constraint such that $x = y$, then solutions $\mathrm{1b}$ and $\mathrm{1c}$ become time-reversal partners of each other and must therefore be degenerate.
    This can be easily verified by inspecting their analytic forms in Table~\ref{tab:allsols}, or by inspecting the $\lvert\alpha^g \beta^g\rvert$ and $\lvert\alpha^u \beta^u\rvert$ panels in Figure~\ref{fig:pathway-a-energy-i} in which the energy curves for solutions $\mathrm{1b}$ and $\mathrm{1c}$ fall exactly on top of each other at all geometries.
    In addition, the fact that $\mathrm{1b}$ and $\mathrm{1c}$ are time-reversal partners but must remain distinct solutions of $\bar{P}\left(\boldsymbol{z}; \boldsymbol{B}\right) = 0$ implies that $\hat{q}\chi_1 \ne \pm\chi_2$ (as $\gls*{op:holoconj} = \id$ for persistently real solutions), and so $\mathrm{1b}$ and $\mathrm{1c}$ must be non-\gls*{acr:rhf} solutions and together span a two-dimensional reducible corepresentation in $\mathcal{T}$.
    On the other hand, $\mathrm{1a}$ and $\mathrm{1d}$ are not only distinct but also unrelated by time-reversal symmetry which implies that each of them is its own time-reversal partner and thus conserves time-reversal symmetry.
    This makes $\mathrm{1a}$ and $\mathrm{1d}$ two qualitatively different \gls*{acr:rhf}-like solutions.

    However, when the spatial extrinsic constraints on $\chi_1$ and $\chi_2$ differ and $x \ne y$, such as for $\lvert\alpha^u \beta^g\rvert$, then all four solutions $\mathrm{1a}$--$\mathrm{1d}$ must be non-\gls*{acr:rhf} whose time-reversal partners live in the space in which $x$ and $y$ are swapped.
    The domain $\mathscr{D}$ in each constraining space $[\hat{R}]'_{xy}$ thus contains time-reversal partners that are symmetric about the diagonal $x = y$.
    The spin-orbital forms in Table~\ref{tab:pathway-a-orbitals-i} can be consulted for concrete examples that illustrate the above description.

    Outside the real regimes, $\gls*{op:holoconj}$ is no longer the identity, and the action of $\gls*{op:timerev}$ depends on the composite behaviors of $\gls*{op:holoconj}$ and $\gls*{op:piyspinrot}$ on $\gls*{wf:hf}$.
    In particular, for \gls*{acr:rhf}-like solutions, there are two possibilities:
    \begin{enumerate*}[label=(\roman*)]
      \item in the special case where $\chi^*_i = -\chi_i$, \textit{i.e.}, the spin-orbitals have purely imaginary coefficients, then $\gls*{wf:hf}$ and $\gls*{op:timerev}\gls*{wf:hf}$ differ only by a phase factor and $\gls*{wf:hf}$ still conserves time-reversal symmetry;
      \item more generally, if $\chi^*_i \ne -\chi_i$, then $\gls*{wf:hf}$ and $\gls*{op:timerev}\gls*{wf:hf}$ are necessarily linearly independent and $\gls*{wf:hf}$ now breaks time-reversal symmetry.
    \end{enumerate*}
    For non-\gls*{acr:rhf} solutions, there are also two cases:
    \begin{enumerate*}[label=(\roman*)]
      \item in the special case where $\hat{q}\chi_1 = \pm \chi^*_2$, \textit{i.e.}, the spatial parts of $\chi_1$ and $\chi_2$ are complex-conjugates of each other, then $\gls*{wf:hf}$ and $\gls*{op:timerev}\gls*{wf:hf}$ are linearly dependent, and so $\gls*{wf:hf}$ conserves time-reversal symmetry;
      \item in general, there is no constraint between $\chi_1$ and $\chi_2$, so $\gls*{wf:hf}$ is time-reversal symmetry-broken.
    \end{enumerate*}
    These various behaviors under time reversal are summarized in Table~\ref{tab:timerevsym}.

    \begin{table}
      \centering
      \caption{Time-reversal symmetry of the \gls*{acr:uhf} solutions for \ce{[H4]^{2+}} in this work.}
      \label{tab:timerevsym}
      \ifdefined\twocolumnmode
        \renewcommand{\arraystretch}{1.3}
        \begin{tabular}{
          >{\raggedright\arraybackslash} p{2.0cm}
          >{\raggedright\arraybackslash} p{3.0cm}
          >{\raggedright\arraybackslash} p{3.0cm}
        }
          \toprule
          coefficients & real & non-real \\
          \midrule
          \gls*{acr:rhf}-like
            & $\hat{q}\chi_1 = \pm \chi_2$,
            & $\chi^*_i = -\chi_i$ \\
            & $\gls*{op:timerev}\gls*{wf:hf} = \pm \gls*{wf:hf}$
            & $\Leftrightarrow \gls*{op:timerev}\gls*{wf:hf} = \pm \gls*{wf:hf}$ \\[6pt]
          non-\gls*{acr:rhf}
            & $\hat{q}\chi_1 \ne \pm \chi_2$,
            & $\hat{q}\chi_1 = \pm \chi^*_2$\\
            & $\gls*{op:timerev}\gls*{wf:hf} \ne \pm \gls*{wf:hf}$
            & $\Leftrightarrow \gls*{op:timerev}\gls*{wf:hf} = \pm \gls*{wf:hf}$ \\
          \bottomrule
        \end{tabular}
      \else
        \begin{tabular}{
          >{\raggedright\arraybackslash} p{3.0cm}
          >{\raggedright\arraybackslash} p{3.3cm}
          >{\raggedright\arraybackslash} p{3.3cm}
        }
          \toprule
          coefficients & real & non-real \\
          \midrule
          \gls*{acr:rhf}-like
            & $\hat{q}\chi_1 = \pm \chi_2$,
            & $\chi^*_i = -\chi_i$ \\
            & $\gls*{op:timerev}\gls*{wf:hf} = \pm \gls*{wf:hf}$
            & $\Leftrightarrow \gls*{op:timerev}\gls*{wf:hf} = \pm \gls*{wf:hf}$ \\[6pt]
          non-\gls*{acr:rhf}
            & $\hat{q}\chi_1 \ne \pm \chi_2$,
            & $\hat{q}\chi_1 = \pm \chi^*_2$\\
            & $\gls*{op:timerev}\gls*{wf:hf} \ne \pm \gls*{wf:hf}$
            & $\Leftrightarrow \gls*{op:timerev}\gls*{wf:hf} = \pm \gls*{wf:hf}$ \\
          \bottomrule
        \end{tabular}
      \fi
    \end{table}

    As a particular solution $\gls*{wf:hf}$ varies smoothly along any molecular symmetry pathway, its spin-orbitals must also vary smoothly due to holomorphicity\cite{article:Burton2016}.
    This thus has implications concerning the time-reversal symmetry of $\gls*{wf:hf}$ along this pathway.
    Consider an \gls*{acr:rhf}-like transiently real solution $\gls*{wf:hf}$, such as the $\mathrm{2a}$ and $\mathrm{2a}'$ solutions of $\lvert\alpha^u\beta^u\rvert$.
    In the real regimes, clearly $\chi^*_i = +\chi_i$ and $\gls*{op:timerev}\gls*{wf:hf} = \pm\gls*{wf:hf}$, so that $\gls*{wf:hf}$ conserves time-reversal symmetry.
    As $\gls*{wf:hf}$ enters the non-real regimes, however, it cannot be that $\chi^*_i = -\chi_i$ abruptly since a smoothly varying normalized spin-orbital as constrained by Equation~\ref{eq:generalnormalisation} cannot switch suddenly from being entirely real to being entirely imaginary.
    Hence, $\gls*{op:timerev}\gls*{wf:hf}$ cannot equal $\pm\gls*{wf:hf}$ and $\gls*{wf:hf}$ now breaks time-reversal symmetry.
    This switch in time-reversal behavior across the real/non-real boundary is thus essential for \gls*{acr:rhf}-like solutions and, more generally, for the \gls*{acr:rhf} spin extrinsic constraint.
    An example of this can be seen in the $\mathrm{2a}$ and $\mathrm{2a}'$ solutions of $\lvert\alpha^u\beta^u\rvert$ which independently span the corepresentation $A$ of $\mathcal{T}$ in the real regime but then switch to spanning $A \oplus B$ together in the non-real regimes (Table~\ref{tab:pathway-a-orbitals-i}).

    If $\gls*{wf:hf}$ is a non-\gls*{acr:rhf} transiently real solution instead, then in general, $\gls*{wf:hf}$ breaks time-reversal symmetry throughout.
    This is certainly always true in the real regimes.
    However, if the relation $\hat{q}\chi_1 = \pm \chi^*_2$ holds in the non-real regimes, then $\gls*{wf:hf}$ and $\gls*{op:timerev}\gls*{wf:hf}$ describe the same state and $\gls*{wf:hf}$ thus conserves time-reversal symmetry (see Table~\ref{tab:timerevsym}).
    As such, this switch in time-reversal behavior across the real/non-real boundary does not necessarily occur unless other symmetries of the spin-orbitals permit the above relation, such as those observed in the $\mathrm{2b}$ and $\mathrm{2b}'$ solutions under the $\lvert\alpha^u\beta^u\rvert$ extrinsic constraint (see Table~\ref{tab:pathway-a-orbitals-i}).
    On the other hand, in the extrinsic constraint $\lvert\alpha^u\beta^g\rvert$, or more generally, in any extrinsic constraint $[\hat{R}]'_{xy}$ with $x \ne y$ (see Table~\ref{tab:extrinsicconstraints}), the relation $\hat{q}\chi_1 = \pm \chi^*_2$ is forbidden by the constraining symmetry element $\hat{R}$, and no conservation of time-reversal symmetry can be observed in the non-real regimes for non-\gls*{acr:rhf} solutions.

    Before moving on, we remark that the discussion so far reveals that the effects of time reversal on holomorphic solutions can be counter-intuitive, particularly because of the antilinearity of the time-reversal operator.
    Two key features stand out.
    Firstly, while the spatial symmetry of a solution remains unchanged throughout the molecular pathway (except at coalescence points; see Section~\ref{subsec:solconnectivity} below), the same cannot be said for time-reversal symmetry.
    This is because the antilinearity of the time-reversal operator captures and reflects any transition between the real and non-real regimes of a solution.
    As a consequence, the normal association of time-reversal symmetry conservation/breaking to restricted/unrestricted spin constraints breaks down outside of real regimes: the antilinearity of the time-reversal operator $\gls*{op:timerev}$ sets it apart from the closely-related, but linear, operator $\gls*{op:piyspinrot}$ (defined in Equation~\ref{eq:totspinrot}) that simply effects spin-flipping.

  \subsection{Symmetry and Solution Connectivity}
  \label{subsec:solconnectivity}

    We mention in Section~\ref{subsubsec:coalescenceboundaries} that, on coalescence boundaries, the two solutions in a transiently real pair become identical to one of the persistently real solutions.
    The two transiently real solutions therefore lose their linear independence and must span an identical (co)representation space to that of the persistently real solution.
    As they move away from the coalescence boundary, they become linearly independent and their (co)representation space must increase in dimensionality, but in so doing, it must include the (co)representation space of the persistently real solution, as proven in Appendix~\ref{app:symvariation}.
    Consequently, along \emph{any} pathway connecting some coalescence boundaries, the (co)representation space spanned by the transiently real solutions must contain those spanned by the persistently real solutions with which they come into coalescence.

    The above property is best illustrated by the $\lvert\alpha^u\beta^g\rvert$ and $\lvert\alpha^u\beta^u\rvert$ solutions: the transiently real solutions connect various persistently real solutions (see the energy curves in Figure~\ref{fig:pathway-a-energy-i}), and the symmetry symbols of the transiently real solutions always include those of the persistently real solutions they connect (see also Table~\ref{tab:pathway-a-orbitals-i}).
    For example, the transiently real $\mathrm{2a}$ and $\mathrm{2a}'$ solutions under $\lvert\alpha^u\beta^g\rvert$ (middle panel in Figure~\ref{fig:pathway-a-energy-i}) have $\prescript{A}{}{B}_{2u} \oplus \prescript{A}{}{B}_{3u} \oplus \prescript{B}{}{B}_{2u} \oplus \prescript{B}{}{B}_{3u}$ symmetry, which includes the $\prescript{A}{}{B}_{2u} \oplus \prescript{B}{}{B}_{2u}$ symmetry and the $\prescript{A}{}{B}_{3u} \oplus \prescript{B}{}{B}_{3u}$ of the persistently real $\mathrm{1a}$ and $\mathrm{1c}$ solutions, respectively.
    Here, the symmetry terms of the $\mathrm{2a}$ and $\mathrm{2a}'$ solutions are a disjoint union of those of the $\mathrm{1a}$ and $\mathrm{1c}$ solutions.
    However, this need not be always the case.
    Consider the $\lvert\alpha^u\beta^u\rvert$ transiently real $\mathrm{2a}$ and $\mathrm{2a}'$ solutions and the persistently real $\mathrm{1a}$ and $\mathrm{1d}$ solutions (right panel in Figure~\ref{fig:pathway-a-energy-i}).
    The symmetry of $\mathrm{2a}$ and $\mathrm{2a}'$ is $\prescript{A}{}{A}_{g} \oplus \prescript{A/B}{}{B}_{1g}$, while the $\mathrm{1a}$ and $\mathrm{1d}$ solutions both have $\prescript{A}{}{A}_{g}$ symmetry.
    The switch in the time-reversal symmetry associated with the $B_{1g}$ spatial component of the $\mathrm{2a}$ and $\mathrm{2a}'$ solutions (discussed in Section~\ref{subsec:timerevsym}) and the necessary lack thereof in the $\mathrm{1a}$ and $\mathrm{1d}$ solutions mean that the $\prescript{A/B}{}{B}_{1g}$ component of the transiently real solutions cannot arise from the two persistently real solutions they connect.
    Nevertheless, this does not violate the property raised in the preceding paragraph since the $\prescript{A}{}{A}_{g}$ symmetry of the two persistently real solutions is certainly included in the symmetry of the connecting transiently real solutions.

    If the persistently real solutions being connected span different (co)representations, then the transiently real solutions that connect them are guaranteed to span reducible (co)representations and therefore be symmetry-broken.
    The spin-orbital forms for the transiently real solutions in Table~\ref{tab:pathway-a-orbitals-i} show that, in the real regimes, the symmetry breaking occurs via the magnitudes of the \gls*{acr:mo} coefficients, whereas in the non-real regimes, the symmetry breaking is due to the coefficient phases instead.
    The origin of this behavior can be traced back to the identification of $\gls*{op:holoconj}$ with $\hat{C}_2^{y}$ in Equation~\ref{eq:timerevunitaryop}.
    More significantly, the forced symmetry breaking of the transiently real solutions implies that, as investigated in our earlier work\cite{article:Huynh2020}, these solutions carry some description of electron correlation and can be symmetry-restored to yield multi-determinantal wavefunctions that have the right symmetry and that also incorporate more electron correlation.
    The solution structure detailed in Table~\ref{tab:allsols} thus means that the constraints in Table~\ref{tab:extrinsicconstraints} necessitate the existence of the symmetry-broken transiently real solutions, even though without appropriate reality requirements, they are non-real and cannot be found by conventional \gls*{acr:hf} methods.

\tikzsetexternalprefix{./h4model-constraintconnections/tikz/}

\section{\ce{[H4]^{2+}} Model: Constraint Connections}
\label{sec:h4analmodel-constraintconnections}

  The extrinsic constraints considered so far ($\left\lvert\alpha^g \beta^g\right\rvert$, $\left\lvert\alpha^u \beta^g\right\rvert$, and $\left\lvert\alpha^u \beta^u\right\rvert$) have been deliberately chosen to target the numerically locatable \gls*{acr:uhf} solutions along molecular symmetry pathway A (Figure~\ref{fig:h42p-prelim}).
  However, it is of theoretical interest to explore other local regions and global structures of the \gls*{acr:scf} landscape by varying the extrinsic and intrinsic constraints, respectively.
  This helps identify hidden connections between solutions that have so far been described in separate extrinsically constrained spaces.

  In this Section, we first discuss solutions obtained under a set of extrinsic constraints induced by $\hat{\sigma}^{xz}$, a different symmetry element that also persists along pathway A.
  This is to further demonstrate the effects of extrinsic constraints on the forms and reality of \gls*{acr:scf} solutions.
  We then show the one-to-one correspondence between solutions with $\hat{i}$ extrinsic constraints and those with $\hat{\sigma}^{xz}$ extrinsic constraints via molecular pathway B, incidentally exploring their strong reality requirements in the vicinity of the high-symmetry $\mathcal{T}_d$ geometry.
  Finally, we explore the symmetry-breaking regions that inter-connect the special extrinsic constraints that have been the focus of the discussion to gain some insight into how relaxing the extrinsic constraints imposed on the molecular orbitals affects solution reality.

  \subsection{\boldmath Special $\hat{\sigma}^{xz}$-Symmetry-Conserved Extrinsic Constraints}
  \label{subsec:a-sigma}

\begin{figure*}
      \centering
      \tikzsetexternalprefix{./h4model-constraintconnections/tikz/energy/}
      \subfloat[\label{subfig:pathway-a-energy-sigma}]{
        \centering
        \ifdefined\twocolumnmode
          \useexternalfile{0.85}{34.92221pt}{7.54836pt}{regimeplot.D2h.rec.sigma.single.all}{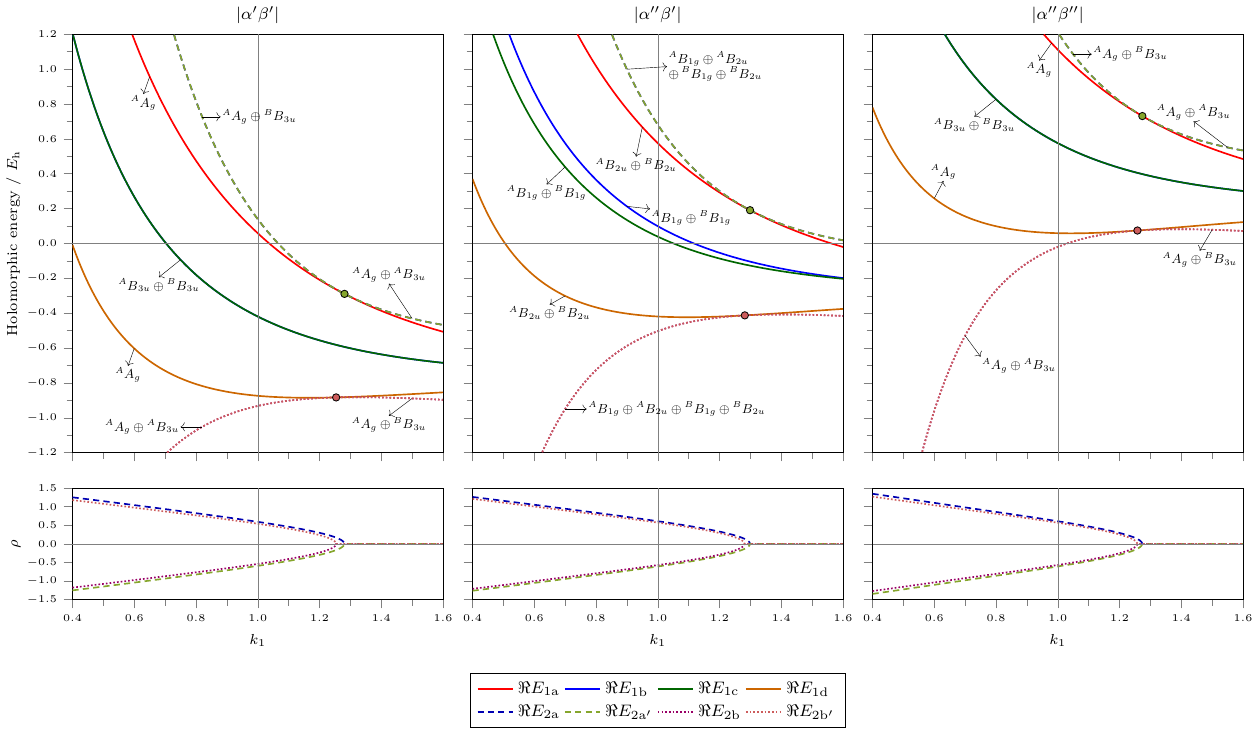}
        \else
          \useexternalfile{0.70}{34.92221pt}{7.54836pt}{regimeplot.D2h.rec.sigma.single.all}{figure6a}
        \fi
      }

      \vspace{12pt}

      \tikzsetexternalprefix{./h4model-constraintconnections/tikz/realityregimes/}
      \subfloat[\label{subfig:pathway-a-realityregimes-sigma}]{
        \centering
        \ifdefined\twocolumnmode
          \useexternalfile{0.85}{29.85779pt}{46.81604pt}{regimeplot.surf.D2h.rec.sigma.all.all}{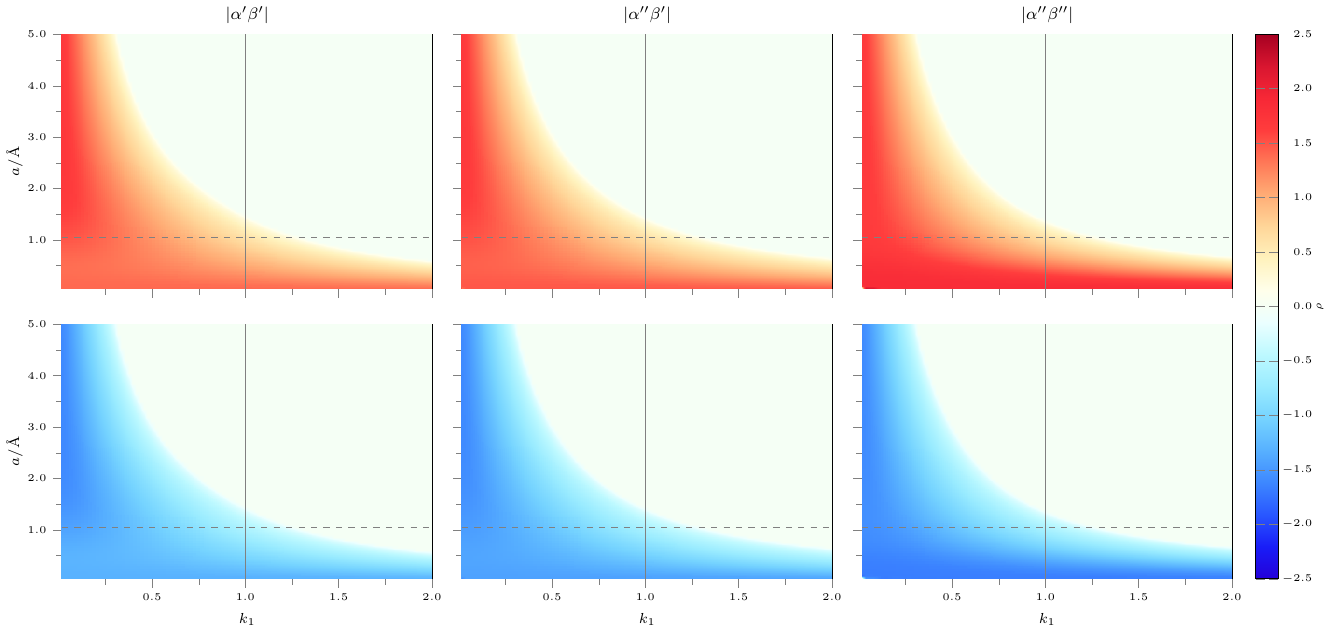}
        \else
          \useexternalfile{0.70}{29.85779pt}{46.81604pt}{regimeplot.surf.D2h.rec.sigma.all.all}{figure6b}
        \fi
      }

      \caption{
        Solutions under special $\hat{\sigma}^{xz}$-symmetry-conserved extrinsic constraints.
        \protect\subref{subfig:pathway-a-energy-sigma} Holomorphic energy, $\mathcal{T} \otimes \mathcal{D}_{2h}$ symmetry, and reality indicators along molecular symmetry pathway A at $a = \SI{1.058350}{\angstrom}$.
        \protect\subref{subfig:pathway-a-realityregimes-sigma} Reality indicators of solutions $\mathrm{2a}$ (top) and $\mathrm{2b}$ (bottom) along pathway A at different scale lengths $a$.
        The vertical line at $k_1 = 1$ indicates $\mathcal{D}_{4h}$ symmetry, and the dashed horizontal line indicates $a = \SI{1.058350}{\angstrom}$.
      }
      \label{fig:pathway-a-energyrealityregimes-sigma}
    \end{figure*}

    We show in Figure~\ref{fig:pathway-a-energyrealityregimes-sigma} the energy, symmetry, and reality regimes of the solutions obtained under the special $\hat{\sigma}^{xz}$-symmetry-conserved extrinsic constraints (see Table~\ref{tab:extrinsicconstraints-special} for their definitions).
    In addition, the forms of their spin-orbitals along pathway A are given in the included animations (Section S-II of the Supplementary Material).
    It can be verified that the persistently real solutions in all cases have already been found with the $\hat{i}$-symmetry-conserved extrinsic constraints.
    The transiently real solutions, however, are all not previously encountered.
    In fact, the symmetry of these solutions (Figure~\ref{subfig:pathway-a-energy-sigma}) shows that they are neither purely $g$ nor $u$ under $\hat{i}$ and thus must be excluded from the $\hat{i}$-symmetry-conserved regions of the \gls*{acr:scf} landscape.

    Figure~\ref{subfig:pathway-a-realityregimes-sigma} reveals that there is now no longer any strong reality requirement at $\mathcal{D}_{4h}$ in any of the special $\hat{\sigma}^{xz}$-symmetry-conserved extrinsic constraints.
    Weak reality requirements, however, are still in effect and these only allow the transiently real solutions to become real at $\mathcal{D}_{4h}$ for scale length $a$ of at least approximately $\SI{1.45}{\angstrom}$.
    This explains why none of these solutions can be found numerically in the vicinity of $\mathcal{D}_{4h}$ at $a = \SI{1.058350}{\angstrom}$, as clearly lacking in Figure~\ref{fig:h42p-prelim}.

    Comparing the energy curves and reality indicators of the $\hat{i}$-symmetry-conserved in Figure~\ref{fig:pathway-a-energy-i} to those of the $\hat{\sigma}^{xz}$-symmetry-conserved solutions in Figure~\ref{subfig:pathway-a-energy-sigma}, we notice that, due to the lack of strong reality requirements in the latter, none of the transiently real solutions appear to coalesce with more than one persistently real solution along pathway A.
    Nevertheless, their corepresentations in $\mathcal{T} \otimes \mathcal{D}_{2h}$ suggest otherwise.
    For example, the $\mathrm{2a}$ and $\mathrm{2a}'$ solutions of $\lvert\alpha''\beta'\rvert$ span $\prescript{A}{}{B}_{1g} \oplus \prescript{A}{}{B}_{2u} \oplus \prescript{B}{}{B}_{1g} \oplus \prescript{B}{}{B}_{2u}$ but only coalesce with the $\mathrm{1a}$ solution that spans $\prescript{A}{}{B}_{2u} \oplus \prescript{B}{}{B}_{2u}$ along pathway A.
    We thus suspect that there exist other pathways along which the unobserved coalescence between $\mathrm{2a}$ and $\mathrm{2a}'$ with another persistently real solution that spans $\prescript{A}{}{B}_{1g} \oplus \prescript{B}{}{B}_{1g}$ ($\mathrm{1b}$ or $\mathrm{1c}$) occurs.

    \tikzexternalenable
    \tikzexternaldisable

  \subsection{\boldmath Connecting $[\hat{i}]'_{xy}$ and $[\hat{\sigma}^{xz}]'_{xy}$ via Pathway B}
  \label{subsec:connectingviapathwayb}

\begin{figure*}
      \centering
      \tikzsetexternalprefix{./h4model-constraintconnections/tikz/energy/}
      \subfloat[\label{subfig:pathway-b-energy-c2z}]{
        \centering
        \ifdefined\twocolumnmode
          \useexternalfile{0.85}{34.92221pt}{7.54836pt}{regimeplot.D2.c2z.single.all}{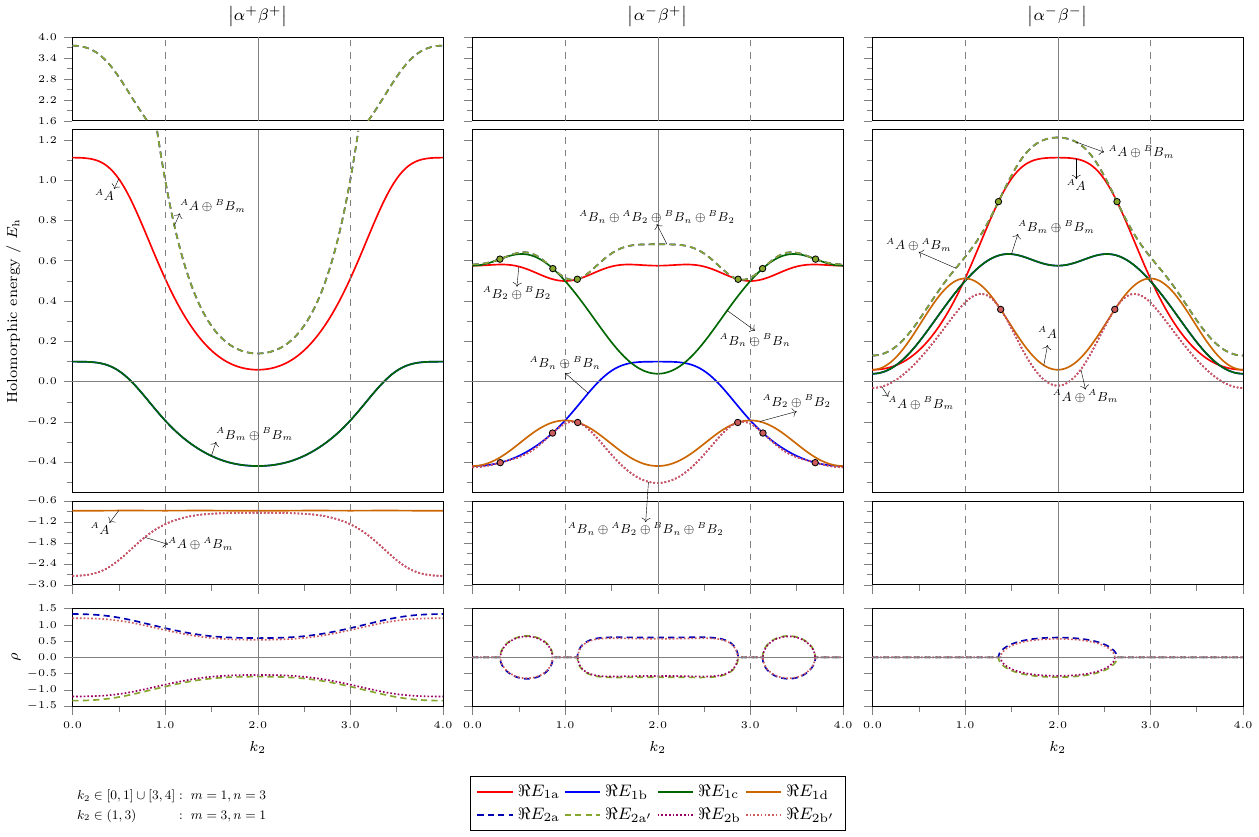}
        \else
          \useexternalfile{0.70}{34.92221pt}{7.54836pt}{regimeplot.D2.c2z.single.all}{figure7a}
        \fi
      }

      \ifdefined\twocolumnmode
        \vspace{12pt}
      \else
        \vspace{3pt}
      \fi

      \tikzsetexternalprefix{./h4model-constraintconnections/tikz/realityregimes/}
      \subfloat[\label{subfig:pathway-b-realityregimes-c2z}]{
        \centering
        \ifdefined\twocolumnmode
          \useexternalfile{0.85}{29.85779pt}{46.952pt}{regimeplot.surf.D2.c2z.all.all}{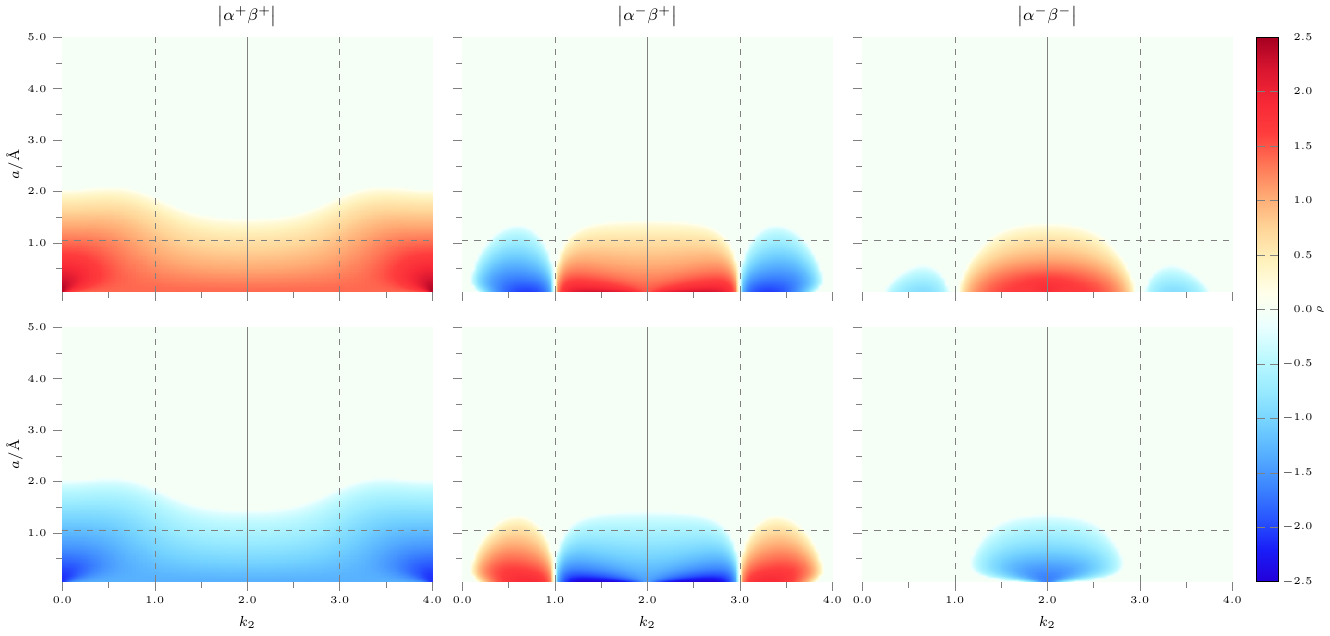}
        \else
          \useexternalfile{0.70}{29.85779pt}{46.952pt}{regimeplot.surf.D2.c2z.all.all}{figure7b}
        \fi
      }

      \caption{
        Solutions under special $\hat{C}^{\clubsuit}$-symmetry-conserved extrinsic constraints.
        \protect\subref{subfig:pathway-b-energy-c2z} Holomorphic energy, $\mathcal{T} \otimes \mathcal{D}_{2}$ symmetry, and reality indicators along molecular symmetry pathway B at $a = \SI{1.058350}{\angstrom}$.
        \protect\subref{subfig:pathway-b-realityregimes-c2z} Reality indicators of solutions $\mathrm{2a}$ (top) and $\mathrm{2b}$ (bottom) along pathway B at different scale lengths $a$.
        The dashed vertical lines indicate $\mathcal{T}_{d}$ symmetry, the solid vertical line indicates the twisted $\mathcal{D}_{4h}$ configuration (see Figure~\ref{fig:symbreakingpathways}), and the dashed horizontal line indicates $a = \SI{1.058350}{\angstrom}$.
      }
      \label{fig:pathway-b-energyrealityregimes-c2z}
    \end{figure*}

    The similarity in the patterns of the energy curves between the $\hat{i}$-symmetry-conserved solutions and the $\hat{\sigma}^{xz}$-symmetry-conserved solutions (Figures~\ref{fig:pathway-a-energy-i}~and~\ref{subfig:pathway-a-energy-sigma}) and the identicality in their analytic forms (Table~\ref{tab:allsols}) both suggest that there is a one-to-one correspondence between these two sets of solutions.
    In fact, it can be seen from Table~\ref{tab:extrinsicconstraints} that, for a particular $(x, y)$, if \ce{H^1} and \ce{H^2} are swapped, then the $\hat{i}$ constraints become $\hat{\sigma}^{xz}$ and \textit{vice versa}.
    This thus suggests that the solutions obtained under the two constraints can be connected by a pathway along which the \ce{H^1}---\ce{H^2} bond is rotated about its mutual perpendicular bisector with the \ce{H^3}---\ce{H^4} bond.
    This is almost pathway B that we introduced earlier (Figure~\ref{subfig:tetrahedron}), except that along pathway B, the perpendicular distance between the two bonds is also varied such that when the symmetry factor $k_2$ equals $1$ or $3$, the system attains a tetrahedral symmetry.
    Incidentally, this enables us to access another high symmetry configuration and examine any strong reality requirements that result.
    The energy variation and reality indicators along pathway B under the special $\hat{C}_2^{\clubsuit}$-symmetry-conserved extrinsic constraints are plotted in Figure~\ref{fig:pathway-b-energyrealityregimes-c2z}.
    The spin-orbital forms of these solutions and their symmetry can be seen in the included animations (see Section S-II of the Supplementary Material).
    By comparing these plots with Figures~\ref{fig:pathway-a-energy-i}, \ref{fig:pathway-a-realityregimes-i}, and \ref{fig:pathway-a-energyrealityregimes-sigma}, we see that the $\hat{C}_2^{\clubsuit}$ constraints have been chosen specifically to connect the $\hat{i}$-symmetry-conserved solutions at $\mathcal{D}_{4h}$ ($k_2 = 0$) to the $\hat{\sigma}^{xz}$-symmetry-conserved solutions at $\mathcal{D}_{4h}$ ($k_2 = 2$).

    We first observe, on the basis of the reality regimes in Figure~\ref{subfig:pathway-b-realityregimes-c2z}, that the $\lvert\alpha^-\beta^+\rvert$ and $\lvert\alpha^-\beta^-\rvert$ extrinsic constraints afford strong reality requirements at $\mathcal{T}_d$ ($k_2 = 1$ or $3$), a property that can be verified analytically by considering vanishing electron-integral coefficients in a similar way to those shown in Table~\ref{tab:vanishingcoeffs-i}.
    This once again demonstrates the forced reality of the transiently real solutions at a high-symmetry point ($\mathcal{T}_d$ in this case) which quickly coalesce with other persistently real solutions and become non-real as the molecular symmetry is descended.
    But more importantly, due to the connectivity of $\hat{i}$ and $\hat{\sigma}^{xz}$ constraints via pathway B, the transiently real solutions under the special $\hat{\sigma}^{xz}$ constraints $\lvert\alpha''\beta'\rvert$ and $\lvert\alpha''\beta''\rvert$ that exhibit no strong reality requirements at $\mathcal{D}_{4h}$ ($k_2 = 2$; see also Figure~\ref{subfig:pathway-a-realityregimes-sigma}) can \emph{always} be located in the strongly required real regimes at $\mathcal{T}_d$ for \emph{any} scale length $a$ and then tracked to the desired geometry.
    Unfortunately, no such guarantee can be made for the $\lvert\alpha^+\beta^+\rvert$ extrinsic constraint which does not exhibit any strong reality requirements along pathway B.

    The connectivity provided by pathway B also helps make sense of the corepresentation spaces spanned by the transiently real solutions of $\lvert\alpha''\beta'\rvert$.
    For example, as mentioned at the very end of Section~\ref{subsec:a-sigma}, the coalescence between the $\mathrm{2a}/\mathrm{2a'}$ solutions and the $\mathrm{1a}$ solution along pathway A only accounts for two of the four symmetry terms of the former.
    However, by tracking these solutions along pathway B, the missing coalescence with the $\mathrm{1c}$ solution which accounts for the remaining two symmetry terms can now be observed (Figure~\ref{subfig:pathway-b-energy-c2z}).

    \tikzexternalenable
    \tikzexternaldisable

  \subsection{Symmetry-Broken Extrinsic Constraints}
  \label{subsec:symbrokenextrinsicconstraints}

    The symmetry-conserved extrinsic constraints examined in detail so far have been deliberately chosen such that the analytic solutions we obtain do correspond to true stationary points of the \gls*{acr:scf} energy landscape.
    In fact, by considering the spin-orbital-dependent terms of the Fock operator $\hat{f}$ in Equation~\ref{eq:fockop} in conjunction with the extrinsic constraints in Table~\ref{tab:extrinsicconstraints}, we see that, by imposing symmetry conservation, \textit{i.e.}, setting $x, y = \pm 1$, the constraining symmetry operations leave $\hat{f}$ invariant under their actions.
    The spin-orbitals that conserve these symmetries are thus allowed to be eigenfunctions of $\hat{f}$ without any contradiction.
    In other words, the symmetry-conserved constraints are compatible with, and hence encompass, \gls*{acr:scf} stationary points.

    The analysis so far makes it clear that both intrinsic and extrinsic constraints must cooperate to strongly force reality on the transiently real solutions.
    Essentially, intrinsic constraints determine the global structure of the \gls*{acr:scf} energy landscape alongside its regimes of both strong and weak reality requirements.
    Extrinsic constraints then enable us to choose and explore stationary points within certain local regions of this landscape, some of which happen to admit strong reality requirements.
    It is therefore expected that there exist pathways in the \gls*{acr:scf} energy landscape that connect regions of extrinsic constraints together and that reveal some local structures of the regimes of reality requirements between stationary points in this landscape.
    We shall henceforth refer to these pathways as ``\gls*{acr:scf} pathways'' to distinguish them from the molecular symmetry pathways A and B that we have been considering.
    These \gls*{acr:scf} pathways shall involve the variations of the $x$ and $y$ constraining parameters over the domain $\mathscr{D}$ that are defined for each constraining space $[\hat{R}]'_{xy}$ in Table~\ref{tab:extrinsicconstraints}.

    We note that, as we move away from the corners of $\mathscr{D}$ in any constraining space $[\hat{R}]'_{xy}$, the spin-orbitals $\chi_i$, and hence the overall determinant $\gls*{wf:hf}$, are constrained to break symmetry under $\hat{R}$.
    This also forces the Fock operator $\hat{f}$ to break symmetry, and consequently, there is no guarantee that $\chi_i$ would be an eigenfunction of $\hat{f}$ because $\chi_i$ and $\hat{f}$ cannot be expected to transform compatibly under $\hat{R}$.
    An optimization procedure under these symmetry-broken constraints thus yields solutions that are not expected to be stationary points of the \gls*{acr:scf} energy landscape in general.
    In fact, as the four corners of $\mathscr{D}$ must correspond to true \gls*{acr:scf} stationary points, stepping away from them into $\mathscr{D}$ in \emph{any} direction must mean traversing along non-stationary paths.
    \tikzexternalenable
    \tikzsetexternalprefix{./h4model-constraintconnections/tikz/symbrokenregimes/}
    \begin{figure*}
      \centering
      \subfloat[]{
        \centering
        \ifdefined\twocolumnmode
          \useexternalfile{0.85}{33.61598pt}{59.18503pt}{regimeplot.vol.D2h.rec.single.all.all.all}{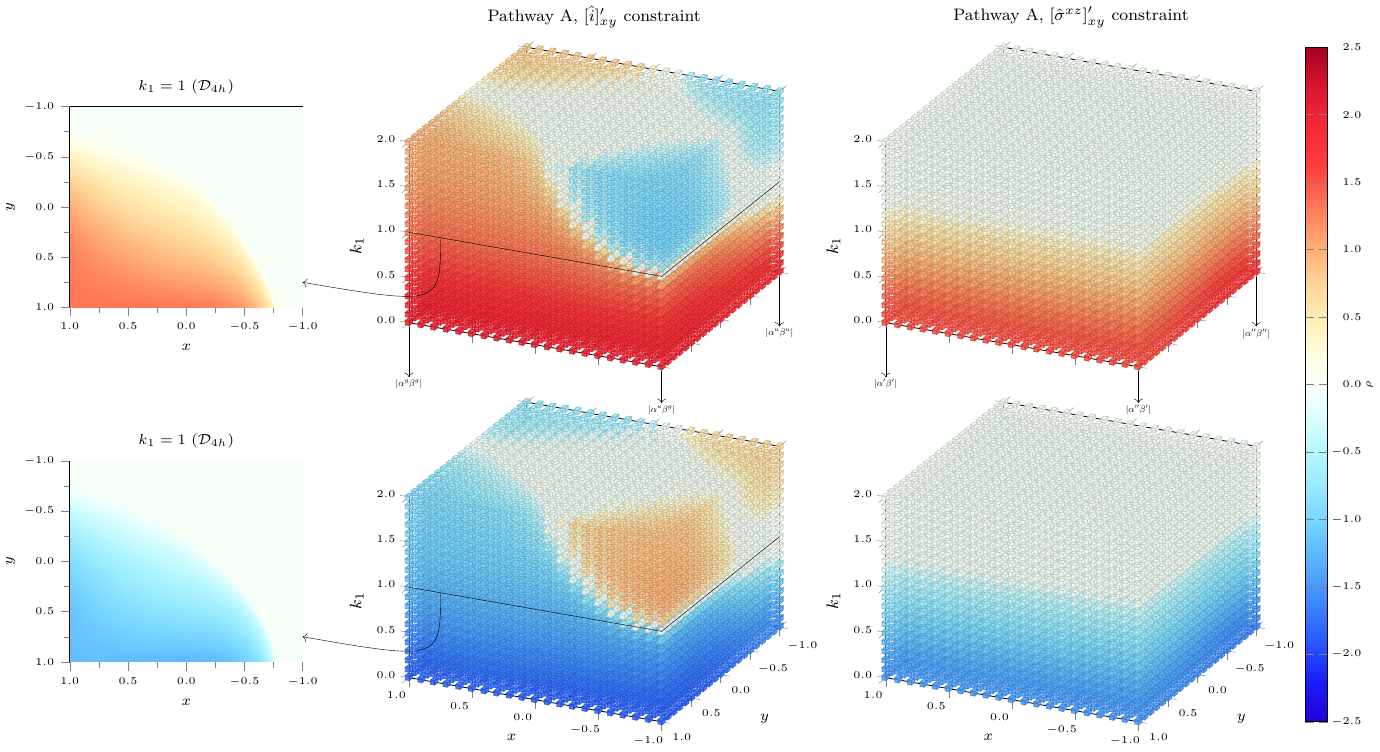}
        \else
          \useexternalfile{0.70}{33.61598pt}{59.18503pt}{regimeplot.vol.D2h.rec.single.all.all.all}{figure8a}
        \fi
      }

      \ifdefined\twocolumnmode
        \vspace{12pt}
      \else
        \vspace{3pt}
      \fi

      \subfloat[]{
        \centering
        \ifdefined\twocolumnmode
          \useexternalfile{0.85}{33.61598pt}{59.18503pt}{regimeplot.vol.D2.tet.single.all.all.all}{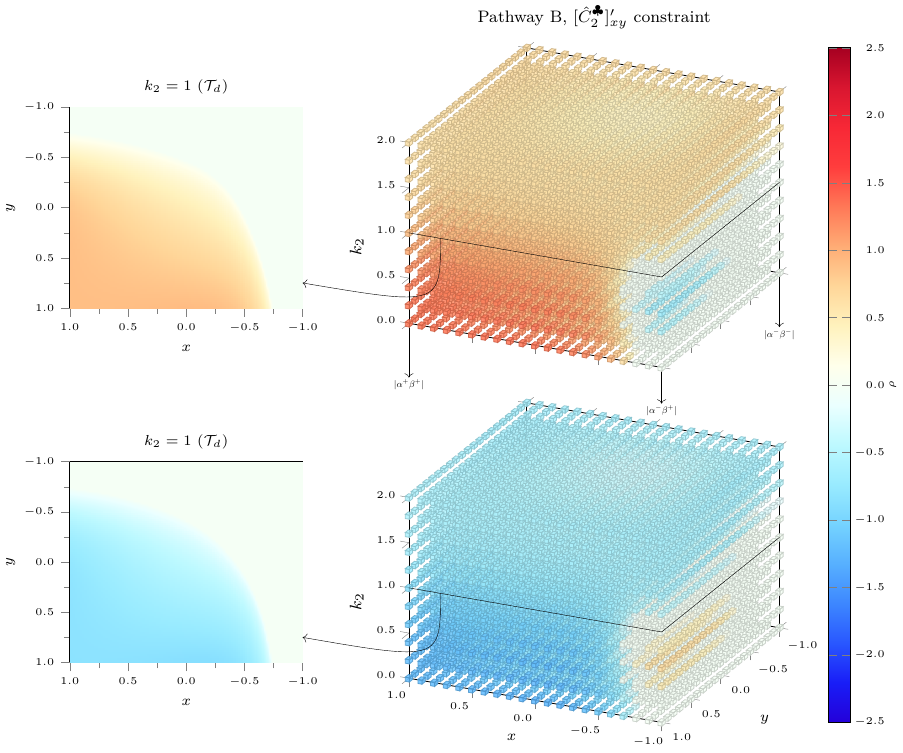}
        \else
          \useexternalfile{0.70}{33.61598pt}{59.18503pt}{regimeplot.vol.D2.tet.single.all.all.all}{figure8b}
        \fi
      }

      \caption{
        Reality indicators of solutions $\mathrm{2a}$ and $\mathrm{2b}$ in the constraining spaces $[\hat{R}]'$, where $\hat{R} \in \lbrace\hat{i}, \hat{\sigma}^{xz}, \hat{C}_2^{\clubsuit}\rbrace$, at a fixed scale length $a = \SI{1.058350}{\angstrom}$ and different symmetry factors $k_i$.
        In each plot, the top panel is for solution $\mathrm{2a}$ and the bottom one for solution $\mathrm{2b}$.
        The domain $\mathscr{D}$ of the parameters $(x, y)$ within each constraining space connects the symmetry-conserved extrinsic constraints located at the corners of $\mathscr{D}$.
      }
      \label{fig:symbrokenregimes}
    \end{figure*}
    \tikzexternaldisable

    In Figure~\ref{fig:symbrokenregimes}, we plot the values of the reality indicator $\rho$ over the domain $\mathscr{D}$ at various symmetry factors $k_i$ while fixing the scale length $a = \SI{1.058350}{\angstrom}$.
    Every horizontal slice in this plot therefore corresponds to a two-dimensional cross-section of the \gls*{acr:scf} landscape cornered by the $\hat{R}$-symmetry-conserved extrinsic constraints within the $[\hat{R}]'_{xy}$ symmetry-broken constraining space.
    We identify the vertical edges at $(x, y) = (\pm1, \pm1)$ with the $\rho$-$k_i$ indicator plots for the various $\hat{R}$-symmetry-conserved extrinsic constraints in Figures~\ref{fig:pathway-a-energy-i}, \ref{subfig:pathway-a-energy-sigma}, and \ref{subfig:pathway-b-energy-c2z}.

    We observe that the points of strong reality requirements for $\lvert \alpha^u \beta^g \rvert$ (and $\lvert \alpha^g \beta^u \rvert$ by time-reversal symmetry) and $\lvert \alpha^u \beta^u\rvert$ at $\mathcal{D}_{4h}$ along pathway A ($k_1 = 1$) and for $\lvert \alpha^- \beta^+\rvert$ (and $\lvert \alpha^+ \beta^-\rvert$ by time-reversal symmetry) and $\lvert \alpha^- \beta^-\rvert$ at $\mathcal{T}_d$ along pathway B ($k_2 = 1$) are actually embedded within local ``seas'' of reality over $\mathscr{D}$.
    In addition, there exist continuously real paths in $\mathscr{D}$ that connect points of strong reality requirements.
    The cross-sections at $k_i = 1$ for the $[\hat{i}]'_{xy}$ and $[\hat{C}_2^{\clubsuit}]'_{xy}$ constraining spaces plotted as insets in Figure~\ref{fig:symbrokenregimes} show these features clearly.
    In fact, for these two constraining spaces, there are two special paths in $\mathscr{D}$, one along the $x = -1$ edge and the other along the $y = -1$ edge, that admit strong reality requirements as verifiable by considering the relevant forms of the electron-integral coefficients shown in Section~S-I of the Supplementary Material in a similar fashion to that described in Section~\ref{subsubsec:strongreality}.

    These special paths reveal that the strong reality requirements observed so far are really the consequence of either $x$ or $y$ being equal to $-1$, but not necessarily both.
    In other words, as long as one of the two spin-orbitals undergoes a phase reversal under $\hat{i}$ at $\mathcal{D}_{4h}$ or $\hat{C}_2^{\clubsuit}$ at $\mathcal{T}_d$, then strong reality requirements are in force.
    In addition, since phase reversals introduce nodes into the spin-orbitals and raise their energy, the transiently real solutions that benefit from strong reality requirements are most likely \emph{not} the ground solutions in the holomorphic formalism.
    This can be seen most clearly in Figure~\ref{fig:pathway-a-energy-i}: the $\lvert\alpha^u\beta^g\rvert$ and $\lvert\alpha^u\beta^u\rvert$ extrinsic constraints admit transiently real solutions in the vicinity of $\mathcal{D}_{4h}$ that are real and fairly high in energy, whereas the $\lvert\alpha^g\beta^g\rvert$ extrinsic constraint admits a pair of much lower-lying transiently real solutions that remain non-real throughout the depicted $k_1$ range.
    A similar observation can be made for the $\hat{C}_2^{\clubsuit}$-symmetry-conserved transiently real solutions at $\mathcal{T}_d$ in Figure~\ref{subfig:pathway-b-energy-c2z}.

\section{Discussion}
  \label{sec:discussion}

  In this article, we discuss the various forms of constraints imposable on the \gls*{acr:hf} theory and then present an analytic investigation of the \gls*{acr:scf} solutions for a model two-electron \ce{[H4]^{2+}} system at different molecular geometries.
  Through this, we gain insight into the roles of symmetry constraints on the reality of holomorphic \gls*{acr:hf} solutions.
  In particular, we discover that intrinsic and extrinsic constraints cooperate intimately to determine local properties of the \gls*{acr:scf} landscape and impose reality requirements on holomorphic \gls*{acr:hf} solutions.
  An \gls*{acr:scf} search based on conventional \gls*{acr:hf} performed within these symmetry constraints must be able to locate these solutions in the real domains of the \gls*{acr:mo} coefficients.
  We also show that these constraints dictate the existence and locations of coalescence boundaries, therefore making them a fundamental feature of the \gls*{acr:scf} landscape.
  Hence, any disappearance of \gls*{acr:scf} solutions in conventional \gls*{acr:hf} must be interpreted as an essential consequence of the \gls*{acr:hf} equations that is governed in part by symmetry.

  The reality requirements analyzed thus far arise from the intrinsic and extrinsic constraints specifically chosen partly to target the numerically obtained solutions shown at the beginning in Figure~\ref{fig:h42p-prelim}, and partly to result in the system $P\left(\boldsymbol{\theta}; \boldsymbol{B}\right) = 0$ (Equation~\ref{eq:Psys}) for which it is possible to obtain closed-form analytic solutions.
  This thus restricts the investigation to a very small system in the minimal basis set.
  Unfortunately, a rigorous generalization to larger systems in larger basis sets based on the results presented so far is not yet possible due to the complicated algebraic structure of the general \gls*{acr:hf} equations.
  Nonetheless, we believe that the qualitative features that have been brought to our attention by the analytic solutions, namely the distinction between persistently real and transiently real solutions and the strong and weak reality requirements in relation to the symmetry of the system and of the solutions, also exist in larger systems, based on the ubiquity of the coalescence and disappearance of certain \gls*{acr:scf} solutions away from high-symmetry geometries in other larger systems that we have observed numerically as described in Section~\ref{sec:intro}.

  Finally, it is conceivable that there are other symmetries that the electron integrals can exhibit that give rise to other types of constraints, and these might very well provide more structure to the observed weak reality requirements of the investigated solutions.
  These symmetries are considered to be ``hidden'' because so far they have not shown any obvious relationships to the molecular and wavefunction symmetries we examine, as evident by the rather arbitrary nature of the coalescence boundaries enclosing regimes of weak reality requirements.
  One example of such symmetries involves the effective nuclear charge felt by the valence electrons which controls the extent to which the one-electron integrals contribute to the algebraic coefficients.
  It is therefore imperative that an investigation into other factors that can affect electron integrals be carried out so that a more general understanding of how symmetry beyond spatial governs the reality of \gls*{acr:scf} solutions can be obtained.
  This will be the focus of a future study.

  We conclude the article with a remark that the analytic model used in this work for \ce{[H4]^{2+}} and the associated constraints can be extended to other valence-isoelectronic systems, such as \ce{H4} with a frozen core, or the $\pi$-framework of the $2\pi$-aromatic \ce{[C4H4]^{2+}}.
  For example, a possible isomorphism for the \ce{H4} case is one in which we choose the two lowest-lying spin-orbitals from certain \gls*{acr:scf} solutions of \ce{H4} to form a totally symmetric frozen core which allows us to determine the effective nuclear charges experienced by the remaining two electrons in the higher-lying spin-orbitals.
  The effective nuclear charges then let us map the actual \ce{H4} \gls*{acr:scf} solutions to those in \ce{[H4]^{2+}} obtained with the one-electron integrals appropriately modified to take into account the effective nuclear charges.
  We believe that this will provide a way to map a subset of \gls*{acr:scf} solutions in \ce{H4} to those that we have discussed in detail for \ce{[H4]^{2+}}.

  The isomorphism proposed above reveals that the effective nuclear charges felt by the valence electrons now become an important factor in the weak reality requirements of the solutions, for they control the extent to which the one-electron integrals contribute to the algebraic coefficients that influence whether the transiently real solutions are real or non-real.
  This will form part of a future investigation in which we hope to understand the roles of other factors and hidden symmetries in governing the reality of solutions.

\section*{Supplementary Material}
  See Supplementary Material for the detailed functional forms of the electron integral coefficients $S$ and $A$ in terms of the one- and two-electron integrals, the video entitled \texttt{spatialsymreality.orbitalplots.mp4} showing the forms of the \glspl*{acr:mo} for the analytic holomorphic solutions of \ce{[H4]^{2+}} in STO-3G and animating their variations as the geometry of the cation changes, and the derivation of the weak reality requirements and the equations describing the coalescence boundaries.

\begin{acknowledgments}
  B.C.H. is grateful for the financial support from Cambridge Commonwealth, European \& International Trust and Peterhouse during the duration of this work throughout the COVID-19 national lockdowns in England.
  A.J.W.T. thanks the Royal Society for a University Research Fellowship (UF110161).
  Both authors thank the anonymous reviewers who provided many critical and constructive comments that helped refine the discussions in this article.
\end{acknowledgments}

\appendix

\section{Non-Vanishing Electron Integrals}
\label{app:nonvanishingeints}

  \subsection{General Formulation}

    Given a set of real \gls*{acr:ao} basis functions $\{\varphi_i\}$ localized on nuclei that are invariant under a certain point group $\mathcal{G}$, the numbers of non-vanishing independent components of the one-electron integral matrix, $\braket{\varphi_{\mu'} | \hat{O} | \varphi_{\mu}}$, two-electron integral tensor, $\braket{\varphi_{\mu'} \varphi_{\nu'} |\varphi_{\mu}\varphi_{\nu}}$, and antisymmetrized two-electron integral tensor, $\braket{\varphi_{\mu'} \varphi_{\nu'} ||\varphi_{\mu}\varphi_{\nu}}$ depend on the constraints imposed by both the permutation symmetries of the integrals and the point-group symmetries under $\mathcal{G}$.

    Let $S = \{\alpha^k \mid k=1,2,\ldots,n\}$ be a set of \emph{function labels} where each function label $\alpha^k$ has associated with it an index $i_{\alpha^k}$ to specify a function $\alpha^k_{i_{\alpha^k}}$.
    For example, if $\alpha^k$ is the label $2p$, then the index $i_{\alpha^k}$ can take on one of $x, y, z$, so that $\alpha^k_{i_{\alpha^k}}$ is one of the three $2p$ hydrogenic orbitals.
    Then, consider $n$ sets of functions $F^k = \{\alpha^k_{i_{\alpha^k}} \mid i_{\alpha^k} = 1, 2, \ldots\}$ where $k = 1, 2, \ldots, n$.
    A general element in an $n$-ary Cartesian product over the $F^k$ is an $n$-tuple that we denote as
    \ifdefined\twocolumnmode
      \begin{align*}
        \Omega^{(n)}
        &= \underbrace{(\alpha^1_{i_{\alpha^1}}, \alpha^2_{i_{\alpha^2}}, \ldots, \alpha^n_{i_{\alpha^n}})}_{\text{$n$-tuple}}\\
        &\equiv \left(\alpha^k_{i_{\alpha^k}}\right)_{k=1}^n
        \in F^1 \times F^2 \times \cdots \times F^n.
      \end{align*}
    \else
      \begin{equation*}
        \Omega^{(n)}
          = \underbrace{(\alpha^1_{i_{\alpha^1}}, \alpha^2_{i_{\alpha^2}}, \ldots, \alpha^n_{i_{\alpha^n}})}_{\text{$n$-tuple}}
          \equiv \left(\alpha^k_{i_{\alpha^k}}\right)_{k=1}^n
          \in F^1 \times F^2 \times \cdots \times F^n.
      \end{equation*}
    \fi
    The notation $\left(\cdot\right)_{k=1}^n$ means an $n$-tuple whose elements are labeled by the index $k$.
    All of the (real) \gls*{acr:ao} electron integrals of interest to us can be considered generically as a multilinear map
      \begin{alignat*}{2}
        M:\ F^1 \times F^2 \times \cdots \times F^n &\rightarrow \mathbb{R}\\
        \Omega^{(n)} &\mapsto M[\Omega^{(n)}]
      \end{alignat*}
    that is linear in each of its arguments.
    The permutation symmetries of the electron integrals then imply that $M$ has some permutation invariance
      \begin{equation}
        M[\Omega^{(n)}] = M[\Omega^{(n)}_{\sigma}],
        \label{eq:perminvariance}
      \end{equation}
    where
    \ifdefined\twocolumnmode
      \begin{align}
        \Omega^{(n)}_{\sigma}
        &= \hat{P}_{\sigma} \Omega^{(n)} \nonumber\\
        &= \hat{P}_{\sigma} \left(\alpha^k_{i_{\alpha^k}}\right)_{k=1}^n
        = \left(\sigma\alpha^k_{i_{\alpha^k}}\right)_{k=1}^n
        \label{eq:permtuple}
      \end{align}
    \else
      \begin{equation}
        \Omega^{(n)}_{\sigma}
        = \hat{P}_{\sigma} \Omega^{(n)}
        = \hat{P}_{\sigma} \left(\alpha^k_{i_{\alpha^k}}\right)_{k=1}^n
        = \left(\sigma\alpha^k_{i_{\alpha^k}}\right)_{k=1}^n
        \label{eq:permtuple}
      \end{equation}
    \fi
    is the $n$-tuple that has been acted on by $\hat{P}_{\sigma} \in \operatorname{Sym}(S)$ which permutes the function labels, sending sending $\alpha^k$ to $\sigma\alpha^k$.

    We now consider the effects of point-group symmetry on a general permutation sum over a subgroup $T$ of $\operatorname{Sym}(S)$ for a particular index combination ${\{i_{\alpha}\}}$ of the multi-linear map $M$:
      \begin{equation}
        A_{\{i_{\alpha}\}}
        = \sum_{\sigma \in T} \lambda_{\sigma} M[\Omega^{(n)}_{\sigma}]
        = \sum_{\sigma \in T} \lambda_{\sigma} \hat{P}_{\sigma} M[\Omega^{(n)}],
        \label{eq:permsum}
      \end{equation}
    where $\lambda_{\sigma}$ is the signature associated with the permutation $\hat{P}_{\sigma}$ and determined by the physics of the problem that gives rise to $A_{\{i_{\alpha}\}}$.
    We require that the set $\{\lambda_{\sigma} \mid \hat{P}_{\sigma} \in T\}$ forms the basis for one of the one-dimensional irreducible representations of $T$ so that
      \begin{equation}
        \hat{P}_{\sigma} \hat{P}_{\sigma'} = \hat{P}_{\sigma''}
        \Leftrightarrow
        \lambda_{\sigma}\lambda_{\sigma'} = \lambda_{\sigma''},
        \label{eq:signaturemult}
      \end{equation}
    where group multiplication under $\operatorname{Sym}(S)$ is denoted by juxtaposition.
    Let $\hat{R}$ be a symmetry element in the group $\mathcal{G}$ with a representation matrix $\boldsymbol{D}^k(\hat{R})$ in the basis of $F^k = \{\alpha^k_{i_{\alpha^k}}\}$:
      \begin{equation}
        \hat{R}\alpha^k_{i_{\alpha^k}} =
        \sum_{i'_{\alpha^k}}
        \alpha^k_{i'_{\alpha^k}} D^k_{i'_{\alpha^k} i_{\alpha^k}} (\hat{R}).
        \label{eq:Rrepmat}
      \end{equation}
    The effect of $\hat{R}$ on a general term $M[\Omega^{(n)}_{\sigma}]$ in $A_{\{i_{\alpha}\}}$ is given by
    \ifdefined\twocolumnmode
      \begin{subequations}
        \begin{align}
          \hat{R}&M[\Omega^{(n)}_{\sigma}]\\
          &= \hat{R} M\left(
                \sigma\alpha^k_{i_{\alpha^k}}
              \right)_{k=1}^n \label{subeq:frompermtuple}\\
          &= M\left(
                \hat{R}\sigma\alpha^k_{i_{\alpha^k}}
              \right)_{k=1}^n \label{subeq:Rintotuple}\\
          &= M\left(
                \sum_{i'_{\sigma\alpha^k}}
                \sigma\alpha^k_{i'_{\sigma\alpha^k}} D^k_{i'_{\sigma\alpha^k} i_{\alpha^k}} (\hat{R})
              \right)_{k=1}^n  \label{subeq:Rrepmatintotuple}\\
          &= \sum_{\{i'_{\sigma\alpha}\}} M\left(
                \sigma\alpha^k_{i'_{\sigma\alpha^k}}
              \right)_{k=1}^n
              \prod_{k=1}^{n}
              D^k_{i'_{\sigma\alpha^k} i_{\alpha^k}} (\hat{R}) \label{subeq:pullingsumout}\\
          &= \sum_{\{i'_\alpha\}} M\left(
                \alpha^k_{i'_{\alpha^k}}
              \right)_{k=1}^n
              \prod_{k=1}^{n}
              D^k_{i'_{\sigma\alpha^k} i_{\alpha^k}} (\hat{R}), \label{subeq:finalresult}
        \end{align}
      \end{subequations}
    \else
      \begin{subequations}
        \begin{align}
          \hat{R}M[\Omega^{(n)}_{\sigma}]
          &= \hat{R} M\left(
                \sigma\alpha^k_{i_{\alpha^k}}
              \right)_{k=1}^n \label{subeq:frompermtuple}\\
          &= M\left(
                \hat{R}\sigma\alpha^k_{i_{\alpha^k}}
              \right)_{k=1}^n \label{subeq:Rintotuple}\\
          &= M\left(
                \sum_{i'_{\sigma\alpha^k}}
                \sigma\alpha^k_{i'_{\sigma\alpha^k}} D^k_{i'_{\sigma\alpha^k} i_{\alpha^k}} (\hat{R})
              \right)_{k=1}^n  \label{subeq:Rrepmatintotuple}\\
          &= \sum_{\{i'_{\sigma\alpha}\}} M\left(
                \sigma\alpha^k_{i'_{\sigma\alpha^k}}
              \right)_{k=1}^n
              \prod_{k=1}^{n}
              D^k_{i'_{\sigma\alpha^k} i_{\alpha^k}} (\hat{R}) \label{subeq:pullingsumout}\\
          &= \sum_{\{i'_\alpha\}} M\left(
                \alpha^k_{i'_{\alpha^k}}
              \right)_{k=1}^n
              \prod_{k=1}^{n}
              D^k_{i'_{\sigma\alpha^k} i_{\alpha^k}} (\hat{R}), \label{subeq:finalresult}
        \end{align}
      \end{subequations}
    \fi
    where we use the definition in Equation~\ref{eq:permtuple} to write Equation~\ref{subeq:frompermtuple}.
    Next, we utilise the fact that $\hat{R}$ acts on each function in the argument of the linear map $M$ separately to move $\hat{R}$ inside the tuple in Equation~\ref{subeq:Rintotuple}.
    The definition of the representation matrices for $\hat{R}$ in Equation~\ref{eq:Rrepmat} is then invoked to expand each term in the tuple to give Equation~\ref{subeq:Rrepmatintotuple}.
    The linearity of $M$ on each of its arguments allows the individual sums in the tuple to be pulled outside as multiple sums over all function indices in Equation~\ref{subeq:pullingsumout}, $\sum_{\{i'_{\sigma\alpha}\}} \equiv \sum_{i'_{\sigma\alpha^1}} \cdots \sum_{i'_{\sigma\alpha^n}}$.
    Finally, the permutation invariance of $M$ (Equation~\ref{eq:perminvariance}) and the closure of $S$ under $\hat{P}_{\sigma}$ yield Equation~\ref{subeq:finalresult}.
    The overall effect of $\hat{R}$ on $A_{\{i_{\alpha}\}}$ is thus
      \begin{equation}
        \hat{R}A_{\{i_{\alpha}\}}
        = \sum_{\{i'_\alpha\}}
          M\left(\alpha^k_{i'_{\alpha^k}}\right)_{k=1}^n
          \sum_{\sigma} \lambda_{\sigma}
          \prod_{k=1}^{n}
            D^k_{i'_{\sigma\alpha^k} i_{\alpha^k}} (\hat{R}).
        \label{eq:RA}
      \end{equation}

    We note that the sum $A_{\{i_{\alpha}\}}$ is invariant up to a phase factor under the permutation subgroup $T$ since
      \begin{align*}
        \forall \hat{P}_{\sigma'} \in T, \quad
        \hat{P}_{\sigma'} A_{\{i_{\alpha}\}}
          &= \sum_{\sigma \in T}
              \lambda_{\sigma}
              \hat{P}_{\sigma'} \hat{P}_{\sigma} M[\Omega^{(n)}] \\
          &= \sum_{\sigma'' \in T}
              \lambda_{\sigma'}^{-1} \lambda_{\sigma''}
              \hat{P}_{\sigma''} M[\Omega^{(n)}] \\
          &= \lambda_{\sigma'}^{-1} A_{\{i_{\alpha}\}},
      \end{align*}
    where we have invoked the closure of $T$ under group multiplication and the requirement for signature multiplication (Equation~\ref{eq:signaturemult}) to ``absorb'' one of the permutation operators into the other for the second equality.
    This allows us to write
    \ifdefined\twocolumnmode
      \begin{align*}
        \hat{R}A_{\{i_{\alpha}\}}
        &= \hat{R}\left(
        \frac{1}{\lvert T\rvert} \sum_{\sigma' \in T} \lambda_{\sigma'} \hat{P}_{\sigma'} A_{\{i_{\alpha}\}}
        \right)\\
        &= \frac{1}{\lvert T\rvert} \sum_{\sigma' \in T} \lambda_{\sigma'} \hat{P}_{\sigma'} \hat{R}A_{\{i_{\alpha}\}}.
      \end{align*}
    \else
      \begin{equation*}
        \hat{R}A_{\{i_{\alpha}\}}
        = \hat{R}\left(
        \frac{1}{\lvert T\rvert} \sum_{\sigma' \in T} \lambda_{\sigma'} \hat{P}_{\sigma'} A_{\{i_{\alpha}\}}
        \right)
        = \frac{1}{\lvert T\rvert} \sum_{\sigma' \in T} \lambda_{\sigma'} \hat{P}_{\sigma'} \hat{R}A_{\{i_{\alpha}\}}.
      \end{equation*}
    \fi
    Substitution of the result in Equation~\ref{eq:RA} for the $\hat{R}A_{\{i_{\alpha}\}}$ term under the summation gives
    \ifdefined\twocolumnmode
      \begin{align*}
        \hat{R}A_{\{i_{\alpha}\}}
          &= \sum_{\{i'_{\alpha}\}}
             \begin{multlined}[t]
               \left[
                 \sum_{\sigma' \in T} \lambda_{\sigma'} \hat{P}_{\sigma'}
                 M\left(\alpha^k_{i'_{\alpha^k}}\right)_{k=1}^n
               \right]\\
               \times \left[
                 \frac{1}{\lvert T\rvert} \sum_{\sigma \in T} \lambda_{\sigma}
                 \prod_{k=1}^{n}
                 D^k_{i'_{\sigma\alpha^k} i_{\alpha^k}} (\hat{R})
               \right]
             \end{multlined} \\
          &= \sum_{\{i'_{\alpha}\}}
               A_{\{i'_{\alpha}\}}
               D^{A_{\{i_{\alpha}\}}}_{\{i'_{\alpha}\}, \{i_{\alpha}\}} (\hat{R}).
      \end{align*}
    \else
      \begin{align*}
        \hat{R}A_{\{i_{\alpha}\}}
          &= \sum_{\{i'_{\alpha}\}}
             \left[
               \sum_{\sigma' \in T} \lambda_{\sigma'} \hat{P}_{\sigma'}
               M\left(\alpha^k_{i'_{\alpha^k}}\right)_{k=1}^n
             \right]
             \left[
               \frac{1}{\lvert T\rvert} \sum_{\sigma \in T} \lambda_{\sigma}
               \prod_{k=1}^{n}
               D^k_{i'_{\sigma\alpha^k} i_{\alpha^k}} (\hat{R})
             \right]\\
          &= \sum_{\{i'_{\alpha}\}}
               A_{\{i'_{\alpha}\}}
               D^{A_{\{i_{\alpha}\}}}_{\{i'_{\alpha}\}, \{i_{\alpha}\}} (\hat{R}).
      \end{align*}
    \fi
    We thus obtain an expression for the representation matrix of the permutation sum $A_{\{i_{\alpha}\}}$ under $\hat{R}$:
      \begin{equation}
        D^{A_{\{i_{\alpha}\}}}_{\{i'_{\alpha}\}, \{i_{\alpha}\}} (\hat{R})
        =
        \frac{1}{\lvert T\rvert} \sum_{\sigma \in T} \lambda_{\sigma}
        \prod_{k=1}^{n}
        D^k_{i'_{\sigma\alpha^k} i_{\alpha^k}} (\hat{R}).
        \label{eq:Dmat}
      \end{equation}
    which gives a practical way to determine how $\hat{R}$ affects the multilinear map $M[\Omega^{(n)}]$ subject to the permutation invariance \ref{eq:perminvariance}.
    For $A_{\{i_{\alpha}\}}$ to be non-vanishing, $\boldsymbol{D}^{A_{\{i_{\alpha}\}}} (\hat{R})$ must contain totally symmetric components.

  \subsection{Applications to Particular Electron Integrals}

    \paragraph{One-Electron Integrals.}
      The map of interest for (real) one-electron integrals is $M = \braket{\cdot|\hat{O}|\cdot}$, where $\hat{O}$ is a one-electron Hermitian operator.
      We consider a set $S$ of two function labels, $S = \{\xi, \eta\}$, which gives rise to two sets of real functions $F^{\xi} = \{\xi_{i_{\xi}}\}$ and $F^{\eta} = \{\eta_{i_{\eta}}\}$ that we insist form \textit{identical} bases for a representation $\Gamma_{\mathcal{G}}$ of the group $\mathcal{G}$.
      The Hermiticity of $\hat{O}$ and the reality of the functions in $F$ imply that the invariant permutation sum we need to consider is
       \begin{equation*}
         A_{\{i_{\xi}i_{\eta}\}} = \braket{\xi_{i_{\xi}}|\hat{O}|\eta_{i_{\eta}}} +  \braket{\eta_{i_{\xi}}|\hat{O}|\xi_{i_{\eta}}},
       \end{equation*}
      which defines a permutation subgroup $T = \{\hat{e}, (\xi\eta)\}$ where both elements have a signature of $+1$.
      From Equation~\ref{eq:Dmat}, the representation matrix for any element $R \in \mathcal{G}$ is
      \ifdefined\twocolumnmode
        \begin{multline*}
          D^{A_{\{i_{\xi}i_{\eta}\}}}_{i'_{\xi}i'_{\eta},i_{\xi}i_{\eta}} (\hat{R})
            =\\
            \frac{1}{2}\left\lbrace
              D^{\Gamma_{\mathcal{G}}}_{i'_{\xi}i_{\xi}}(\hat{R}) D^{\Gamma_{\mathcal{G}}}_{i'_{\eta}i_{\eta}}(\hat{R})
              +
              D^{\Gamma_{\mathcal{G}}}_{i'_{\eta}i_{\xi}}(\hat{R}) D^{\Gamma_{\mathcal{G}}}_{i'_{\xi}i_{\eta}}(\hat{R})
            \right\rbrace.
        \end{multline*}
      \else
        \begin{equation*}
          D^{A_{\{i_{\xi}i_{\eta}\}}}_{i'_{\xi}i'_{\eta},i_{\xi}i_{\eta}} (\hat{R})
            =
            \frac{1}{2}\left\lbrace
              D^{\Gamma_{\mathcal{G}}}_{i'_{\xi}i_{\xi}}(\hat{R}) D^{\Gamma_{\mathcal{G}}}_{i'_{\eta}i_{\eta}}(\hat{R})
              +
              D^{\Gamma_{\mathcal{G}}}_{i'_{\eta}i_{\xi}}(\hat{R}) D^{\Gamma_{\mathcal{G}}}_{i'_{\xi}i_{\eta}}(\hat{R})
            \right\rbrace.
        \end{equation*}
      \fi
      Setting $i'_{\xi} = i_{\xi}$ and $i'_{\eta} = i_{\eta}$ and summing over, we get
        \begin{equation*}
          \chi^{A_{\{i_{\xi}i_{\eta}\}}} (\hat{R})
            =
            \frac{1}{2}\left\lbrace
              \left[\chi^{\Gamma_{\mathcal{G}}}(\hat{R})\right]^2
              + \chi^{\Gamma_{\mathcal{G}}}(\hat{R}^2)
            \right\rbrace,
        \end{equation*}
      which is the character of $\hat{R}$ in the familiar symmetrized square of $\Gamma_{\mathcal{G}}$\cite{book:Wigner1959,book:Griffith1961}.
      Applying the reduction formula for the totally symmetric irreducible representation of $\mathcal{G}$ then gives the number of totally symmetric components contained in the representation spanned by $A_{\{i_{\xi}i_{\eta}\}}$, and thus the number of non-vanishing components:
      \ifdefined\twocolumnmode
        \begin{multline}
          n_1(\gls*{struct:rep}_{\gls*{struct:gengroup}}, \gls*{struct:gengroup})\\
            = \frac{1}{\lvert\gls*{struct:gengroup}\rvert}
                \sum_{\hat{R} \in \gls*{struct:gengroup}}
                \frac{1}{2}
                  \left\lbrace
                    \left[\gls*{bas:spinorb}^{\gls*{struct:rep}_{\gls*{struct:gengroup}}}(\hat{R})\right]^2
                    + \gls*{bas:spinorb}^{\gls*{struct:rep}_{\gls*{struct:gengroup}}}\left(\hat{R}^2\right)
                  \right\rbrace.
        \end{multline}
      \else
        \begin{equation}
          n_1(\gls*{struct:rep}_{\gls*{struct:gengroup}}, \gls*{struct:gengroup})
            = \frac{1}{\lvert\gls*{struct:gengroup}\rvert}
              \sum_{\hat{R} \in \gls*{struct:gengroup}}
                \frac{1}{2}
                \left\lbrace
                  \left[\gls*{bas:spinorb}^{\gls*{struct:rep}_{\gls*{struct:gengroup}}}(\hat{R})\right]^2
                  + \gls*{bas:spinorb}^{\gls*{struct:rep}_{\gls*{struct:gengroup}}}\left(\hat{R}^2\right)
                \right\rbrace.
        \end{equation}
      \fi

    \paragraph{Two-Electron Integrals.}
      The map of interest for (real) two-electron integrals is now $M = \braket{\cdot\cdot|\cdot\cdot}$ with the function labels $S = \{\alpha, \beta, \gamma, \delta\}$ giving rise to four sets of functions $F^{\alpha} = \{\alpha_{i_{\alpha}}\}$, $F^{\beta} = \{\beta_{i_{\beta}}\}$, $F^{\gamma} = \{\gamma_{i_{\gamma}}\}$, and $F^{\delta} = \{\delta_{i_{\delta}}\}$ forming \textit{identical} bases for a representation $\Gamma_{\mathcal{G}}$ in $\mathcal{G}$.
      The relevant invariant permutation sum $A_{\{i_{\alpha}i_{\beta}i_{\gamma}i_{\delta}\}}$ can be constructed from the permutation subgroup
      \ifdefined\twocolumnmode
        \begin{align*}
          T ={} \{
            &\hat{e},\\
            &(\alpha\gamma)(\beta\delta), (\alpha\beta)(\gamma\delta), (\alpha\delta)(\beta\gamma),\\
            &(\alpha\gamma), (\beta\delta),\\
            &(\alpha\beta\gamma\delta), (\alpha\delta\gamma\beta)
          \}
        \end{align*}
      \else
        \begin{equation*}
          T = \{
          \hat{e},
          (\alpha\gamma)(\beta\delta), (\alpha\beta)(\gamma\delta), (\alpha\delta)(\beta\gamma),
          (\alpha\gamma), (\beta\delta),
          (\alpha\beta\gamma\delta), (\alpha\delta\gamma\beta)
          \}
        \end{equation*}
      \fi
      where all eight elements have a signature of $+1$.
      Following the same approach as above, we obtain the character formula for $\hat{R} \in \mathcal{G}$,
      \ifdefined\twocolumnmode
        \begin{multline*}
          \chi^{A_{\{i_{\alpha}i_{\beta}i_{\gamma}i_{\delta}\}}} (\hat{R})
          =\\
          \frac{1}{8}\left\lbrace
            \left[\chi^{\Gamma_{\mathcal{G}}}(\hat{R})\right]^4
              + 3\left[\chi^{\Gamma_{\mathcal{G}}}(\hat{R}^2)\right]^2
          \right.\\
          \left.
            +\ 2\chi^{\Gamma_{\mathcal{G}}}(\hat{R}^2) \left[\chi^{\Gamma_{\mathcal{G}}}(\hat{R})\right]^2
            + 2\chi^{\Gamma_{\mathcal{G}}}(\hat{R}^4)
          \right\rbrace.
        \end{multline*}
      \else
        \begin{equation*}
          \chi^{A_{\{i_{\alpha}i_{\beta}i_{\gamma}i_{\delta}\}}} (\hat{R})
          =
          \frac{1}{8}\left\lbrace
            \left[\chi^{\Gamma_{\mathcal{G}}}(\hat{R})\right]^4
            + 3\left[\chi^{\Gamma_{\mathcal{G}}}(\hat{R}^2)\right]^2
            + 2\chi^{\Gamma_{\mathcal{G}}}(\hat{R}^2) \left[\chi^{\Gamma_{\mathcal{G}}}(\hat{R})\right]^2
            + 2\chi^{\Gamma_{\mathcal{G}}}(\hat{R}^4)
          \right\rbrace.
        \end{equation*}
      \fi
      An application of the reduction formula for the totally symmetric irreducible representation of $\mathcal{G}$ thus yields
      \ifdefined\twocolumnmode
        \begin{multline}
          n_2(\gls*{struct:rep}_{\gls*{struct:gengroup}}, \gls*{struct:gengroup})\\
            = \frac{1}{\lvert\gls*{struct:gengroup}\rvert}
              \sum_{\hat{R} \in \gls*{struct:gengroup}}
                \frac{1}{8}
                  \left\lbrace
                    \left[\gls*{bas:spinorb}^{\gls*{struct:rep}_{\gls*{struct:gengroup}}}(\hat{R})\right]^4
                    + 3\left[\gls*{bas:spinorb}^{\gls*{struct:rep}_{\gls*{struct:gengroup}}}(\hat{R}^2)\right]^2
                  \right. \\
                  \left.
                    +\ 2\gls*{bas:spinorb}^{\gls*{struct:rep}_{\gls*{struct:gengroup}}}(\hat{R}^2) \left[\gls*{bas:spinorb}^{\gls*{struct:rep}_{\gls*{struct:gengroup}}}(\hat{R})\right]^2
                    + 2\gls*{bas:spinorb}^{\gls*{struct:rep}_{\gls*{struct:gengroup}}}(\hat{R}^4)
                  \right\rbrace.
        \end{multline}
      \else
        \begin{equation}
          n_2(\gls*{struct:rep}_{\gls*{struct:gengroup}}, \gls*{struct:gengroup})
          = \frac{1}{\lvert\gls*{struct:gengroup}\rvert}
            \sum_{\hat{R} \in \gls*{struct:gengroup}}
              \frac{1}{8}
              \left\lbrace
                \left[\gls*{bas:spinorb}^{\gls*{struct:rep}_{\gls*{struct:gengroup}}}(\hat{R})\right]^4
                + 3\left[\gls*{bas:spinorb}^{\gls*{struct:rep}_{\gls*{struct:gengroup}}}(\hat{R}^2)\right]^2
                + 2\gls*{bas:spinorb}^{\gls*{struct:rep}_{\gls*{struct:gengroup}}}(\hat{R}^2) \left[\gls*{bas:spinorb}^{\gls*{struct:rep}_{\gls*{struct:gengroup}}}(\hat{R})\right]^2
                + 2\gls*{bas:spinorb}^{\gls*{struct:rep}_{\gls*{struct:gengroup}}}(\hat{R}^4)
              \right\rbrace.
        \end{equation}
      \fi

    \paragraph{Anti-Symmetrized Two-Electron Integrals.}
      The map of interest is now $M = \braket{\cdot\cdot||\cdot\cdot}$.
      Since $\braket{\alpha\beta||\gamma\delta} = \braket{\alpha\beta|\gamma\delta} - \braket{\alpha\beta|\delta\gamma}$, the relevant permutation subgroup is now
      \ifdefined\twocolumnmode
        \begin{align*}
          T ={} \{
            &\hat{e},\\
            &(\alpha\beta)(\gamma\delta), (\alpha\gamma)(\beta\delta), (\alpha\delta)(\beta\gamma),\\
            &(\alpha\beta), (\gamma\delta),\\
            &(\alpha\delta\beta\gamma), (\alpha\gamma\beta\delta)
          \},
        \end{align*}
      \else
        \begin{equation*}
          T = \{
          \hat{e},
          (\alpha\beta)(\gamma\delta), (\alpha\gamma)(\beta\delta), (\alpha\delta)(\beta\gamma),
          (\alpha\beta), (\gamma\delta),
          (\alpha\delta\beta\gamma), (\alpha\gamma\beta\delta)
          \},
        \end{equation*}
      \fi
      but the signatures of the last four elements are $-1$.
      This thus leads to the character formula
      \ifdefined\twocolumnmode
        \begin{multline*}
          \chi^{A_{\{i_{\alpha}i_{\beta}i_{\gamma}i_{\delta}\}}} (\hat{R})
          =\\
          \frac{1}{8}\left\lbrace
            \left[\chi^{\Gamma_{\mathcal{G}}}(\hat{R})\right]^4
            + 3\left[\chi^{\Gamma_{\mathcal{G}}}(\hat{R}^2)\right]^2
          \right.\\
          \left.
            - 2\chi^{\Gamma_{\mathcal{G}}}(\hat{R}^2) \left[\chi(\hat{R})\right]^2
            - 2\chi^{\Gamma_{\mathcal{G}}}(\hat{R}^4)
          \right\rbrace.
        \end{multline*}
      \else
        \begin{equation*}
          \chi^{A_{\{i_{\alpha}i_{\beta}i_{\gamma}i_{\delta}\}}} (\hat{R})
          =
          \frac{1}{8}\left\lbrace
          \left[\chi^{\Gamma_{\mathcal{G}}}(\hat{R})\right]^4
          + 3\left[\chi^{\Gamma_{\mathcal{G}}}(\hat{R}^2)\right]^2
          - 2\chi^{\Gamma_{\mathcal{G}}}(\hat{R}^2) \left[\chi(\hat{R})\right]^2
          - 2\chi^{\Gamma_{\mathcal{G}}}(\hat{R}^4)
          \right\rbrace.
        \end{equation*}
      \fi

\section{The Time-Reversal Group}
\label{app:timerevgroup}

  Let $\mathcal{T}$ be the abstract cyclic group of order $2$: $\mathcal{T} = \lbrace e, t \mid t^2 = e \rbrace$, and let $\rho$ be a linear-antilinear representation\cite{article:Garola1981} (also known as corepresentation\cite{book:Wigner1959}) of $\mathcal{T}$ on a complex Hilbert space $\mathcal{H}$ so that the familiar identity operator (which is linear) and the time-reversal operator (which is antilinear) on this space can be identified as
  \begin{equation*}
    \hat{E} \equiv \rho_e, \qquad \hat{\Theta} \equiv \rho_t,
  \end{equation*}
  respectively.
  Due to the homomorphicity of $\rho$, we can speak of $\mathcal{T}$ as the concrete group $\lbrace \hat{E}, \hat{\Theta} \rbrace$ instead of the abstract cyclic group of order $2$.
  The characters for $\mathcal{T}$ for its two corepresentations $A$ and $B$ are given in Table~\ref{tab:Tcharactertable}.
  These corepresentations are constructed based on the procedure described by Wigner\cite{book:Wigner1959} and then later summarized by Cracknell\cite{article:Cracknell1965}.

  \begin{table}
    \centering
    \caption{Character table for the time-reversal group.}
    \label{tab:Tcharactertable}
    \begin{tabular}[t]{
        >{\raggedright\arraybackslash}m{0.4cm} | *{2}{>{\centering\arraybackslash}m{0.7cm}}
      }
      \toprule
      $\mathcal{T}$ & $\hat{E}$ & $\hat{\Theta}$ \\
      \midrule
      $A$ & $1$ & $\hphantom{-}1\hphantom{-}$ \\
      $B$ & $1$ & $-1\hphantom{-}$ \\
      \bottomrule
    \end{tabular}
  \end{table}

\section{Variation of Representations}
\label{app:symvariation}

  In this Appendix, we state and prove a proposition that allows us to relate the symmetry of a wavefunction (or a set of symmetry-related degenerate wavefunctions) at a coalescence point with that in the vicinity.
  This relation implies that coalescing wavefunctions must contain common (co)representations.
  For example, a set of degenerate wavefunctions $\{\gls*{wf:gen}_i\}$ spanning a (co)representation $\gls*{struct:rep}_1 \oplus \gls*{struct:rep}_2 \oplus \gls*{struct:rep}_3 \oplus \gls*{struct:rep}_4$ can coalesce with another set of degenerate wavefunctions $\{\gls*{wf:gen}'_j\}$ spanning a (co)representation $\gls*{struct:rep}_2 \oplus \gls*{struct:rep}_3 \oplus \gls*{struct:rep}_5$ ($\gls*{struct:rep}_i$ being irreducible representations of a certain group), where at the coalescence point, wavefunctions from both sets become identical and span either $\gls*{struct:rep}_2$, $\gls*{struct:rep}_3$, or $\gls*{struct:rep}_2 \oplus \gls*{struct:rep}_3$.
  As a consequence, in the cases discussed in this article where the symmetry of persistently real solutions does not change upon coalescence with transiently real solutions, the transiently real solutions must then contain the (co)representations of the persistently real solutions with which they come into coalescence.

  Let $f$ and $g$ be two \emph{linearly independent} wavefunctions in a certain Hilbert space $\mathcal{H}$.
  We shall be interested in the linear or antilinear actions of a symmetry group $\mathcal{G}$ on the above wavefunctions.
  Let us consider the identity-containing subsets $\mathcal{G}_{f} \subseteq \mathcal{G}$ and $\mathcal{G}_{g} \subseteq \mathcal{G}$ such that each of the sets
  \begin{equation*}
    \mathcal{G}_{f}\cdot f = \{ \hat{R}_i f \mid \hat{R}_i \in \mathcal{G}_{f} \},\qquad
    \mathcal{G}_{g}\cdot g = \{ \hat{R}_j g \mid \hat{R}_j \in \mathcal{G}_{g} \}
  \end{equation*}
  contains all possible linearly independent elements so that for any $\hat{R}_m \in \mathcal{G}$, we can always write
  \begin{equation}
    \hat{R}_m f = \sum_{\hat{R}_i \in \mathcal{G}_f} \lambda^f_{im} \hat{R}_i f, \qquad
    \hat{R}_m g = \sum_{\hat{R}_j \in \mathcal{G}_g} \lambda^g_{jm} \hat{R}_j g
    \label{eq:linindeptrelations}
  \end{equation}
  for some $\lambda^f_{im}, \lambda^g_{jm} \in \mathbb{C}$.
  We further require that the elements in the two sets $\mathcal{G}_{f}\cdot f$ and $\mathcal{G}_{g}\cdot g$ are linearly independent of one another.
  These two sets thus form bases for two linear subspaces of $\mathcal{H}$ which we denote $\Gamma_f$ and $\Gamma_g$, respectively.
  Clearly, $\Gamma_f$ and $\Gamma_g$ are guaranteed by construction to be invariant under the linear actions of $\mathcal{G}$ and are therefore also representation spaces of $\mathcal{G}$.
  For brevity, in this Appendix, we shall refer to both linear representations and linear-antilinear corepresentations\cite{book:Wigner1959, article:Garola1981} simply as representations.

  Let us now consider a wavefunction $\psi$ in the same Hilbert space $\mathcal{H}$ given by
  \begin{equation}
    \psi = f + g
    \label{eq:psidef}
  \end{equation}
  and the corresponding subset $\mathcal{G}_{\psi} \subseteq \mathcal{G}$ for the linearly independent set
  \begin{equation*}
    \mathcal{G}_{\psi}\cdot \psi = \{ \hat{R}_k \psi \mid \hat{R}_k \in \mathcal{G}_{\psi} \}
  \end{equation*}
  that forms a basis for the representation space $\Gamma_{\psi}$.
  We now state and prove a key proposition.

  \begin{proposition}
    Take $\Gamma_f$ and $\Gamma_g$ to be irreducible representations of $\mathcal{G}$.
    If $\Gamma_f$ and $\Gamma_g$ are not equivalent to each other, or if they are equivalent but their representation matrices in the bases $\mathcal{G}_{f}\cdot f$ and $\mathcal{G}_{g}\cdot g$ are not identical, then $\Gamma_{\psi}$ is equivalent to $\Gamma_f \oplus \Gamma_g$, which we denote $\Gamma_{\psi} \sim \Gamma_f \oplus \Gamma_g$.
    On the other hand, if $\Gamma_f$ and $\Gamma_g$ are equivalent and have identical representation matrices, then $\Gamma_{\psi} \sim \Gamma_f \sim \Gamma_g$.
    \label{prop:irrepsum}
  \end{proposition}
  \begin{proof}
    For notational convenience, we gather the linearly independent elements of $\mathcal{G}_{f}\cdot f$, $\mathcal{G}_{g}\cdot g$, and $\mathcal{G}_{\psi}\cdot \psi$ into the corresponding column vectors $\boldsymbol{f}$, $\boldsymbol{g}$, and $\boldsymbol{\psi}$:
    \begin{alignat*}{2}
      f_i &= \hat{R}_i f,      \quad &&\hat{R}_i \in \mathcal{G}_{f},\\
      g_j &= \hat{R}_j g,      \quad &&\hat{R}_j \in \mathcal{G}_{g},\\
      \psi_k &= \hat{R}_k\psi, \quad &&\hat{R}_k \in \mathcal{G}_{\psi}.
    \end{alignat*}
    Let $\hat{R}$ be an element in $\mathcal{G}$.
    The representation matrices of $\hat{R}$ in the spaces $\Gamma_f$, $\Gamma_g$, and $\Gamma_{\psi}$ are given by
    \begin{subequations}
      \begin{alignat}{4}
        \hat{R} f_i &= \sum_{i' = 1}^{\lvert\mathcal{G}_f\rvert} f_{i'} D^{\Gamma_f}_{i'i}(\hat{R})
          \quad
          &\Leftrightarrow
          \quad
          &\hat{R} \boldsymbol{f}^{\T} &&= \boldsymbol{f}^{\T} \boldsymbol{D}^{\Gamma_f}(\hat{R}), \label{eq:Dfdef}\\
        \hat{R} g_j &= \sum_{j' = 1}^{\lvert\mathcal{G}_g\rvert} g_{j'} D^{\Gamma_g}_{j'j}(\hat{R})
          \quad
          &\Leftrightarrow
          \quad
          &\hat{R} \boldsymbol{g}^{\T} &&= \boldsymbol{g}^{\T} \boldsymbol{D}^{\Gamma_g}(\hat{R}), \label{eq:Dgdef}\\
        \hat{R} \psi_k &= \sum_{k' = 1}^{\lvert\mathcal{G}_{\psi}\rvert} \psi_{k'} D^{\Gamma_{\psi}}_{k'k}(\hat{R})
          \quad
          &\Leftrightarrow
          \quad
          &\hat{R} \boldsymbol{\psi}^{\T} &&= \boldsymbol{\psi}^{\T} \boldsymbol{D}^{\Gamma_{\psi}}(\hat{R}).
          \label{eq:Dpsidef}
      \end{alignat}%
      \label{eq:repmatdefs}%
    \end{subequations}

    To prove the proposition, we seek a relation between these representation matrices.
    From the definition of $\psi$ in (\ref{eq:psidef}) and the expansions in (\ref{eq:linindeptrelations}), we can write
    \begin{subequations}
      \begin{align}
        \psi_k
          &= \hat{R}_k f + \hat{R}_k g \nonumber\\
          &= \sum_{i = 1}^{\lvert\mathcal{G}_f\rvert} f_i\lambda^f_{ik}
          +
          \sum_{j = 1}^{\lvert\mathcal{G}_g\rvert} g_j\lambda^g_{jk},
          \label{eq:psikexpansion}
      \end{align}
      or more compactly,
      \begin{equation}
        \boldsymbol{\psi}^{\T} =
          \begin{pmatrix}
            \boldsymbol{f}^{\T} & \boldsymbol{g}^{\T}
          \end{pmatrix}
          \begin{pmatrix}
            \boldsymbol{\lambda}^f \\ \boldsymbol{\lambda}^g
          \end{pmatrix}
          =
          \begin{pmatrix}
            \boldsymbol{f}^{\T} & \boldsymbol{g}^{\T}
          \end{pmatrix}
          \boldsymbol{\lambda}^{\psi},
        \label{eq:psikexpansionmat}%
      \end{equation}%
    \end{subequations}
    where the second equality defines the matrix $\boldsymbol{\lambda}^{\psi}$ with dimensions $(\lvert\mathcal{G}_f\rvert + \lvert\mathcal{G}_g\rvert) \times \lvert\mathcal{G}_{\psi}\rvert$.
    The linear independence of $\psi_k$ implies that the columns of $\boldsymbol{\lambda}^{\psi}$ are linearly independent.
    In addition,  the linear independence of $\{f_i\}$ and $\{g_j\}$ requires that $\lvert\mathcal{G}_{\psi}\rvert$, the number of columns in $\boldsymbol{\lambda}^{\psi}$, must satisfy
    \begin{equation*}
      \max(\lvert\mathcal{G}_f\rvert, \lvert\mathcal{G}_g\rvert)
      \le
      \lvert\mathcal{G}_{\psi}\rvert
      \le
      \lvert\mathcal{G}_f\rvert + \lvert\mathcal{G}_g\rvert.
    \end{equation*}
    Combining (\ref{eq:Dpsidef}) and (\ref{eq:psikexpansionmat}), we get
    \begin{subequations}
      \begin{equation}
        \hat{R} \boldsymbol{\psi}^{\T} =
          \begin{pmatrix}
            \boldsymbol{f}^{\T} & \boldsymbol{g}^{\T}
          \end{pmatrix}
          \boldsymbol{\lambda}^{\psi}
          \boldsymbol{D}^{\Gamma_{\psi}}(\hat{R}),
          \label{eq:Rpsipsi}
      \end{equation}
      for any $\hat{R} \in \mathcal{G}$.
      But from (\ref{eq:Dfdef}) and (\ref{eq:Dgdef}), we can also write
      \begin{align}
        \hat{R} \boldsymbol{\psi}^{\T}
        &= \begin{pmatrix}
            \hat{R}\boldsymbol{f}^{\T} & \hat{R}\boldsymbol{g}^{\T}
           \end{pmatrix}
          \boldsymbol{\lambda}^{\psi} \nonumber\\
        &= \begin{pmatrix}
            \boldsymbol{f}^{\T} & \boldsymbol{g}^{\T}
           \end{pmatrix}
          \begin{pmatrix}
            \boldsymbol{D}^{\Gamma_f}(\hat{R}) & \boldsymbol{0} \\
            \boldsymbol{0}                     & \boldsymbol{D}^{\Gamma_g}(\hat{R})
          \end{pmatrix}
          \boldsymbol{\lambda}^{\psi}. \label{eq:Rpsifg}
      \end{align}%
      \label{eq:Rpsi}%
    \end{subequations}
    Comparing (\ref{eq:Rpsipsi}) with (\ref{eq:Rpsifg}) and making use of the linear independence of the functions in $\boldsymbol{f}$ and $\boldsymbol{g}$, we deduce that
    \begin{equation}
      \boldsymbol{\lambda}^{\psi} \boldsymbol{D}^{\Gamma_{\psi}}(\hat{R})
      =
      \begin{pmatrix}
        \boldsymbol{D}^{\Gamma_f}(\hat{R}) & \boldsymbol{0} \\
        \boldsymbol{0}                     & \boldsymbol{D}^{\Gamma_g}(\hat{R})
      \end{pmatrix}
      \boldsymbol{\lambda}^{\psi}.
      \label{eq:Drelation}
    \end{equation}

    To proceed, we now need to condition $\boldsymbol{\lambda}^{\psi}$.
    Without loss of generality, let us choose $\hat{R}_1$ to be the identity of $\mathcal{G}$.
    The definition of $\psi$ in (\ref{eq:psidef}) and the fact that $\hat{R}_1$ is also a member of $\mathcal{G}_f$, $\mathcal{G}_g$, and $\mathcal{G}_{\psi}$ by construction imply that
    \begin{equation*}
      \lambda^f_{i1} = \delta_{i1}, \qquad \lambda^g_{j1} = \delta_{j1}.
    \end{equation*}
    Then, writing $\hat{R}_k = \hat{R}_k\hat{R}_1$ for any $\hat{R}_k \in \mathcal{G}_{\psi}$ and applying the expansion in (\ref{eq:psikexpansion}), we get
    \begin{align*}
      \psi_k
      &=
        \hat{R}_k \hat{R}_1 f +
        \hat{R}_k \hat{R}_1 g\\
      &=
        \hat{R}_k \sum_{i = 1}^{\lvert\mathcal{G}_f\rvert} f_i\lambda^f_{i1} +
        \hat{R}_k \sum_{j = 1}^{\lvert\mathcal{G}_g\rvert} g_j\lambda^g_{j1}\\
      &=
        \sum_{i = 1}^{\lvert\mathcal{G}_f\rvert} \hat{R}_k f_i\delta_{i1} +
        \sum_{j = 1}^{\lvert\mathcal{G}_g\rvert} \hat{R}_k g_j\delta_{j1},
    \end{align*}
    where in the last equality we have used the fact that $\hat{R}_k$ is linear or antilinear and that the Kronecker deltas are real.
    Using the definition of the representation matrices in (\ref{eq:repmatdefs}) for $\hat{R} = \hat{R}_k$, we obtain
    \begin{align}
      \psi_k
      &=
        \sum_{i,i' = 1}^{\lvert\mathcal{G}_f\rvert} f_{i'} D^{\Gamma_f}_{i'i}(\hat{R}_k) \delta_{i1} +
        \sum_{j,j' = 1}^{\lvert\mathcal{G}_g\rvert} g_{j'} D^{\Gamma_g}_{j'j}(\hat{R}_k) \delta_{j1} \nonumber\\
      &=
        \sum_{i = 1}^{\lvert\mathcal{G}_f\rvert} f_{i} D^{\Gamma_f}_{i1}(\hat{R}_k) +
        \sum_{j = 1}^{\lvert\mathcal{G}_g\rvert} g_{j} D^{\Gamma_g}_{j1}(\hat{R}_k),
        \label{eq:psikexpansionasD}
    \end{align}
    where we have relabeled the dummy indices $i'$ to $i$ and $j'$ to $j$ in the second equality.
    Comparing (\ref{eq:psikexpansionasD}) to (\ref{eq:psikexpansion}), we deduce that
    \begin{equation*}
      \lambda^f_{ik} = D^{\Gamma_f}_{i1}(\hat{R}_k), \qquad
      \lambda^g_{jk} = D^{\Gamma_g}_{j1}(\hat{R}_k),
    \end{equation*}
    which says that the $k$\textsuperscript{th} columns of $\boldsymbol{\lambda}^f$ and $\boldsymbol{\lambda}^g$ are given by the first columns of the representation matrices for $\hat{R}_k$ in the bases $\mathcal{G}_f\cdot f$ and $\mathcal{G}_g\cdot g$, respectively.

    If $\Gamma_f$ and $\Gamma_g$ have non-identical representation matrices, then there must exist at least one $\hat{R}_k$ such that $\boldsymbol{D}^{\Gamma_f}(\hat{R}_k) \ne \boldsymbol{D}^{\Gamma_g}(\hat{R}_k)$.
    Consequently, $\boldsymbol{\lambda}^f \ne \boldsymbol{\lambda}^g$ and there is thus no constraint between $\boldsymbol{\lambda}^f$ and $\boldsymbol{\lambda}^g$.
    Therefore, we are guaranteed to be able to find $\mathcal{G}_{\psi}$ such that $\lvert\mathcal{G}_{\psi}\rvert = \lvert\mathcal{G}_f\rvert + \lvert\mathcal{G}_g\rvert$ and $\boldsymbol{\lambda}^{\psi}$ is a square invertible matrix.
    From (\ref{eq:Drelation}), we obtain
    \begin{equation*}
      \boldsymbol{D}^{\Gamma_{\psi}}(\hat{R})
      =
      (\boldsymbol{\lambda}^{\psi})^{-1}
      \begin{pmatrix}
        \boldsymbol{D}^{\Gamma_f}(\hat{R}) & \boldsymbol{0} \\
        \boldsymbol{0}                     & \boldsymbol{D}^{\Gamma_g}(\hat{R})
      \end{pmatrix}
      \boldsymbol{\lambda}^{\psi},
    \end{equation*}
    and subsequently,
    \begin{equation*}
      \tr \boldsymbol{D}^{\Gamma_{\psi}}(\hat{R}) =
      \tr \boldsymbol{D}^{\Gamma_f}(\hat{R})
      +
      \tr \boldsymbol{D}^{\Gamma_g}(\hat{R}),
    \end{equation*}
    or equivalently,
    \begin{equation*}
      \chi^{\Gamma_{\psi}}(\hat{R}) =
        \chi^{\Gamma_f}(\hat{R})
        +
        \chi^{\Gamma_g}(\hat{R}).
    \end{equation*}
    The above equation holds for all $\hat{R} \in \mathcal{G}$, from which it must follow that $\Gamma_{\psi} \sim \Gamma_f \oplus \Gamma_g$.

    On the other hand, if $\Gamma_f$ and $\Gamma_g$ have identical representation matrices in the sense that $\boldsymbol{D}^{\Gamma_f}(\hat{R}_k) = \boldsymbol{D}^{\Gamma_g}(\hat{R}_k) \equiv \boldsymbol{D}^{\Gamma}(\hat{R}_k)$ for all $\hat{R}_k \in \mathcal{G}_{\psi}$, then $\boldsymbol{\lambda}^f = \boldsymbol{\lambda}^g$ and it follows that the linear independence of the columns of $\boldsymbol{\lambda}^{\psi}$ is constrained by the linear independence of the columns of $\boldsymbol{\lambda}^f$ or $\boldsymbol{\lambda}^g$.
    Hence, $\lvert\mathcal{G}_{\psi}\rvert = \lvert\mathcal{G}_f\rvert = \lvert\mathcal{G}_g\rvert$ so that $\boldsymbol{\lambda}^{\psi}$ is rectangular and non-invertible.
    However, the identicality between $\boldsymbol{\lambda}^f$ and  $\boldsymbol{\lambda}^g$ enables us to write $\boldsymbol{\psi}^{\T}$ in (\ref{eq:psikexpansionmat}) as
    \begin{equation*}
      \boldsymbol{\psi}^{\T} =
        \begin{pmatrix}
          \boldsymbol{f}^{\T} + \boldsymbol{g}^{\T}
        \end{pmatrix}
        \bar{\boldsymbol{\lambda}}^{\psi}
    \end{equation*}
    where $\bar{\boldsymbol{\lambda}}^{\psi} = \boldsymbol{\lambda}^f = \boldsymbol{\lambda}^g$ is now a square invertible matrix.
    This gives two equations analogous to (\ref{eq:Rpsi}):
    \begin{align*}
      \hat{R}\boldsymbol{\psi}^{\T}
        &=
        \begin{pmatrix}
          \boldsymbol{f}^{\T} + \boldsymbol{g}^{\T}
        \end{pmatrix}
        \bar{\boldsymbol{\lambda}}^{\psi}
        \boldsymbol{D}^{\Gamma_{\psi}}(\hat{R}), \\
      \hat{R}\boldsymbol{\psi}^{\T}
        &=
        \begin{pmatrix}
          \boldsymbol{f}^{\T} + \boldsymbol{g}^{\T}
        \end{pmatrix}
        \boldsymbol{D}^{\Gamma}(\hat{R})
        \bar{\boldsymbol{\lambda}}^{\psi},
    \end{align*}
    the comparison of which results in
    \begin{equation*}
      \boldsymbol{D}^{\Gamma_{\psi}}(\hat{R}) =
      (\bar{\boldsymbol{\lambda}}^{\psi})^{-1}
      \boldsymbol{D}^{\Gamma}(\hat{R})
      \bar{\boldsymbol{\lambda}}^{\psi},
    \end{equation*}
    so that
    \begin{equation*}
      \chi^{\Gamma_{\psi}}(\hat{R}) = \chi^{\Gamma}(\hat{R}) = \chi^{\Gamma_f}(\hat{R}) = \chi^{\Gamma_g}(\hat{R})
    \end{equation*}
    for all $\hat{R} \in \mathcal{G}$.
    Hence, $\Gamma_{\psi} \sim \Gamma_f \sim \Gamma_g$.
  \end{proof}

  \begin{remark}
    If either one or both of $\Gamma_f$ and $\Gamma_g$ are reducible, they can be decomposed into irreducible components to which Proposition~\ref{prop:irrepsum} can be applied.
    It is then trivial to see that, in all cases, $\Gamma_{\psi}$ must contain two possibly non-disjoint subrepresentations, one of which is equivalent to $\Gamma_f$ and the other to $\Gamma_g$.
  \end{remark}

  Let us now consider a certain wavefunction $\Psi_0$ in a Hilbert space $\mathcal{H}$ and and another wavefunction $\Psi_1$ in its immediate neighborhood such that there exists $\delta\Psi$ that is linearly independent of $\Psi_0$ and that allows us to write
  \begin{equation*}
    \Psi_1 = c_0\Psi_0 + c_{\delta}\delta\Psi
  \end{equation*}
  where the coefficients $c_0$ and $c_{\delta}$ ensure that $\Psi_1$ is normalized.
  If $\Gamma_1$ and $\Gamma_0$ are the representations spanned by all symmetry-equivalent partners of $\Psi_1$ and $\Psi_0$ respectively, then, from the above proposition and the remark that follows, $\Gamma_1$ must contain a subrepresentation that is equivalent to $\Gamma_0$, \textit{i.e.}, $\Gamma_1 \supseteq \Gamma'_0 \sim \Gamma_0$.
  Consequently, as $\Psi_0$ varies smoothly along any pathway, if its representation changes, then it is either restricted to one of its subrepresentations or included as a subrepresentation of a larger representation.

\bibliography{bib/spatialsymreality}

\end{document}



  \title{On Symmetry and the Reality of Holomorphic Hartree--Fock Wavefunctions\\Supplementary Material}

  \author{Bang C. Huynh}
    \email{bang.huynh@nottingham.ac.uk}
    \affiliation{School of Chemistry, University of Nottingham, University Park, Nottingham NG7 2RD, United Kingdom}
    \affiliation{Yusuf Hamied Department of Chemistry, Lensfield Road, Cambridge CB2 1EW, United Kingdom}
  \author{Alex J. W. Thom}
    \affiliation{Yusuf Hamied Department of Chemistry, Lensfield Road, Cambridge CB2 1EW, United Kingdom}

  \date{\today}
  \maketitle

  \tableofcontents
  \clearpage


\section{Electron Integral Coefficients}

  \subsection{Overlap Integrals}

    \begin{table*}[h]
      \centering
      \caption{
        Functional forms of the overlap integral coefficients $S_1^{\pm}$ and $S_2^{\pm}$ in terms of the overlap integrals.
        Refer to Figure~{\ref{fig:symbreakingpathways}} in the main text for the labels of the hydrogen atoms, which are also the labels for the $1s$ atomic orbitals (\glspl*{acr:ao}) in STO-3G used in the integral notations.
        For example, $\braket{\mu|\nu}$ means $\braket{\varphi_{\mu}|\varphi_{\nu}}$, where $\varphi_{\mu}$ is the sole $1s$ \gls*{acr:ao} localised on hydrogen $\mu$.
      }
      \label{tab:overlapintegrals}
      \renewcommand{\arraystretch}{1.3}
      \resizebox{\textwidth}{!}{%
        \begin{tabular}{llll}
  \toprule
  Pathway & Constraining space & \multicolumn{2}{l}{Overlap integral coefficients}\\
  \midrule
  A
    & $\Psi\{\hat{s}_z, 0\}[\hat{i}]'$
    & $S_1^{-} = \frac{1}{2}(x^2+1)\braket{1|1} + x\braket{1|3}$
    & $S_2^{-} = \frac{1}{2}(x^2+1)\braket{1|2} + x\braket{1|4}$
  \\
   && $S_1^{+} = \frac{1}{2}(y^2+1)\braket{1|1} + y\braket{1|3}$
    & $S_2^{+} = \frac{1}{2}(y^2+1)\braket{1|2} + y\braket{1|4}$
  \\
  \addlinespace[6pt]
    & $\Psi\{\hat{s}_z, 0\}[\hat{\sigma}^{xz}]'$
    & $S_1^{-} = \frac{1}{2}(x^2+1)\braket{1|1} + x\braket{1|4}$
    & $S_2^{-} = \frac{1}{2}(x^2+1)\braket{1|2} + x\braket{1|3}$
  \\
   && $S_1^{+} = \frac{1}{2}(y^2+1)\braket{1|1} + y\braket{1|4}$
    & $S_2^{+} = \frac{1}{2}(y^2+1)\braket{1|2} + y\braket{1|3}$
  \\
  \addlinespace[12pt]
  B
    & $\Psi\{\hat{s}_z, 0\}[\hat{C}_2^{\clubsuit}]'$
    & $S_1^{-} = \frac{1}{2}(x^2+1)\braket{1|1} + x\braket{1|3}$
    & $S_2^{-} = \frac{1}{2}(x^2+1)\braket{1|2} + x\braket{1|4}$
  \\
   && $S_1^{+} = \frac{1}{2}(y^2+1)\braket{1|1} + y\braket{1|3}$
    & $S_2^{+} = \frac{1}{2}(y^2+1)\braket{1|2} + y\braket{1|4}$
  \\
  \bottomrule
\end{tabular}
      }
    \end{table*}

  \subsection{One-Electron Integrals}

    \begin{table*}[h]
      \centering
      \caption{
        Functional forms of the one-electron integral coefficients $A_1^{\pm}$ and $A_2^{\pm}$ in terms of the one-electron integrals.
        Refer to Figure~{\ref{fig:symbreakingpathways}} in the main text for the labels of the hydrogen atoms, which are also the labels for the $1s$ \glspl*{acr:ao} in STO-3G used in the integral notations.
        For example, $\braket{\mu|\hat{h}|\nu}$ means $\braket{\varphi_{\mu}|\hat{h}|\varphi_{\nu}}$, where $\varphi_{\mu}$ is the sole $1s$ \gls*{acr:ao} localised on hydrogen $\mu$.
      }
      \label{tab:oneelectronintegrals}
      \renewcommand{\arraystretch}{1.3}
      \resizebox{\textwidth}{!}{%
        \begin{tabular}{llll}
  \toprule
  Pathway & Constraining space & \multicolumn{2}{l}{One-electron integral coefficients}\\
  \midrule
  A
    & $\Psi\{\hat{s}_z, 0\}[\hat{i}]'$
    & $A_1^{-} = (x^2+1)\braket{1|\hat{h}|1} + 2x\braket{1|\hat{h}|3}$
    & $A_2^{-} = (x^2+1)\braket{1|\hat{h}|2} + 2x\braket{1|\hat{h}|4}$
  \\
   && $A_1^{+} = (y^2+1)\braket{1|\hat{h}|1} + 2y\braket{1|\hat{h}|3}$
    & $A_2^{+} = (y^2+1)\braket{1|\hat{h}|2} + 2y\braket{1|\hat{h}|4}$
  \\
  \addlinespace[6pt]
    & $\Psi\{\hat{s}_z, 0\}[\hat{\sigma}^{xz}]'$
    & $A_1^{-} = (x^2+1)\braket{1|\hat{h}|1} + 2x\braket{1|\hat{h}|4}$
    & $A_2^{-} = (x^2+1)\braket{1|\hat{h}|2} + 2x\braket{1|\hat{h}|3}$
  \\
   && $A_1^{+} = (y^2+1)\braket{1|\hat{h}|1} + 2y\braket{1|\hat{h}|4}$
    & $A_2^{+} = (y^2+1)\braket{1|\hat{h}|2} + 2y\braket{1|\hat{h}|3}$
  \\
  \addlinespace[12pt]
  B
    & $\Psi\{\hat{s}_z, 0\}[\hat{C}_2^{\clubsuit}]'$
    & $A_1^{-} = (x^2+1)\braket{1|\hat{h}|1} + 2x\braket{1|\hat{h}|3}$
    & $A_2^{-} = (x^2+1)\braket{1|\hat{h}|2} + 2x\braket{1|\hat{h}|4}$
  \\
   && $A_1^{+} = (y^2+1)\braket{1|\hat{h}|1} + 2y\braket{1|\hat{h}|3}$
    & $A_2^{+} = (y^2+1)\braket{1|\hat{h}|2} + 2y\braket{1|\hat{h}|4}$
  \\
  \bottomrule
\end{tabular}
      }
    \end{table*}
    \clearpage

    \subsection{Two-Electron Integrals}
      \clearpage
      \begin{turnpage}
        \begin{table*}[h]
          \centering
          \caption{
            Functional forms of the two-electron integral coefficients $A_3$, $A_4^{\pm}$, $A_5$, and $A_6$ in terms of the two-electron integrals.
            Refer to Figure~{\ref{fig:symbreakingpathways}} in the main text for the labels of the hydrogen atoms, which are also the labels for the $1s$ \glspl*{acr:ao} in STO-3G used in the integral notations.
            For example, $\braket{\mu\nu|\mu'\nu'}$ means $\braket{\varphi_{\mu}\varphi_{\nu}|\varphi_{\mu'}\varphi_{\nu'}}$, where $\varphi_{\mu}$ is the sole $1s$ \gls*{acr:ao} localised on hydrogen $\mu$.
            Physicists' notation is used throughout.
          }
          \label{tab:twoelectronintegrals}
          \renewcommand{\arraystretch}{1.3}
          \resizebox{\linewidth}{!}{%
            \begin{tabular}{%
    lll%
  }
  \toprule
  Path. & Constraining space & Two-electron integral coefficients\\
  \midrule
  A
    & $\Psi\{\hat{s}_z, 0\}[\hat{i}]'_{xy}$
    & $A_3 =
          (x^2y^2+1)  \braket{11|11}
        + 2(x+y)(xy+1)\braket{11|13}
        + 4xy         \braket{11|33}
        + (x^2+y^2)   \braket{13|13}
      $
  \\
   && $A_4^{+} =
          (x^2y^2+1)\braket{11|12}
        + (x^2+1)y  \braket{11|14}
        + (y^2+1)x^2\braket{11|23}
        + 4xy       \braket{11|34}
        + (x^2+1)y  \braket{12|13}
        + (x^2+y^2) \braket{13|14}
      $
  \\
   && $A_4^{-} =
          (x^2y^2+1)\braket{11|12}
        + (y^2+1)x  \braket{11|14}
        + (x^2+1)y^2\braket{11|23}
        + 4xy       \braket{11|34}
        + (y^2+1)x  \braket{12|13}
        + (x^2+y^2) \braket{13|14}
      $
  \\
   && $A_5 =
          (x^2y^2+1)  \braket{12|12}
        + 2(x+y)(xy+1)\braket{12|14}
        + 4xy         \braket{12|34}
        + (x^2+y^2)   \braket{14|14}
      $
  \\
   && $A_6 =
          (x^2y^2+1)  \braket{11|22}
        + 2(x+y)(xy+1)\braket{11|24}
        + 2xy         \braket{11|44}
        + 2xy         \braket{12|43}
        + (x^2+y^2)   \braket{13|24}
      $
  \\
  \addlinespace[6pt]
    & $\Psi\{\hat{s}_z, 0\}[\hat{\sigma}^{xz}]'_{xy}$
    & $A_3 =
          (x^2y^2+1)  \braket{11|11}
        + 2(x+y)(xy+1)\braket{11|14}
        + 4xy         \braket{11|33}
        + (x^2+y^2)   \braket{14|14}
      $
  \\
   && $A_4^{+} =
          (x^2y^2+1)\braket{11|12}
        + (x^2+1)y  \braket{11|13}
        + (y^2+1)x^2\braket{11|24}
        + 4xy       \braket{11|34}
        + (x^2+1)y  \braket{12|14}
        + (x^2+y^2) \braket{13|14}
      $
  \\
   && $A_4^{-} =
          (x^2y^2+1)\braket{11|12}
        + (y^2+1)x  \braket{11|13}
        + (x^2+1)y^2\braket{11|24}
        + 4xy       \braket{11|34}
        + (y^2+1)x  \braket{12|14}
        + (x^2+y^2) \braket{13|14}
      $
  \\
   && $A_5 =
          (x^2y^2+1)  \braket{12|12}
        + 2(x+y)(xy+1)\braket{12|13}
        + 4xy         \braket{12|43}
        + (x^2+y^2)   \braket{13|13}
      $
  \\
   && $A_6 =
          (x^2y^2+1)  \braket{11|22}
        + 2(x+y)(xy+1)\braket{11|23}
        + 2xy         \braket{11|33}
        + 2xy         \braket{12|34}
        + (x^2+y^2)   \braket{13|24}
      $
  \\
  \addlinespace[12pt]
  B
    & $\Psi\{\hat{s}_z, 0\}[\hat{C}_2^{\clubsuit}]'_{xy}$
    & $A_3 =
          (x^2y^2+1)  \braket{11|11}
        + 2(x+y)(xy+1)\braket{11|13}
        + 4xy         \braket{11|33}
        + (x^2+y^2)   \braket{13|13}
      $
  \\
   && $A_4^{+} =
          (x^2y^2+1)\braket{11|12}
        + (x^2+1)y  \braket{11|14}
        + (y^2+1)x^2\braket{11|23}
        + 4xy       \braket{11|34}
        + (x^2+1)y  \braket{12|13}
        + (x^2+y^2) \braket{13|14}
      $
  \\
   && $A_4^{-} =
          (x^2y^2+1)\braket{11|12}
        + (y^2+1)x  \braket{11|14}
        + (x^2+1)y^2\braket{11|23}
        + 4xy       \braket{11|34}
        + (y^2+1)x  \braket{12|13}
        + (x^2+y^2) \braket{13|14}
      $
  \\
   && $A_5 =
          (x^2y^2+1)  \braket{12|12}
        + 2(x+y)(xy+1)\braket{12|14}
        + 4xy         \braket{12|34}
        + (x^2+y^2)   \braket{14|14}
      $
  \\
   && $A_6 =
          (x^2y^2+1)  \braket{11|22}
        + 2(x+y)(xy+1)\braket{11|24}
        + 2xy         \braket{11|44}
        + 2xy         \braket{12|43}
        + (x^2+y^2)   \braket{13|24}
      $
  \\
  \bottomrule
\end{tabular}
          }
        \end{table*}
      \end{turnpage}
      \clearpage
      \global\pdfpageattr\expandafter{\the\pdfpageattr/Rotate 90}


\section{Molecular Orbitals and Symmetry of Solutions}

  The evolutions of the \glspl*{acr:mo} for the eight solutions under the extrinsic constraints along the two molecular symmetry pathways considered are animated in the chapters of the included \texttt{mp4} video titled \texttt{spatialsymreality.orbitalplots.mp4}.
  Table~\ref{tab:orbsymvideos} relates the chapter names to the special extrinsic constraints discussed in the main article.
  Each \gls*{acr:mo} is a linear combination of four $1s$ \glspl*{acr:ao} localized on the four hydrogen atoms.
  In these animations, the magnitudes of the linear combination coefficients are proportional to the radii of the circles used to represent the $1s$ \glspl*{acr:ao}.
  The phases of these coefficients are depicted by the color hues in the HSV model, and also by the angular position of a black dot on the circumferences of the \gls*{acr:ao} circles.
  When the coefficient is entirely real, the circumference is bolded, and when the coefficient is entirely imaginary, the circumference is dotted.

  \begin{table*}[h]
    \centering
    \caption{
      Video chapters of the animations for the \glsxtrshort*{acr:mo} evolution of the analytic solutions obtained under the symmetry-conserved extrinsic constraints considered in this work.
      These chapters have been encoded individually by \texttt{ffmpeg} in the \texttt{H.264} standard using the \texttt{x264} encoder with the \texttt{yuv420p} pixel format, and then merged together as part of a single video file entitled \texttt{spatialsymreality.orbitalplots.mp4}.
      This video can be opened with any common video viewer, but for the best viewing experience, the open-source VLC Media Player should be used which allows for the easy navigation between video chapters and which preserves the potentially different aspect ratios of the different chapters.
      During playback, the embedded subtitles can be turned on or off to show or hide the chapter names.
      See Table~\ref{tab:pathway-a-orbitals-i} in the main article for an explanation of how to interpret the areas and colors of the orbital plots.
    }
    \label{tab:orbsymvideos}
    \renewcommand{\arraystretch}{1.3}
    \resizebox{0.95\textwidth}{!}{%
      \begin{tabular}{llll}
        \toprule
        Pathway &  Extrinsic constraint & Video chapter name & Start time\\
        \midrule
        A
          & $\left\lvert \alpha^g \beta^g \right\rvert$
          & \texttt{chapter1.D2h.rectangle.i.gg.orbitalplots}
          & 00:00:00.000\\
          & $\left\lvert \alpha^u \beta^g \right\rvert$
          & \texttt{chapter2.D2h.rectangle.i.ug.orbitalplots}
          & 00:00:18.101\\
          & $\left\lvert \alpha^u \beta^u \right\rvert$
          & \texttt{chapter3.D2h.rectangle.i.uu.orbitalplots}
          & 00:00:36.201\\
        \addlinespace[6pt]
          & $\left\lvert \alpha' \beta' \right\rvert$
          & \texttt{chapter4.D2h.rectangle.sigma.dashdash.orbitalplots}
          & 00:00:54.301\\
          & $\left\lvert \alpha'' \beta' \right\rvert$
          & \texttt{chapter5.D2h.rectangle.sigma.ddashdash.orbitalplots}
          & 00:01:12.401\\
          & $\left\lvert \alpha'' \beta'' \right\rvert$
          & \texttt{chapter6.D2h.rectangle.sigma.ddashddash.orbitalplots}
          & 00:01:30.501\\
        \addlinespace[12pt]
        B
          & $\left\lvert \alpha^+ \beta^+ \right\rvert$
          & \texttt{chapter7.D2.C2z-like.pp.orbitalplots}
          & 00:01:48.601\\
          & $\left\lvert \alpha^- \beta^+ \right\rvert$
          & \texttt{chapter8.D2.C2z-like.mp.orbitalplots}
          & 00:02:08.701\\
          & $\left\lvert \alpha^- \beta^- \right\rvert$
          & \texttt{chapter9.D2.C2z-like.mm.orbitalplots}
          & 00:02:28.801\\
        \bottomrule
      \end{tabular}
    }
  \end{table*}
\clearpage
\global\pdfpageattr\expandafter{\the\pdfpageattr/Rotate 00}

\section{Weak Reality Requirements and Coalescence Boundary Locations}

  Under the condition that $\Delta$ is non-negative, which holds for all cases investigated in this work, the radical forms of $u_{\sigma}$ and $v_{\sigma}$ (Table~\ref{tab:allsols} in the main text) imply that they are either entirely real or entirely imaginary.
  This considerably simplifies the expression for the reality indicator $\rho \equiv \rho_{\alpha}$ (Equation~\ref{eq:indicator} in the main text).
  For real \gls*{acr:scf} solutions, we require that $\rho = 0$, or equivalently, $e^{-\rho} = 1$, which we shall use from here on to simplify the notations.
  There are four cases to consider.

  \begin{enumerate}[label=\textbf{Case \arabic*:}, ref=\arabic*, align=left]
    \item $u_{\alpha}$ and $v_{\alpha}$ are both real.
      Straightforwardly, we can write $\Re u_{\alpha} = u_{\alpha}$ and $\Re v_{\alpha} = v_{\alpha}$ while $\Im u_{\alpha} = 0$ and $\Im v_{\alpha} = 0$.
      Furthermore, from Table~\ref{tab:allsols} in the main text, we identify $u_{\alpha}$ with $u^{-\mp}$ and $v_{\alpha}$ with $v^{-\pm}$ and the reality indicator simplifies to
        \begin{equation*}
          e^{-\rho}
            = u_{\alpha}^2 + v_{\alpha}^2
            = (u^{-\mp})^2 + (v^{-\pm})^2
            =   \frac{D_1^{-+} \mp D_2^{-}\sqrt{\Delta}}{D_1^{--} + D_1^{-+}}
              + \frac{D_1^{--} \pm D_2^{-}\sqrt{\Delta}}{D_1^{--} + D_1^{-+}}
            = 1,
        \end{equation*}
      where we have substituted the expressions for $u^{-\mp}$ and $v^{-\pm}$ from Table~\ref{tab:allsols} in the main text for the third equality.
      This identity holds whatever the values of the electron integral coefficients, thus requiring the \gls*{acr:scf} solutions to be identically real.

    \item $u_{\alpha}$ is real, $v_{\alpha}$ is imaginary.
      This implies that $\Re u_{\alpha} = u_{\alpha}$ and $\Im v_{\alpha} = -iv_{\alpha}$ while $\Im u_{\alpha} = 0$ and $\Re v_{\alpha} = 0$, and subsequently, $e^{-\rho} = (u_{\alpha} {\color{red}\ \pm\ } iv_{\alpha})^2$.
      We note that, for the red ${\color{red}\pm}$, the top sign corresponds to the unprimed variant of the transiently real solutions, and the bottom sign the primed variant.
      We then  have
        \begin{align*}
          e^{-\rho}
            &= (u^{-\mp})^2 - (v^{-\pm})^2 {\color{red}\ \pm\ } 2iu^{-\mp}v^{-\pm}\\
            &=   \frac{D_1^{-+} \mp D_2^{-}\sqrt{\Delta}}{D_1^{--} + D_1^{-+}}
               - \frac{D_1^{--} \pm D_2^{-}\sqrt{\Delta}}{D_1^{--} + D_1^{-+}}
               {\color{red}\ \pm\ } 2i \frac{
                      \sqrt{D_1^{-+} \mp D_2^{-}\sqrt{\Delta}}
                      \sqrt{D_1^{--} \pm D_2^{-}\sqrt{\Delta}}
                    }{\sqrt{D_1^{--} + D_1^{-+}}\sqrt{D_1^{--} + D_1^{-+}}}\\
            &=   \frac{
                   D_1^{-+} - D_1^{--} \mp 2D_2^{-}\sqrt{\Delta}
                   {\color{red}\ \pm\ } 2i \sqrt{(D_1^{-+} \mp D_2^{-}\sqrt{\Delta})
                              (D_1^{--} \pm D_2^{-}\sqrt{\Delta})}
                 }{D_1^{--} + D_1^{-+}}.
        \end{align*}
      For \gls*{acr:scf} solution reality, we require
        \begin{equation*}
          D_1^{-+} - D_1^{--} \mp 2D_2^{-}\sqrt{\Delta}
          {\color{red}\ \pm\ } 2i \sqrt{(D_1^{-+} \mp D_2^{-}\sqrt{\Delta})
                     (D_1^{--} \pm D_2^{-}\sqrt{\Delta})}
          = D_1^{--} + D_1^{-+}
        \end{equation*}
      with $D_1^{--} + D_1^{-+} \ne 0$, or equivalently,
        \begin{equation}
          {\color{red}\pm}i \sqrt{(D_1^{-+} \mp D_2^{-}\sqrt{\Delta})
                  (D_1^{--} \pm D_2^{-}\sqrt{\Delta})}
          = D_1^{--} \pm D_2^{-}\sqrt{\Delta}.
          \label{eq:urealvimagconds}
        \end{equation}
      Squaring both sides and factorising out common terms, we obtain
        \begin{equation*}
          D_1^{--} \pm D_2^{-}\sqrt{\Delta} = 0
        \end{equation*}
      which satisfies both ${\color{red}\pm}$ variations of Equation~\ref{eq:urealvimagconds} and when substituted into the expressions for $u^{-\mp}$ and $v^{-\pm}$ yields
        \begin{equation}
          u^{-\mp} = +1 \quad \textrm{and} \quad v^{-\pm} = 0
          \label{eq:urealvimagconds-uv}
        \end{equation}
      for both unprimed and primed variants.
      \label{case:urealvimag}

    \item $u_{\alpha}$ is imaginary, $v_{\alpha}$ is real.
      Here, we have $e^{-\rho} = (-iu_{\alpha} {\color{red}\ \pm\ } v_{\alpha})^2$.
      Following the same approach as in Case~\ref{case:urealvimag}, we get
        \begin{equation*}
          {\color{red}\pm}i \sqrt{(D_1^{-+} \mp D_2^{-}\sqrt{\Delta})
                  (D_1^{--} \pm D_2^{-}\sqrt{\Delta})}
          = -(D_1^{-+} \mp D_2^{-}\sqrt{\Delta}),
        \end{equation*}
      with $D_1^{--} + D_1^{-+} \ne 0$, which implies that
        \begin{equation}
          u^{-\mp} = 0 \quad \textrm{and} \quad v^{-\pm} = +1
          \label{eq:uimagvrealconds-uv}
        \end{equation}
      for both unprimed and primed variants.

    \item $u_{\alpha}$ and $v_{\alpha}$ are both imaginary.
      The reality indicator now becomes
        \begin{equation*}
          e^{-\rho}
            = (-iu^{-\mp})^2 + (-iv^{-\pm})^2
            = - \frac{D_1^{-+} \mp D_2^{-}\sqrt{\Delta}}{D_1^{--} + D_1^{-+}}
              - \frac{D_1^{--} \pm D_2^{-}\sqrt{\Delta}}{D_1^{--} + D_1^{-+}}
            = -1,
        \end{equation*}
      which is forbidden for real $\rho$.
      This case is therefore impossible.
      In other words, we cannot have
        \begin{equation*}
          \frac{D_1^{-+} \mp D_2^{-}\sqrt{\Delta}}{D_1^{--} + D_1^{-+}} < 0
          \quad \textrm{and} \quad
          \frac{D_1^{--} \pm D_2^{-}\sqrt{\Delta}}{D_1^{--} + D_1^{-+}} < 0
        \end{equation*}
      simultaneously for real $D$ and non-negative $\Delta$.
  \end{enumerate}

  Combining the results from the four cases, we deduce that, for the transiently real solutions to be real, $u_{\alpha}$ and $v_{\alpha}$ must both be real, and if one of them is $+1$ then the other must vanish.
  This is achieved when
  \begin{equation}
    \frac{D_{1}^{-+} \mp D_{2}^{-}\sqrt{\Delta}}{D_{1}^{--} + D_{1}^{-+}} \ge 0 \quad \textrm{and} \quad
    \frac{D_{1}^{--} \pm D_{2}^{-}\sqrt{\Delta}}{D_{1}^{--} + D_{1}^{-+}} \ge 0
  \end{equation}
  These are in fact the conditions given by Equation~\ref{eq:weakrealityineqs} in the main text.

  In addition, the real--non-real transition takes place when either $u_{\alpha}$ or $v_{\alpha}$ switches from being entirely real to being entirely imaginary.
  If the $D$ and $\Delta$ coefficients vary smoothly such that $u_{\alpha}$ and $v_{\alpha}$ also vary smoothly, then this switch can only occur at
    \begin{equation*}
      u_{\alpha} = 0 \quad \textrm{or} \quad v_{\alpha} = 0,
    \end{equation*}
  which is equivalent to
    \begin{equation*}
      D_1^{-+} \mp D_2^{-}\sqrt{\Delta} = 0
      \quad \textrm{or} \quad
      D_1^{--} \pm D_2^{-}\sqrt{\Delta} = 0.
    \end{equation*}
  This gives implicit equations for the locations of the coalescence boundaries as stated in Equation~\ref{eq:boundariesDcoeffs} in the main text.
